\documentclass[%
 superscriptaddress,
 amsmath,amssymb,twocolumn,
 aps,
pra,
]{revtex4-2}

\usepackage{graphicx}
\usepackage{dcolumn}
\usepackage{bm}
\usepackage[dvipsnames]{xcolor}
\usepackage[colorlinks=true,linkcolor=blue,citecolor={blue}]{hyperref}   

\begin{document}
\newcommand{\vP}{\mathbf{P}}
\newcommand{\vers}{\mathbf{r''}}
\newcommand{\vRs}{\mathbf{R''}}
\newcommand{\verp}{\mathbf{r'}}
\newcommand{\vRp}{\mathbf{R'}}
\newcommand{\Wcm}{\;\mathrm{W/cm}^2}
\newcommand{\eV}{\;\mathrm{eV}}
\newcommand{\Rep}{\mathrm{Re}\,}
\newcommand{\Imp}{\mathrm{Im}\,}
\newcommand{\vk}{{\mathbf{k}}}
\newcommand{\vkst}{\mathbf{k}_\textrm{st}}
\newcommand{\vqst}{\mathbf{q}_\textrm{st}}
\newcommand{\vi}{\hat{\mathbf{i}}}
\newcommand{\vj}{\hat{\mathbf{j}}}
\newcommand{\vS}{{\mathbf S}}
\newcommand{\vH}{\mathbf{H}}
\newcommand{\vv}{\mathbf{v}}
\newcommand{\ve}{\hat{\mathbf{e}}}
\newcommand{\0}{\mathbf{0}}
\newcommand{\vE}{\mathbf{E}}
\newcommand{\vA}{\mathbf{A}}
\newcommand{\vT}{\mathbf{T}}
\newcommand{\ver}{\mathbf{r}}
\newcommand{\vd}{\mathbf{d}}
\newcommand{\va}{\mathbf{a}}
\newcommand{\vD}{\mathbf{D}}
\newcommand{\vp}{\mathbf{p}}
\newcommand{\vR}{\mathbf{R}}
\newcommand{\vq}{\mathbf{q}}
\newcommand{\vrho}{\mbox{\boldmath{$\rho$}}}
\newcommand{\del}{\mbox{\boldmath{$\nabla$}}}
\newcommand{\valpha}{\mbox{\boldmath{$\alpha$}}}
\newcommand{\vRR}{\{\vR\}}
\newcommand{\vRRp}{\{\vRp\}}
\newcommand{\Ip}{I_\mathrm{p}}
\newcommand{\Up}{U_\mathrm{p}}
\newcommand{\calT}{\mathbf{\cal T}}
\newcommand{\calF}{\mathbf{\cal F}}
\newcommand{\calU}{\mathbf{\cal U}}
\newcommand{\et}{\tilde{e}}
\newcommand{\cm}{\mathrm{c.m.}}
\newcommand{\vro}{\mathbf{\rho}}
\newcommand{\tos}{t_{0s}}
\newcommand{\ts}{t_{s}}
\newcommand{\Ep}{E_{\mathbf{p}}}


\title{Energy-conservation conditions in saddle-point approximation for the strong-field-ionization of atoms}

\author{T. Rook}
\affiliation{Department of Physics and Astronomy, University College London, Gower Street, London, WC1E 6BT, United Kingdom}
\affiliation{Department of Physics, University of Oxford, Clarendon Laboratory, Parks Road, Oxford, OX1 3PU, United Kingdom}
\author{D. Habibovi\'{c}}
\affiliation{University of Sarajevo, Faculty of Science, Zmaja od Bosne 35, 71000 Sarajevo, Bosnia and Herzegovina}
\author{C. Figueira de Morisson Faria}
\affiliation{Department of Physics and Astronomy, University College London, Gower Street, London, WC1E 6BT, United Kingdom}

\date{\today}

\begin{abstract}

Orbit-based methods are widespread in strong-field laser-matter interaction. They provide a framework in which photoelectron momentum distributions can be interpreted as the quantum interference between different semi-classical pathways the electron can take on its way to the detector, which brings with it great predictive power.  The transition amplitude of an electron going from a bound state to a final continuum state is often written as multiple integrals, which can be computed either numerically, or by employing the saddle-point method. 
If one computes the momentum distribution via a saddle-point method, 
the obtained distribution is highly dependent on the time window from which the saddle points are selected for inclusion in the "sum over paths". In many cases, this leads to the distributions not even satisfying the basic symmetry requirements and often containing many more oscillations and interference fringes than their numerically integrated counterparts. Using the strong-field approximation, we find that the manual enforcement of the energy-conservation condition on the momentum distribution calculated via the saddle-point method provides a unique momentum distribution which satisfies the symmetry requirements of the system and which is in a good agreement with the numerical results. We illustrate our findings using the example of the Ar atom ionized by a selection of monochromatic and bichromatic linearly polarized fields.    
\end{abstract}

\maketitle

\section{\label{sec:intro}Introduction}
When atoms are exposed to the laser field with an intensity comparable to the intensity of the Coulomb field experienced by the electron, various nonlinear processes may occur. The archetypal example of these processes is above-threshold ionization (ATI) where the electron is released from the atom either through tunnelling or through multiphoton ionization, absorbing more photons than is necessary for ionization \cite{Agostini1979}. This phenomenon has been modelled by a myriad of methods, which relate the quantum mechanical pathways in ATI to interfering electron orbits. Examples of orbit-based approaches that retain quantum interference and tunneling are the Strong-Field Approximation (SFA) \cite{Milosevic2003,Popruzhenko2014a,Amini2019}, the Trajectory-based Coulomb SFA (TC-SFA) \cite{Yan2010,Yan2012}, the Quantum Trajectory Monte Carlo (QTMC) \cite{Li2014,Geng2014,Li2014c,richter2015streaking,Xie2016,Li2016PRA,Liu2016}, the Semiclassical Two-Step Model (SCTM) \cite{Lein2016,shvetsov2021semiclassical,Shilovski2018}, and the Coulomb Quantum-Orbit Strong-Field Approximation (CQSFA) \cite{Lai2017,Maxwell2017,Maxwell2017a,Maxwell2018,Maxwell2018b,Rook2022,rodriguez2023,Rook2024}. The most traditional and widespread of the orbit-based models employed in strong-field physics is the SFA (for reviews see \cite{Milosevic2003,Popruzhenko2014a,Amini2019}). The SFA is based on the assumption that the freed electron does not interact with the parent ion upon eventual rescattering \cite{Milosevic2003}. The photoelectrons which do not exhibit additional interaction with the core are denoted as the direct electrons, while the electrons which exhibit rescattering are denoted as the rescattered electrons.  

Within the above-mentioned framework, the quantum-mechanical transition amplitudes for ATI and other strong-field processes manifest as multidimensional integrals of highly oscillatory functions (for reviews on strong-field processes see Refs.~\cite{Becker2002,Agostini2004,Scrinzi2006,Lein2007,Krausz2009,Popruzhenko2014a,Amini2019,Faria2020}). These integrals can be solved numerically or through the utilization of asymptotic methods. The later approach allows to identify the partial contributions of different quantum pathways which correspond to different solutions of the saddle-point equations \cite{Salieres2001,Milosevic2024TR}, and also simplifies the numerical computations considerably. During the last three decades, the saddle-point method was thoroughly used to calculate the transition amplitudes for different strong-field processes.  

Many examples are found in the SFA. For instance, for the high-order harmonic generation, the SFA transition amplitude is five-dimensional and the saddle-point solutions for the ionization and recombination times appear in pairs [well-known long and short orbits introduced in Ref~\cite{Lewenstein1994}]. The classification of these solutions was introduced in \cite{Milosevic2002} for linearly polarized monochromatic field and in  \cite{Odzak2005} for the case when an additional static field is present. Similar classification can be used for the saddle-point solutions of the ATI process with one additional rescattering if the driving field is linearly polarized monochromatic field. For the driving fields which evolve in the plane, this classification has to be improved as explained in Secs.~6 and 7 in Ref.~\cite{Milosevic2024TR}. Using the saddle-point method, the characteristics of the rescattering photoelectron spectra were analyzed by many authors and groups \cite{Zhou2021,Guo2022,Boroumand2022}, as well as the properties of the generated high-order harmonics \cite{Habibovic2020,Fang2021,Milosevic2023}. Furthermore, this method is particularly beneficial for studying correlated two-electron problems, such as inconsequential double ionization. It was applied to the electron-impact ionization \cite{Faria2004,Faria2005} and to the recollision-excitation with subsequent tunneling ionization mechanism of the nonsequential double ionization, in which case the correlated transition amplitude includes five and six-dimensional integrals, respectively \cite{Shaaran2010,Shaaran2012,Maxwell2015,Hashim2024}.   Finally, saddle-point methods have been used to calculate transition amplitudes for  strong-field ionization in a field parameter range for which the dipole approximation becomes questionable, providing a clear picture of the nondipole effects \cite{Madsen2022a,Madsen2022b,Kahvedzic2022,Jasarevic2024}.

Beyond the SFA, the saddle-point method was extensively employed in orbit-based approaches such as the Coulomb-Corrected Strong-Field Approximation (CCSFA) \cite{Popruzhenko2008}, the Eikonal-Volkov Approximation (EVA) \cite{Torlina2012,Kaushal2013} the TC-SFA \cite{Yan2010,Yan2012}, the SCTM \cite{Lein2016,shvetsov2021semiclassical,brennecke2020gouy} and the CQSFA \cite{Lai2015,Lai2017,Maxwell2017,Maxwell2017a,Maxwell2018,Maxwell2018b,Rook2022,rodriguez2023,Rook2024,Carlsen2024}. In particular, it has provided unprecedented insight into how holographic patterns form, by enabling the analysis of specific interfering quantum pathways \cite{Lai2017,Maxwell2017,Maxwell2017a}. These predictions have been backed by subsequent experiments \cite{Maxwell2019,Kang2020,Werby2021,Werby2022}.
Nonetheless, the patterns encountered using saddle-point method exhibited ambiguities and artefacts, that arose from the need for considering a finite interval for the ionization times that started and ended at specific values \cite{Arbo2012,Maxwell2017,Maxwell2018}. These ambiguities were particularly critical if a single cycle was taken into consideration, but would eventually be eliminated once several cycles were incorporated in the calculation. However, including several cycles leads to above-threshold ionization rings that obfuscate the remaining patterns. Problems may also arise if, for a specific ionization time window, the resulting spectra or photoelectron momentum distributions (PMDs) do not fulfil the expected physical symmetries that are inferred from the field.  Several strategies have been devised to tackle these issues. One may, for instance, choose an ionization time window consistent with the field symmetries, but this choice remains arbitrary. A more sophisticated procedure is to average incoherently over ionization probabilities computed using different unit cells for the ionization time windows \cite{Werby2021,Rook2022}. Still, the question remains whether there are underlying physical constraints that can be used instead to ensure compliance with the numerical solution of the multi-dimensional integrals.   

When the saddle-point method is employed to calculate the corresponding transition amplitude, the photoelectron energy is treated as a continuous parameter. For the long driving field with a flat envelope, different saddle-point solutions for the ionization time appear within one optical cycle. The number of these solutions depends on the type of the applied field \cite{Jasarevic2020}. In this paper, we investigate how the partial contributions of different saddle-point solutions should be combined to obtain the results which are in agreement with those obtained by the numerical integration. We show that, in order for the agreement to be good, the energy-conservation condition has to be imposed even though it is not naturally incorporated in the saddle-point method. Here, we focus on ATI without rescattering events and employ the SFA. This is the simplest possible scenario, which facilitates a comparison between the outcome of the saddle-point methods and the numerical integration of the transition amplitudes.  

The paper is organized as follows. In Sec.~\ref{sec:theory} we recall the SFA theory and discuss the saddle-point method. In Secs.~\ref{sec:mono} and \ref{sec:bi} we present our numerical results, in  Sec.~\ref{sec:mono}  for a monochromatic linearly polarized field, and in Sec.~\ref{sec:bi} for bichromatic linearly polarized fields. Finally, in Sec.~\ref{sec:conc} we present our main conclusions. Atomic units are used  unless otherwise stated. 

\section{\label{sec:theory}Theory}

\subsection{\label{sub:sfa}Strong-Field Approximation}
Our driving field is the infinitely long laser field 
\begin{equation}
\vE(t)=[E_1\sin(r\omega t)+E_2\sin(s\omega t+\phi)]\ve_x,
\end{equation}
where $E_j$, $j=1,2$ are the amplitudes of the laser-field components, $r$ and $s$ are integer, $\omega$ is the fundamental frequency, and  $\phi$ is the relative phase. We also define the ratio of the intensities of the field components as $\xi=E_2^2/E_1^2$. The  unit vector $\ve_x$ defines the laser-field polarization  direction.

In the framework of the SFA, the probability amplitude for the transition from the initial bound state $|\psi_0(t)\rangle$ to the continuum state $|\psi_{\vp}(t)\rangle$ with final momentum $\vp$ is
\begin{equation}
 M_\vp=\lim_{\substack{t\rightarrow \infty \\ t'\rightarrow -\infty}}\langle\psi_{\vp}(t)|U(t,t')|\psi_0(t')\rangle,
\end{equation}
where $U(t,t')$ is the evolution operator which corresponds to the total Hamiltonian $H(t)=H_0+H_{\text{int}}(t)$ with the time-independent part $H_0$ and the time-dependent part $H_{\text{int}}(t)=\ver\cdot\vE(t)$ which is present due to the interaction of the electron with a laser field. Here $\vE(t)$ is the electric field and the interaction is treated in the length gauge. Using the Dyson equation 
\begin{equation}
U(t,t')=U_0(t,t')-i\int_{t'}^t dt_0U(t,t_0)H_{\text{int}}(t_0)U_0(t_0,t'),
\end{equation}
where $U_0(t,t')$ corresponds to the Hamiltonian $H_0$, the probability amplitude becomes
\begin{equation}
 M_\vp=-i\lim_{t\rightarrow \infty}\int_{-\infty}^t dt_0\langle\psi_{\vp}(t)|U(t,t_0)\ver\cdot\vE(t_0)|\psi_0(t_0)\rangle.
\end{equation}
The evolution operator $U(t,t_0)$ also satisfies 
\begin{equation}\label{eq:Dysexp}
U(t,t_0)=U_{\text{las}}(t,t_0)-i\int_{t_0}^tdt_1 U(t,t_1)V(\ver)U_{\text{las}}(t_1,t_0),
\end{equation}
where $U_{\text{las}}(t,t_0)$ corresponds to the Hamiltonian of the free electron in the laser field and $V(\ver)$ is the atomic potential. Taking into account only the first term of the expansion \eqref{eq:Dysexp}, we obtain the probability amplitude for the direct electrons
\begin{equation}
 M_\vp^{(0)}=-i\int_{-\infty}^{\infty} dt_0\langle\chi_{\vp}(t_0)|\ver\cdot\vE(t_0)|\psi_0(t_0)\rangle,
\end{equation}
where the final states are the Volkov states $|\chi_{\vp}(t)\rangle=|\vp+\vA(t)\rangle e^{-iS_{\vp}(t)}$ in which $\vA(t)=-\int^t\vE(t')dt'$  is the vector potential and $|\vq\rangle$ is the plane wave. Furthermore, 
\begin{equation}
S_{\vp}(t)=\int^tdt'[\vp+\vA(t')]^2/2=(E_{\vp}+U_p)t+\vp\cdot\valpha+\mathcal{U}_1(t),
\end{equation}
where  $U_p$ is the ponderomotive energy, $E_{\vp}=\vp^2/2$, $\valpha=\int^tdt'\vA(t')$, and $\mathcal{U}_1(t)=\mathcal{U}(t)-U_pt$ with $\mathcal{U}(t)=\int^tdt'\vA^2(t')/2$. Taking into account that $|\psi_0(t)\rangle=|\psi_0\rangle e^{iI_pt}$, with $I_p$ being the ionization potential, the probability amplitude becomes
\begin{eqnarray}
 M_\vp^{(0)}&=&-i\int_{-\infty}^{\infty} dt_0\langle\vp+\vA(t_0)|\ver\cdot\vE(t_0)|\psi_0\rangle \nonumber\\
 &&\times e^{i(E_\vp+U_p+I_p)t_0}e^{i[\vp\cdot\valpha(t_0)+\mathcal{U}_1(t_0)]}.
\end{eqnarray}
The $T$-periodic part of the probability amplitude can be written as 
\begin{equation}
\mathcal{T}_\vp^{(0)}(t)=\sum_{n=-\infty}^{\infty}T_\vp^{(0)}(n)e^{-in\omega t},
\end{equation}
with $T_\vp^{(0)}(n)=\int_0^Tdt\mathcal{T}_\vp^{(0)}(t)e^{in\omega t}/T$ being the  expansion coefficients, and $\omega$ and $T=2\pi/\omega$ being the fundamental frequency and laser-field period, respectively. Using the expression 
\begin{equation}
\sum_ne^{in(E_\vp+U_p+I_p)T}=\omega\sum_n\delta(E_\vp+U_p+I_p-n\omega),
\end{equation}
the amplitude becomes 
\begin{equation}
M_\vp^{(0)}=-2\pi i\sum_n\delta(E_\vp+U_p+I_p-n\omega)T_\vp^{(0)}(n), \label{eq:deltacomb}
\end{equation}
where the $\delta$-function corresponds to the energy-conservation condition and $n$ is the number of absorbed photons. The energy of the absorbed photons is spent on overcoming the ionization and ponderomotive potentials and increasing the kinetic energy of the released electron. The $T_\vp^{(0)}(n)$ is usually denoted as the $T$-matrix element. Finally, the probability density per unit time is $w_{\vp}(n)\propto|T_\vp^{(0)}(n)|^2$ with $n\omega=E_\vp+U_p+I_p$ where $n$ is an  integer. This energy-conservation condition gives the ATI peaks and for our infinitely long driving field, only the photoelectrons with these energies can be expected in the spectrum. In the remainder of the paper we state that the  energy-conservation condition holds only for those values of  the photoelectron  energy  for which $n\omega=E_\vp+U_p+I_p$ is satisfied with integer $n$.

\subsection{\label{sub:spm}Saddle-Point Method}

The integral which appears in the probability amplitude $ M_\vp^{(0)}$ [and also in the $T$-matrix element $T_\vp^{(0)}(n)$] can be solved numerically or by using the saddle-point method. In the latter approach, the stationarity condition $\partial S(\vp;t_0)/\partial t_0=0$ for the action $S(\vp;t_0)=S_\vp(t_0)+I_pt_0$ gives  the saddle-point equation
\begin{equation}\label{spati}
[\vp+\vA(t_0)]^2=-2I_p. 
\end{equation}
The $T$-matrix element $T_\vp^{(0)}(n)$ can be written as
\begin{eqnarray}
T_\vp^{(0)}(n)&=&\frac{1}{T}\sum_{t_{0s}}\sqrt{\frac{2\pi  i}{S''(\vp;t_{0s})}}\nonumber  \\
&& \times\langle\vp+\vA(t_{0s})|\ver\cdot\vE(t_{0s})|\psi_0\rangle e^{iS(\vp;t_{0s})}, 
\end{eqnarray}
where $S''=\partial^2 S(\vp;t_0)/\partial t_0^2=-\vE(t_{0s})\cdot[\vp+\vA(t_{0s})]$, i.e., as the sum of the contributions of the saddle-point solutions. In the saddle-point method, the number $n$ is a continuous parameter. This means that the photoelectrons with the energy which is not in agreement with $n\omega=E_\vp+U_p+I_p$ ($n$ integer) can be expected. This remains true for all saddle-point solutions. Consequently, the interference between the contributions of different saddle-point solutions leads to the nonvanishing probability density for these unphysical values of the photoelectron energy. 

The number of solutions of the saddle-point equation \eqref{spati} depends on the type of the driving field \cite{Jasarevic2020}. In particular, for the $r\omega$--$s\omega$ bichromatic linearly polarized field, equation \eqref{spati} has $2s$ solutions, while for the monochromatic linearly polarized field there are two solutions of the saddle-point equation \eqref{spati} \cite{Jasarevic2020}.

\subsection{\label{sec:unitcell}Unit-Cell Averaging}

One technique to overcome ambiguities that arise from taking a finite range of ionization times is known as unit-cell averaging. A unit cell is the interval $t_{0\min}\leq t_0 \leq t_{0\max}$ for the ionization times $t_0$ employed in saddle-point calculations. Typically, $t_{0\max}=t_{0\min}+T$, so that the unit cell is restricted to a single cycle.  One should note that $t_{0\min}$ and $t_{0\max}$ are arbitrary boundaries for the ionization time window to be taken into consideration. Considering infinitely many cycles eliminates this arbitrariness and renders the boundaries irrelevant, but a coherent sum leads to strong ATI rings, which obfuscate the remaining interference patterns. This is expected as, according to Eq.~\eqref{eq:deltacomb}, these rings are infinitely sharp in the limit of a continuous wave, and are an obstacle if one is interested in intra-cycle interference. 

For a general polychromatic linearly polarized electric field
\begin{equation}
    E(t)=\sum_nE_nf_n(t),
\end{equation}
of amplitudes $E_n$ and time profiles $f_n(t)$, shifting the unit cell is equivalent to taking $f_n(t) \rightarrow f_n(t+\varphi)$ in the above equation, where $\varphi$ is an offset phase which is used to control when the field starts within the unit cell.  Thus, instead of shifting the temporal boundaries, one may shift the field and keep the temporal window fixed. 
The probability densities computed from the non-$T$ periodic part of the amplitude $M^{(0)}_\vp$ must then be calculated for different values of $0<\varphi\leq 2 \pi$ within a single cycle and $\varphi$ must be integrated over. A detailed explanation of this method is provided in \cite{Werby2021} for the CQSFA.

\section{\label{sec:mono}Monochromatic Fields}
We start our analysis by investigating the case of a monochromatic linearly polarized field with intensity $I=E_1^2=1.5\times 10^{14}$~W/cm$^2$ and wavelength of 800~nm. We use the argon atom as a target ($I_p=15.76$~eV). We define the emission angle $\theta_e$ as the angle between the laser-field polarization direction $\ve_x$ and the final photoelectron momentum $\vp$. 

\subsection{\label{sub:ambig}Photoelectron-momentum-distribution ambiguity: asymmetry and unit-cell averaging}
\begin{figure}[!htbp]
    \centering
   \includegraphics[trim={0.3cm 1.cm 2.cm 2.cm},clip,width=0.47\textwidth]{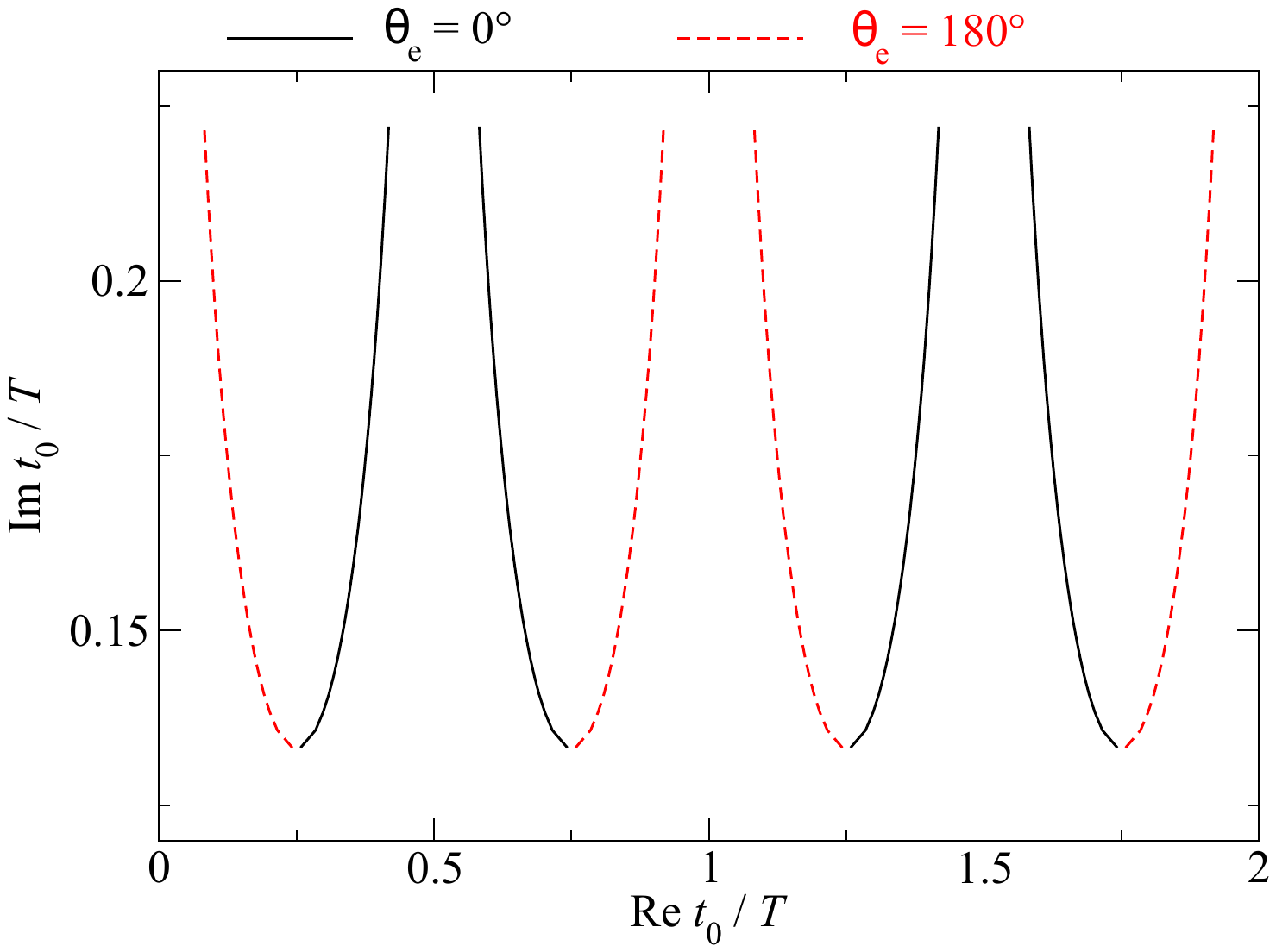}
    \caption{Saddle-point solutions for the ATI induced by a monochromatic linearly polarized field with intensity $E_1^2=1.5\times 10^{14}$~W/cm$^2$ and wavelength of 800~nm. The photoelectrons are emitted in the directions $\theta_e=0^\circ$ (black solid lines) and $\theta_e=180^\circ$ (red dashed lines). The solutions with $T\leq\operatorname{Re}t_0\leq  2T$ are the same as the solutions with $0\leq\operatorname{Re}t_0\leq  T$.}
    \label{fig:monosol}
\end{figure}
In Fig.~\ref{fig:monosol} we present the solutions of the saddle-point equation \eqref{spati} for the photoelectron emission in the directions $\theta_e=0^\circ$ (black solid lines) and $\theta_e=180^\circ$ (red dashed lines) induced by the monochromatic linearly polarized field. Different solutions of the saddle-point equation \eqref{spati} appear within one optical cycle, and if $t_{0s}$ is the solution of Eq.~\eqref{spati} then $t_{0s}+T$ is also a solution of this equation. For our monochromatic linearly polarized field, there are two solutions per laser-field period, and the probability density can be obtained as a coherent sum of the partial contributions of these solutions. One should be cautious when calculating the coherent sum of the partial contributions of these solutions to the probability amplitude. In particular, the resulting photoelectron momentum distributions may not exhibit symmetry properties expected for a given driving field. In addition, the quantitative values of the probability density may not be in agreement with the results obtained by the numerical integration. One may rectify these issues by, for example, shifting the time window from which the saddle-point solutions are chosen,  by averaging the results obtained using the saddle-point solutions from two (or more) time windows, or by performing the unit-cell averaging outlined in Sec.~\ref{sec:unitcell}. 

\begin{figure}[!htbp]
    \centering
    \includegraphics[width=\linewidth]{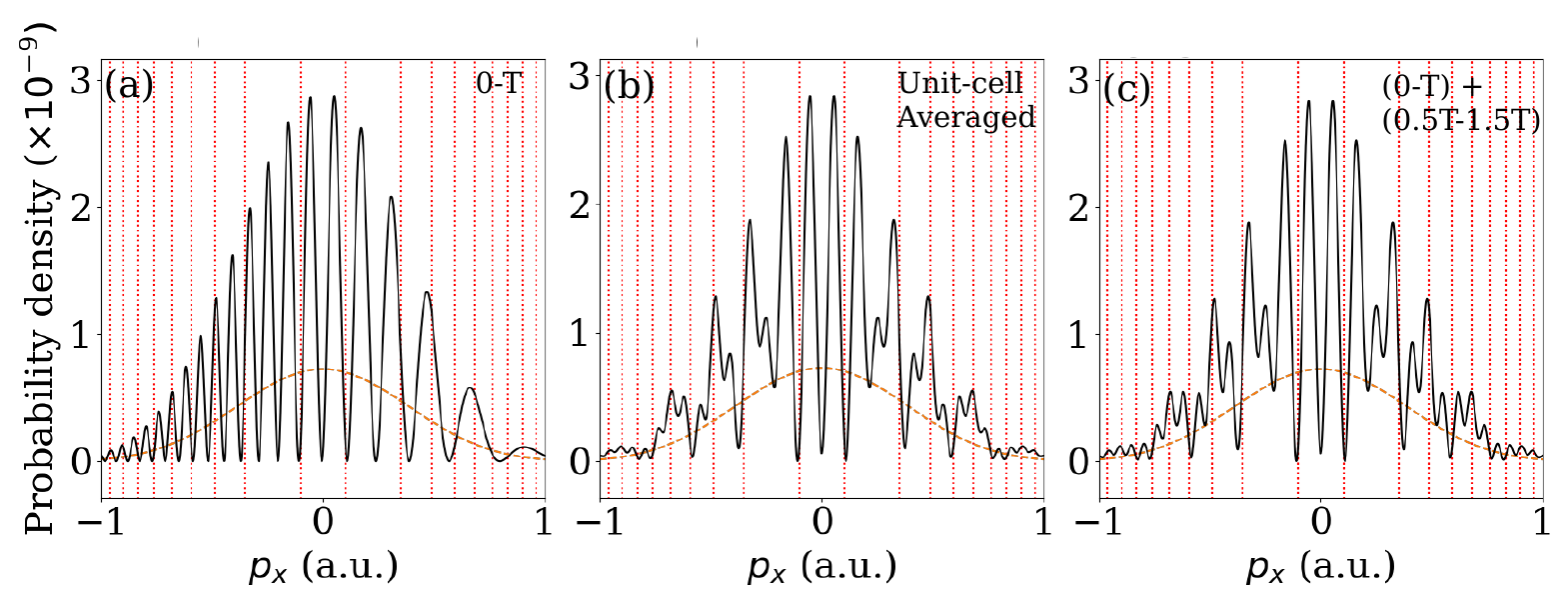}
    \caption{Panel (a): Probability density calculated as a coherent sum of the partial contributions of the saddle-point solutions for $0\leq\operatorname{Re}t_0\leq  T$. Panel (b): Probability density calculated as a coherent sum of the partial contributions of the saddle-point solutions for $0\leq\operatorname{Re}t_0\leq  T$ and by performing unit-sell averaging. Panel(c): Probability density calculated by averaging the coherent sums of the partial contributions of the saddle-point solutions for $0\leq\operatorname{Re}t_0\leq  T$ and $T/2\leq\operatorname{Re}t_0\leq 3T/2$. Red dotted lines represent the values of the momentum $p_x$ allowed by the energy-conservation condition $n\omega=E_\vp+U_p+I_p$ with integer $n$. The orthogonal component of the momentum is $p_y=0$. Orange dashed lines correspond to the contributions of the individual saddle-point solutions. The driving-field parameters are the same as in Fig.~\ref{fig:monosol}.}
    \label{fig:monoCycAv}
\end{figure}
Let us now illustrate the mentioned methods. In Fig.~\ref{fig:monoCycAv}(a) we present the partial contributions of the two saddle-point solutions (orange dashed line) together with their coherent sum (black solid line) for the saddle-point solutions with $0\leq\operatorname{Re}t_0\leq  T$. The orthogonal momentum is $p_y=0$. The red dotted lines represent the values of the momentum allowed by the energy-conservation condition $n\omega=E_\vp+U_p+I_p$ with integer $n$. For the linearly polarized monochromatic field, the partial contributions of the two saddle-point solutions are equal [see the orange dashed lines in Fig.~\ref{fig:monoCycAv}], and the probability density which coherently takes into account the contributions of both solutions should be symmetric with respect to the transformation $p_x\rightarrow -p_x$. However, this is not the case for the coherent sum displayed in Fig.~\ref{fig:monoCycAv}(a) (see the black solid line).  Apart from the incorrect symmetry property, many nonnegligible values of the probability density correspond to the values of the momentum $p_x$ which are not allowed by the energy-conservation condition [see the peaks which appear between the red dotted lines in Fig.~\ref{fig:monoCycAv}(a)].
In order to regain the symmetry of the probability density with respect to the transformation $p_x\rightarrow -p_x$, the averaging should be performed. One protocol to recover the proper symmetry property of the probability density is the unit cell averaging protocol introduced in \cite{Werby2021} (see Eq.~6 and the Appendix in \cite{Werby2021} and Sec.~\ref{sec:unitcell}). The results obtained using this procedure are presented in Fig.~\ref{fig:monoCycAv}(b), and the probability density exhibits the correct symmetry property. However, the peaks for the values of the photoelectron energy which are not in correspondence with  $n\omega=E_\vp+U_p+I_p$ with integer $n$ are still present. More specifically, the probability density is significant for many values of the photoelectron energy which are not in accordance with the energy-conservation condition. 

Another method to obtain the probability density with a proper symmetry property is to combine the coherent sums of the saddle-point solutions from different time windows. The choice of the time windows depends on the type of the driving field. In particular, for the monochromatic linearly polarized field, we use the saddle-point solutions with $0\leq\operatorname{Re}t_0\leq  T$ and $T/2\leq\operatorname{Re}t_0\leq  3T/2$. We use these two windows because the probability density in the momentum region $p_x>0$ ($p_x<0$) obtained using the saddle-point solutions with $0\leq\operatorname{Re}t_0\leq  T$ is the same as the probability density in the momentum region $p_x<0$ ($p_x>0$) obtained using the saddle-point solutions with $T/2\leq\operatorname{Re}t_0\leq 3T/2$. The results obtained using this averaging procedure are presented in Fig.~\ref{fig:monoCycAv}(c). They are similar but not identical to the results obtained using the unit-cell averaging procedure. In particular, the probability density is invariant with respect to the transformation $p_x\rightarrow -p_x$ as it should be, but the unphysical peaks which correspond to the values of energy not allowed by the energy-conservation condition are still present. 

Finally, we mention one additional procedure which can be performed to recover the invariance of the probability density with respect to the reflection $p_x\rightarrow -p_x$. For the time window $0\leq\operatorname{Re}t_0\leq  T$, the difference between the real parts of the two saddle-point solutions in $p_x>0$ part of the momentum plane is not the same as the corresponding difference in the $p_x<0$ part of the momentum plane (cf. the distance between the black solid and red dashed lines in Fig.~\ref{fig:monosol}). However, by choosing the time window $T/4\leq\operatorname{Re}t_0\leq 5T/4$ instead of  $0\leq\operatorname{Re}t_0\leq T$, the difference between the real parts of the two saddle-point solutions becomes equal (for a given photoelectron energy) in both halves of the momentum plane. The coherent sum of the partial contributions calculated using the time window $T/4\leq\operatorname{Re}t_0\leq 5T/4$ exhibits the symmetry $p_x\rightarrow -p_x$. However, due to the presence of the peaks for the energy values forbidden by the energy-conservation law, the agreement with the results calculated using the numerical integration is not expected.  

\subsection{\label{sub:ECC}Energy-conservation condition: recovering a unique momentum distribution}
\begin{figure}[!htbbp]
    \centering
    \includegraphics[width=\linewidth]{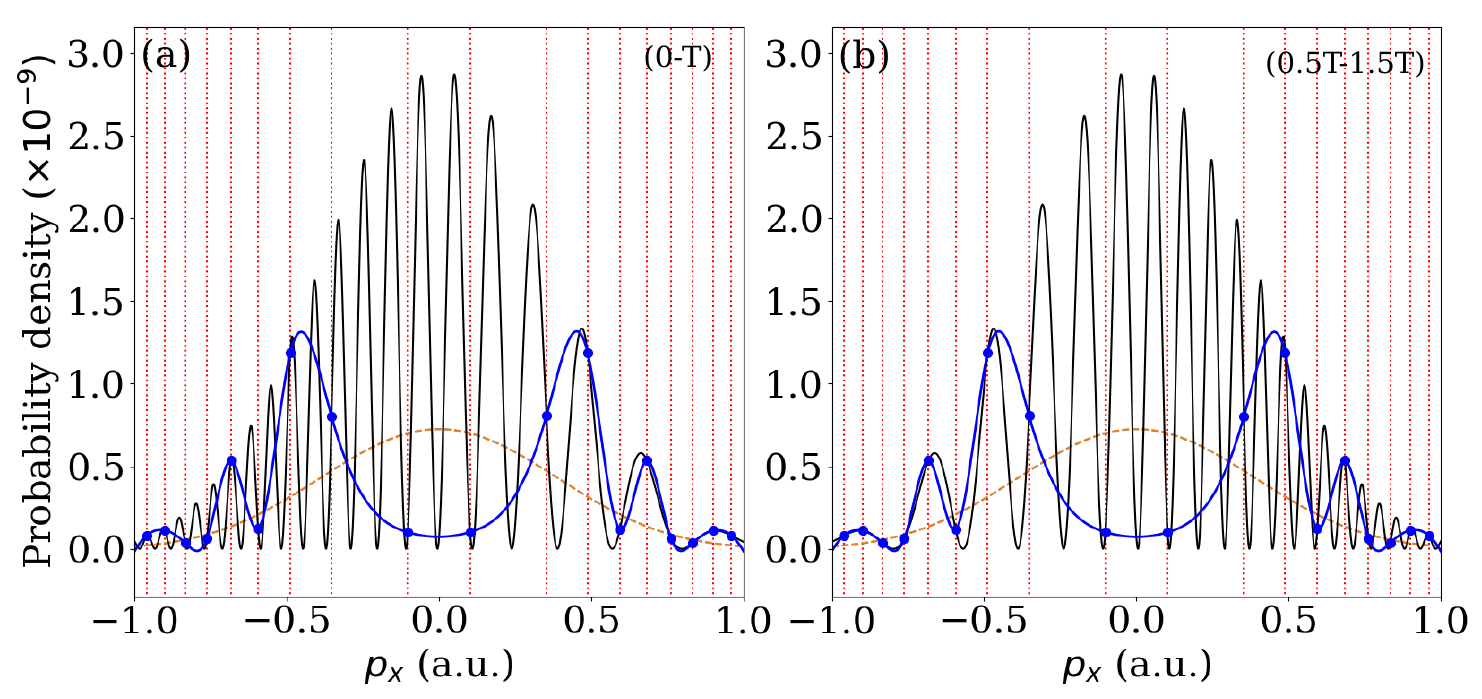}
    \caption{Probability density calculated as a coherent sum of the partial contributions of the saddle-point solutions with $0\leq\operatorname{Re}t_0\leq  T$ (a) and $T/2\leq\operatorname{Re}t_0\leq 3T/2$ (b) (black solid lines). The values of the probability density allowed by the energy-conservation condition (blue dots) are connected by the interpolating curve (blue solid line). Red dotted lines represent the values of the momentum $p_x$ allowed by the energy-conservation condition $n\omega=E_\vp+U_p+I_p$ with integer $n$. The orthogonal component of the momentum is $p_y=0$. Orange dashed lines correspond to the contributions of the individual saddle-point solutions. The driving-field parameters are the same as in Fig.~\ref{fig:monosol}.}
    \label{fig:monoECC}
\end{figure}
Let us now investigate how the energy-conservation condition can be imposed on the results obtained by the saddle-point method and how this affects the probability density. In Fig.~\ref{fig:monoECC} we present the probability density calculated as a coherent sum of the partial contributions of the saddle-point solutions with $0\leq\operatorname{Re}t_0\leq  T$ (a) and $T/2\leq\operatorname{Re}t_0\leq 3T/2$ (b) (black solid lines). The partial contributions of the individual solutions are displayed by the orange dashed lines. In Sec.~\ref{sub:ambig} we have concluded that, by averaging the coherent sums of the saddle-point solutions from these time windows, the probability density invariant with respect to the transformation $p_x\rightarrow -p_x$ can be obtained. The values of the probability density allowed by the energy-conservation condition are denoted by the blue dots. They appear at the intersection of the black solid line (which represents the probability density) and the red dotted lines (which correspond to the values of the momentum allowed by the energy-conservation condition). In other words, the act of imposing the energy-conservation condition assumes that the probability density is calculated only for those values of the photoelectron energy for which $n\omega=E_\vp+U_p+I_p$ is satisfied with integer $n$.  When the energy-conservation condition is imposed, the probability density is the same regardless of the time window we chose to calculate the coherent sum [see the blue solid lines in Figs.~\ref{fig:monoECC}(a) and (b)]. In addition, the same result can be obtained if the energy-conservation condition is imposed on the result obtained using the unit-cell averaging procedure. This means that, if the energy-conservation condition is imposed, the additional averaging procedure is not necessary. This result is rather natural in the sense that, for an infinitely long driving field, only the photoelectrons with energy in agreement with the energy-conservation condition can be expected. The existence of the photoelectron with other values of the energy is the artefact of the saddle-point method, i.e., these electrons can not appear in the experiment.

\begin{figure*}[!htbp]
\centerline{\includegraphics[trim={0.cm 0.cm 0.cm 0.cm},clip,width=0.47\textwidth]{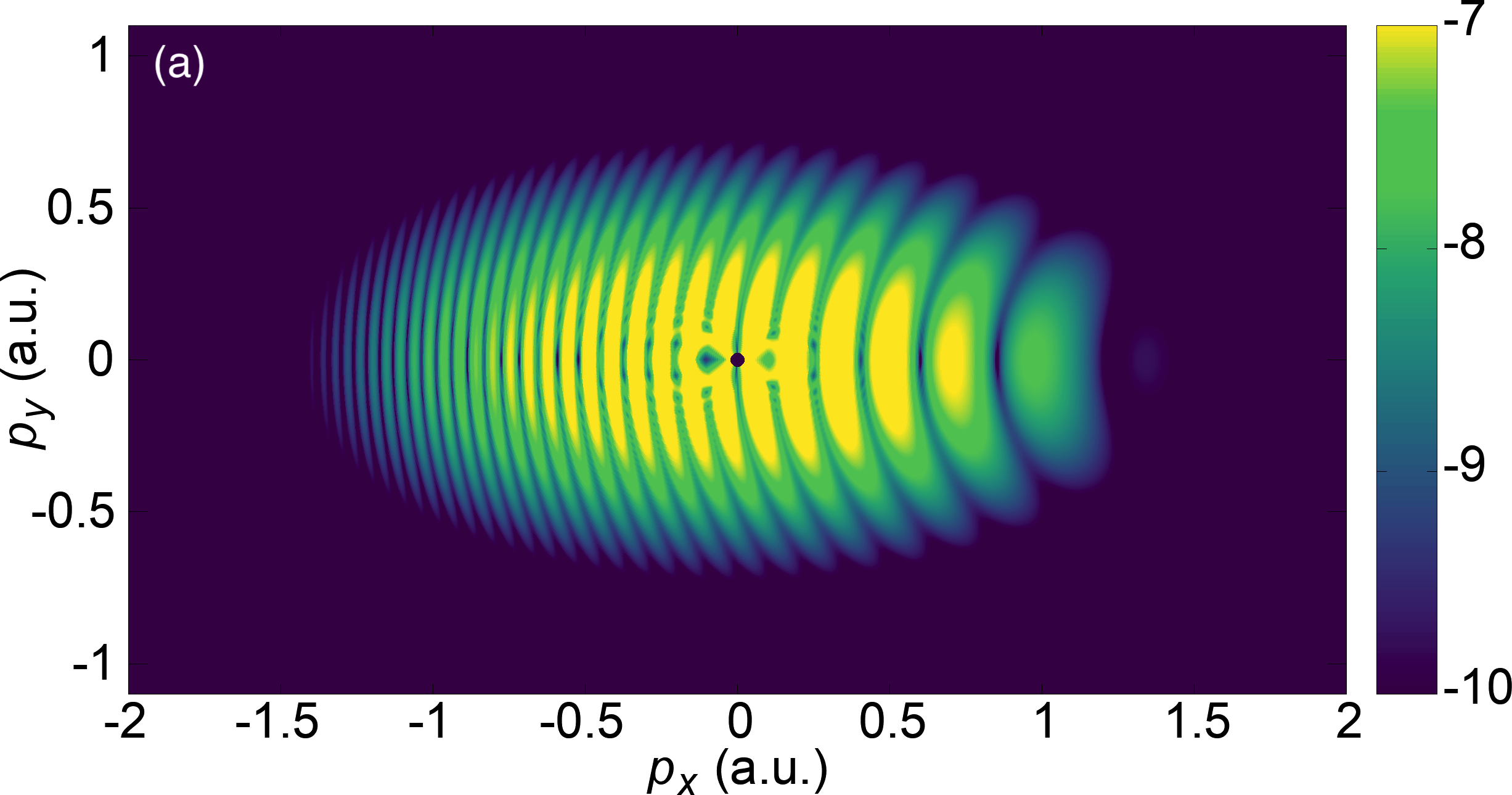}\quad \includegraphics[trim={0.cm 0.cm 0.cm 0.cm},clip,width=0.47\textwidth]{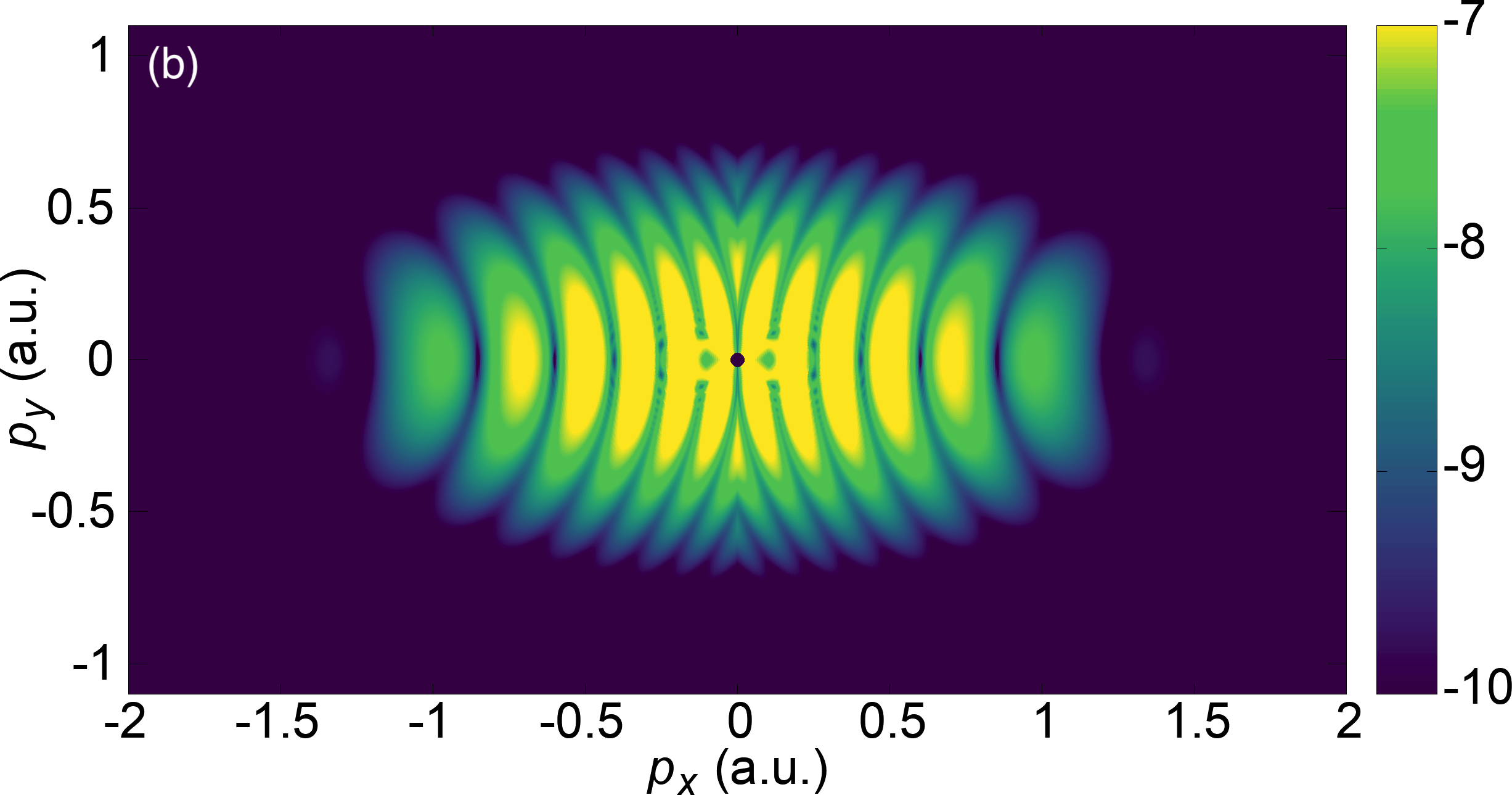}}

\centerline{\includegraphics[trim={0.cm 0.cm 0.cm 0.cm},clip,width=0.47\textwidth]{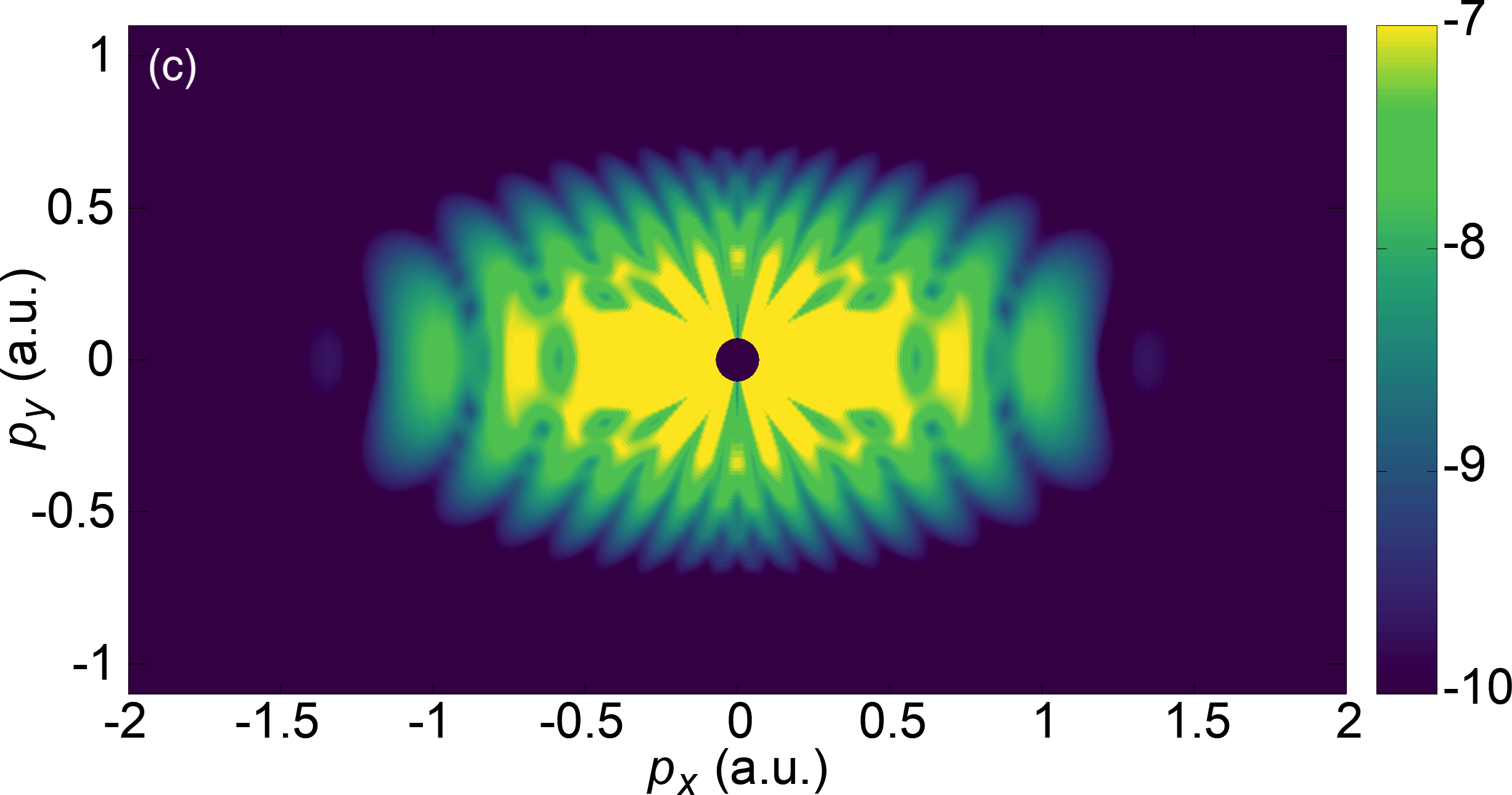}\quad\includegraphics[trim={0.cm 0.cm 0.cm 0.cm},clip,width=0.47\textwidth]{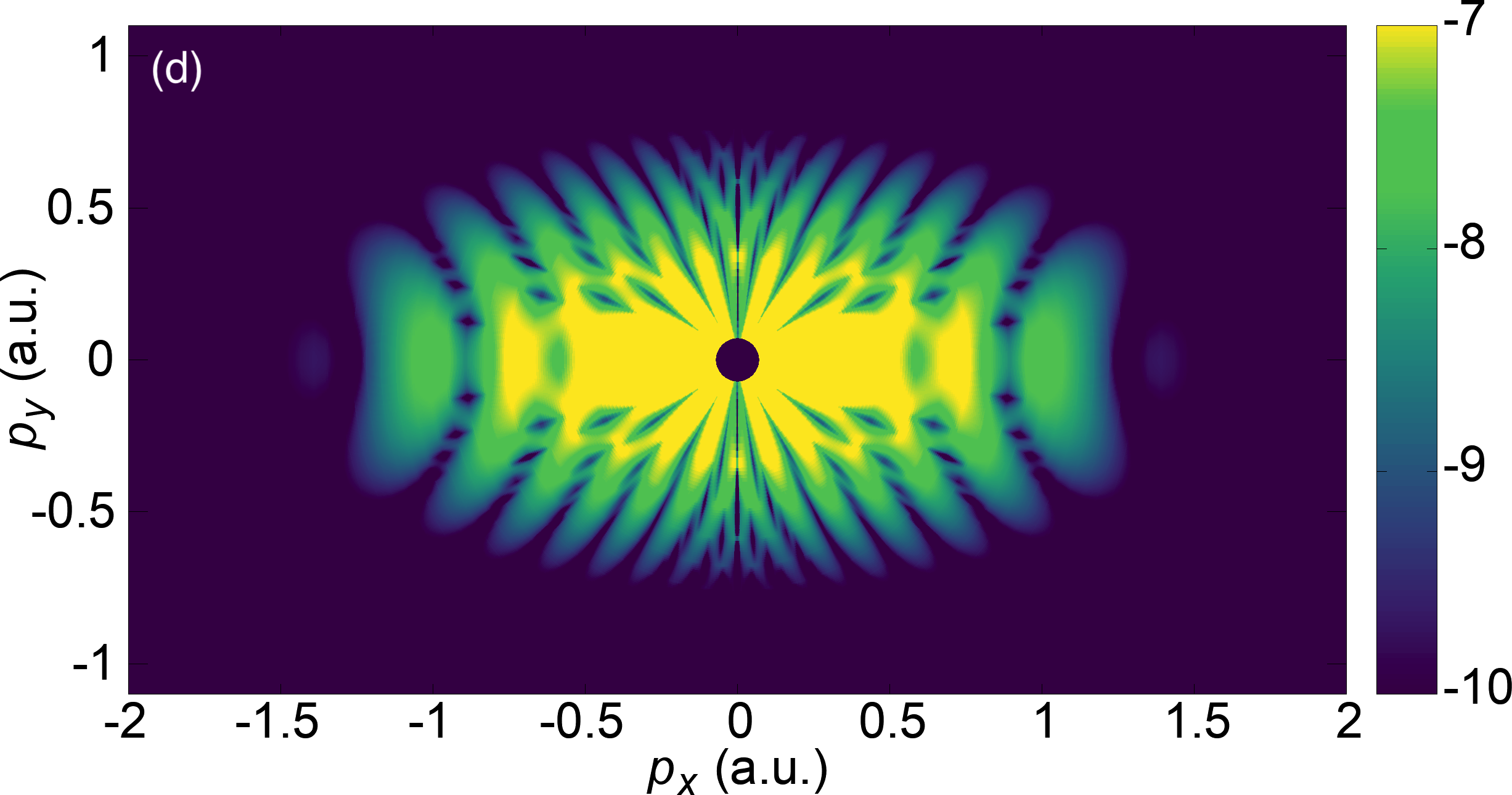}}
\caption{Photoelectron momentum distribution calculated as a coherent sum of the partial contributions of the saddle-point solutions with $0\leq\operatorname{Re}t_0\leq  T$ (a) and $T/4\leq\operatorname{Re}t_0\leq 5T/4$ (b) without imposing the energy-conservation condition, together with the analogous result obtained by imposing the energy-conservation condition (c). Photoelectron momentum distribution calculated using the numerical integration (d). The driving field is the same as in Fig.~\ref{fig:monosol}.} \label{fig:monopmd}
\end{figure*} 
After analyzing how the energy-conservation condition affects the photoelecton spectra for $p_y=0$ we now turn our attention to the photoelectron momentum distributions and to the comparison of the results calculated using the saddle-point method and the numerical integration. In Fig.~\ref{fig:monopmd}(a) we present the photoelectron momentum distribution calculated  as a coherent sum of the partial contributions of the saddle-point solutions with $0\leq\operatorname{Re}t_0\leq  T$ and without imposing the energy-conservation condition. The corresponding results for the time window $T/2\leq\operatorname{Re}t_0\leq 3T/2$ are related to those presented in Fig.~\ref{fig:monopmd}(a) by the reflection $p_x\rightarrow -p_x$. Figure \ref{fig:monopmd}(b) displays the analogous results for the time window $T/4\leq\operatorname{Re}t_0\leq 5T/4$. The results presented in Fig.~\ref{fig:monopmd}(c) are obtained by imposing the energy conservation condition and they do not depend on the time window we have chosen. This condition is imposed manually by calculating the probability density only for those values of the photoelectron energy for which $n\omega=E_\vp+U_p+I_p$ is satisfied with integer $n$. Finally, Fig.~\ref{fig:monopmd}(d) displays the photoelectron momentum distribution calculated via numerical integration and we refer to this result as exact. The difference between the distributions presented in Figs.~\ref{fig:monopmd}(a) and (b), and in Fig.~\ref{fig:monopmd}(d) are significant. Not only that the spectra obtained by the saddle-point method and without imposing the energy-conservation condition [presented in Figs.~\ref{fig:monopmd}(a)] do not obey the symmetry property expected for a monochromatic linearly polarized field (invariance with respect to the  transformation $p_x\rightarrow -p_x$), but also the oscillatory pattern is different. By shifting the time window by $T/4$, the invariance with respect  to the $p_x\rightarrow -p_x$ is recovered [see Fig.~\ref{fig:monopmd}(b)], but the agreement with the exact result is not good [cf. Fig.~\ref{fig:monopmd}(b) with Fig.~\ref{fig:monopmd}(d)]. The shift of the time window by $T/4$ recovers the invariance with respect to $p_x\rightarrow -p_x$ due  to the fact that, for $T/4\leq\operatorname{Re}t_0\leq 5T/4$, the saddle-point solutions for $p_x<0$ and $p_x>0$ are in the same position relative to each other [for $p_y=0$ compare the mutual positions of the black solid and red dashed lines in Fig.~\ref{fig:monosol}]. On the other hand, when the energy-conservation condition is imposed [see Fig.~\ref{fig:monopmd}(c)], the agreement between the results obtained using the saddle-point method and by the numerical integration is excellent [cf. Fig.~\ref{fig:monopmd}(c) with Fig.~\ref{fig:monopmd}(d)]. Also, when the energy-conservation condition is imposed, the time window from which the solutions are chosen is not relevant as long as all solutions are taken into consideration.

It should be noted, that there is some disagreement between the plotted two-dimensional PMDs in Fig.~\ref{fig:monopmd}(c) and Fig.~\ref{fig:monopmd}(d), and the one-dimensional PMDs, given by the blue lines, in Fig.~\ref{fig:monoECC}. In particular, the interference minimum around zero momentum is conspicuously absent in Fig.~\ref{fig:monopmd}(c) and Fig.~\ref{fig:monopmd}(d). These problematic points are a product of the polynomial interpolation between the values of the probability density at momenta where the energy conservation condition is satisfied. There are two possible reasons for the discrepancy; the increased dimensionality of the spectrum itself in Fig.~\ref{fig:monopmd} will influence the polynomial fit as well as the fact that Fig.~\ref{fig:monoECC} is plotted on a linear scale, while Fig.~\ref{fig:monopmd} uses a logarithmic scale. That being said, we will not pay much attention to this issue. One of the main points of this section is to illustrate that the value of the probability density, for momenta not satisfying the energy-conservation condition, should always be treated with caution and we have utilized interpolation in these regions mainly as a means of guiding the eye.

\section{\label{sec:bi}Bichromatic fields}
After analyzing a monochromatic linearly polarized field, we now turn our attention to the case of a bichromatic linearly polarized field. We assume that the intensity of the first field component is $E_1^2=1.5\times 10^{14}$~W/cm$^2$, while the fundamental wavelength is 800~nm. The argon atom is employed as a target.  

\subsection{$\omega$--$2\omega$ case}\label{sub:2w}
First, we investigate the case of the $\omega$--$2\omega$ field with the ratio of the component intensities $\xi=0.7$ and the relative phase $\phi=0^\circ$. The $\omega$--$2\omega$ field configuration does not possess the half-cycle symmetry so that the photoelectron momentum distribution is not invariant with respect to $p_x\rightarrow -p_x$.
\begin{figure}[!htbp]
    \centering
    \includegraphics[width=\linewidth]{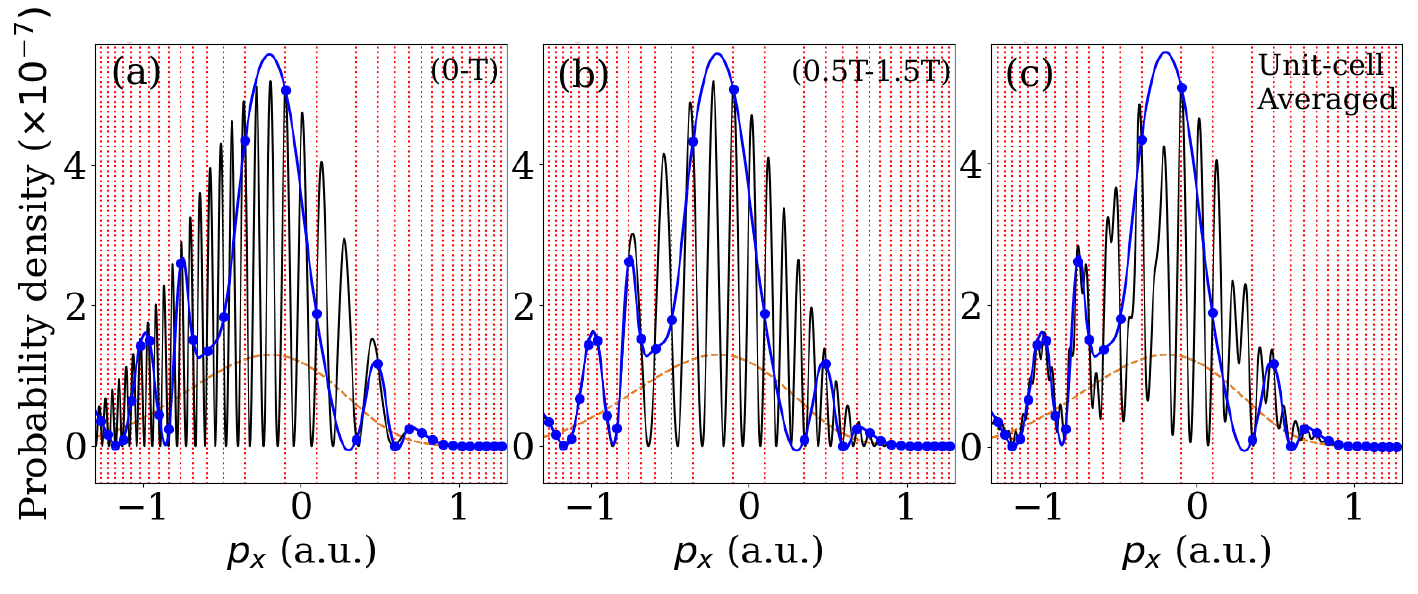}
    \caption{Probability density calculated as a coherent sum of the partial contributions of the saddle-point solutions with $0\leq\operatorname{Re}t_0\leq  T$ (a), $T/2\leq\operatorname{Re}t_0\leq 3T/2$ (b), and by unit-cell averaging (c) (black solid lines). The values of the probability density allowed by the energy-conservation condition (blue dots) are connected by the interpolating curve (blue solid line). Red dotted lines represent the values of the momentum allowed by the energy-conservation condition $n\omega=E_\vp+U_p+I_p$ with integer $n$. The orthogonal component of the momentum is $p_y=0$. Orange dashed lines represent the partial contributions of the saddle-point solutions to the probability density. The driving field is $\omega$--$2\omega$ bichromatic linearly polarized field with intensity $E_1^2=1.5\times 10^{14}$~W/cm$^2$, fundamental wavelength of 800~nm, and the ratio of the component intensities $\xi=0.7$. The relative phase is $\phi=0^\circ$.}
    \label{fig:bi2}
\end{figure}
The saddle-point equation \eqref{spati} has four solutions, two of which lead to nonnegligible contributions to the probability density. In Figs.~\ref{fig:bi2}(a) and (b) we present the probability density calculated as a coherent sum of the partial contributions of the saddle-point solutions with $0\leq\operatorname{Re}t_0\leq  T$ and $T/2\leq\operatorname{Re}t_0\leq 3T/2$, respectively (black solid lines). Red dotted lines represent the values of the momentum allowed by the energy-conservation condition $n\omega=E_\vp+U_p+I_p$ ($n$ integer). Orange dashed lines represent the partial contributions of individual saddle-point solutions. The probability density does not exhibit the reflection symmetry with respect to $p_x\rightarrow -p_x$ which is expected because the $\omega$--$2\omega$ bichromatic linearly polarized driving field does not satisfy the half-cycle symmetry \cite{Rook2022,Habibovic2021}. In addition, the probability density obtained using the saddle-point solutions from the time window $0\leq\operatorname{Re}t_0\leq  T$ and the probability density calculated for $T/2\leq\operatorname{Re}t_0\leq 3T/2$ are related via reflection $p_x\rightarrow -p_x$. In similarity with the monochromatic linearly polarized driving field, there are many nonnegligible values of the probability density which correspond to the values of the momentum for which the energy-conservation condition is not satisfied. Figure.~\ref{fig:bi2}(c) displays  the results obtained using the unit-cell averaging (black solid line) and similar conclusions hold as for the results presented in Figs.~\ref{fig:bi2}(a) and (b). By isolating only the values of the photoelectron momentum which are in agreement with the energy-conservation condition (these values appear at the intersection of the black solid and red dotted lines) and connecting the corresponding values of the probability density by the interpolation polynomial, we obtain the same curve regardless of whether we started from the coherent sum of the solutions from the time window $0\leq\operatorname{Re}t_0\leq  T$ or $T/2\leq\operatorname{Re}t_0\leq 3T/2$, or from the unit-cell-averaged result. In addition, any other time interval of length $T$ can be used as well. The obtained curve corresponds to the results which should be compared with the results calculated by the numerical integration.

\begin{figure}[!htbp]
\centerline{\includegraphics[trim={0.cm 0.cm 0.cm 0.cm},clip,width=0.47\textwidth]{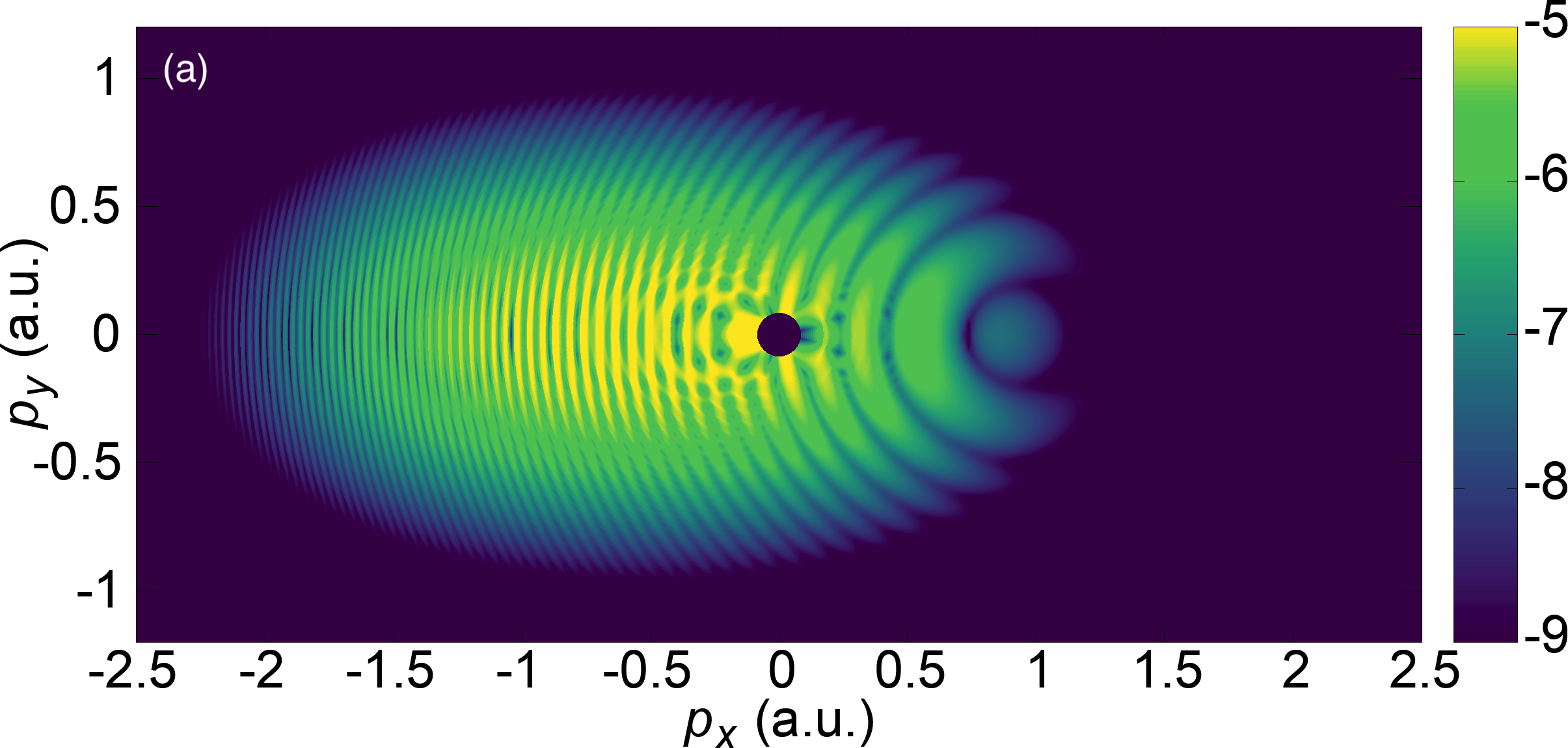}}
\centerline{\includegraphics[trim={0.cm 0.cm 0.cm 0.cm},clip,width=0.47\textwidth]{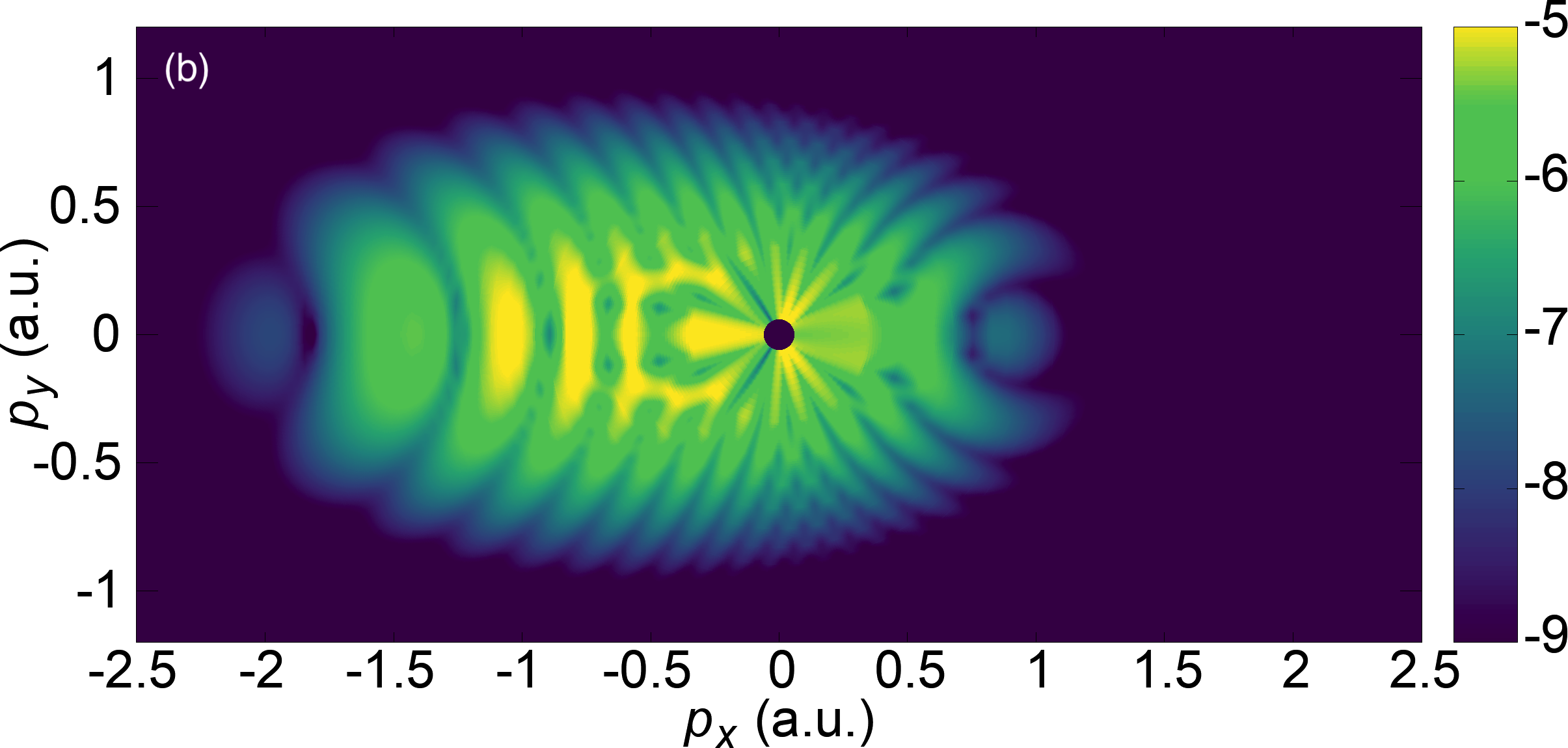}}
\centerline{\includegraphics[trim={0.cm 0.cm 0.cm 0.cm},clip,width=0.47\textwidth]{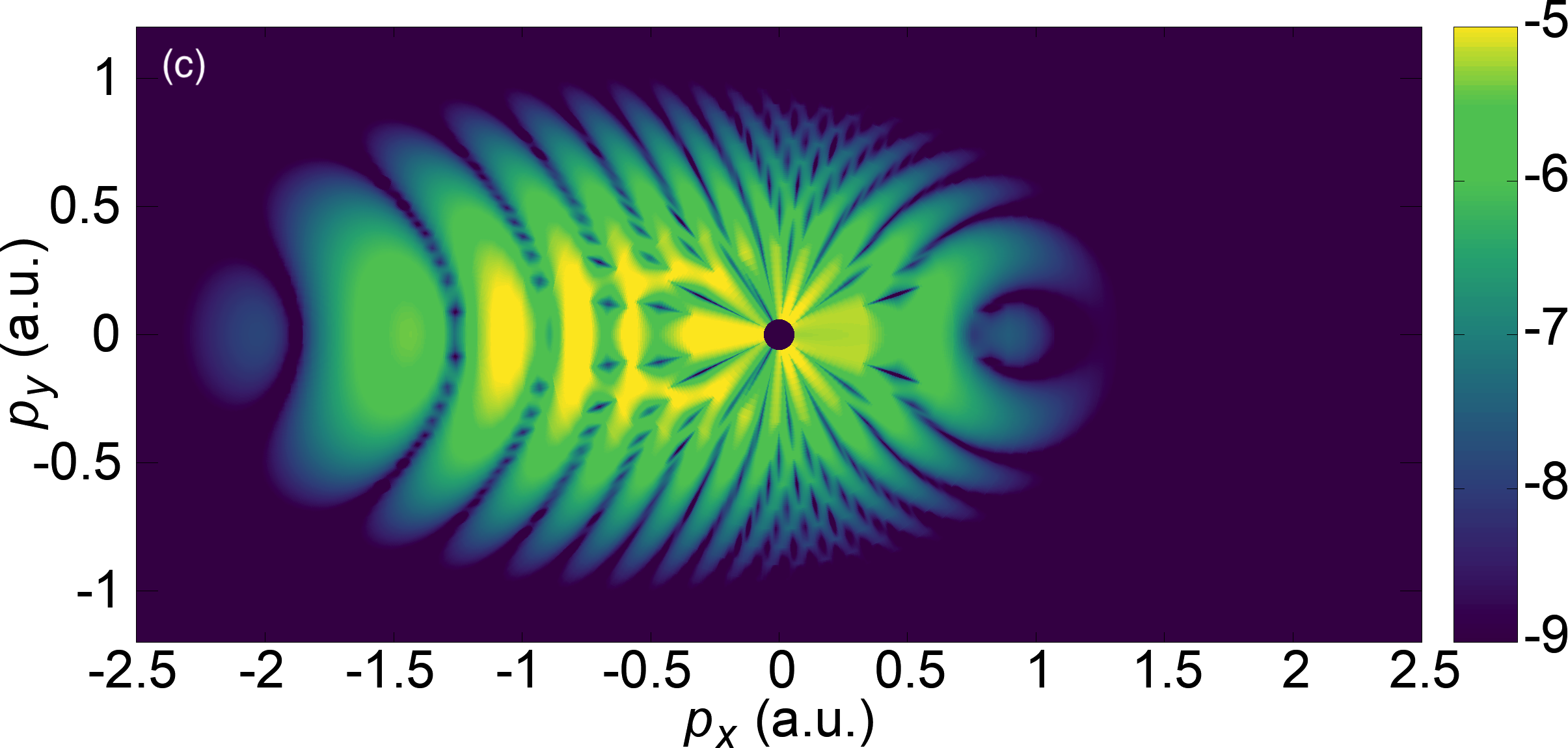}}
\caption{Photoelectron momentum distribution calculated as a coherent sum of the partial contributions of the saddle-point solutions with $0\leq\operatorname{Re}t_0\leq  T$ without imposing the energy-conservation condition (a) and by imposing the energy-conservation condition (b), together with the analogous results obtained using the numerical integration (c). The driving field is the same as in Fig.~\ref{fig:bi2}.} \label{fig:w2wpmd}
\end{figure}
In Fig.~\ref{fig:w2wpmd} we present the photoelectron momentum distributions calculated as a coherent sum of the contributions of the saddle-point solutions from the time window $0\leq\operatorname{Re}t_0\leq T$ without imposing the energy-conservation condition (a) and the analogous results obtained by imposing the energy-conservation condition (b). In addition, panel (c) displays the results calculated by the numerical integration. The results obtained without imposing the energy-conservation condition are not in agreement with the results calculated by the numerical integration [cf. the panels (a) and (c) in Fig.~\ref{fig:w2wpmd}], particularly in the region $p_x<0$. In this momentum region, the oscillatory pattern in Fig.~\ref{fig:w2wpmd}(a) produces convex ring-structures, while the corresponding pattern in the exact momentum distribution [Fig.~\ref{fig:w2wpmd}(c)] produces concave ring-structures. Moreover, the number of rings is different in these two cases. The shift of the time window will produce different results in comparison with those presented in Fig.~\ref{fig:w2wpmd}(a) which may be more or less in agreement with the results calculated by the numerical integration. However, the only way to correctly reproduce the exact calculations is to manually impose the energy-conservation condition, in which case the choice of the time interval becomes irrelevant. When the energy conservation condition is imposed, the agreement between the exact results and the results calculated via the saddle-point method is excellent [cf. the panels (b) and (c) in Fig.~\ref{fig:w2wpmd}]. Finally, we stress that the results which are in agreement with the numerically obtained results cannot be obtained by choosing a different time window. Only better agreement with the numerical results can be achieved.

\subsection{$\omega$--$3\omega$ case}\label{sub:3w}

\begin{figure}[!htbp]
    \centering
    \includegraphics[width=\linewidth]{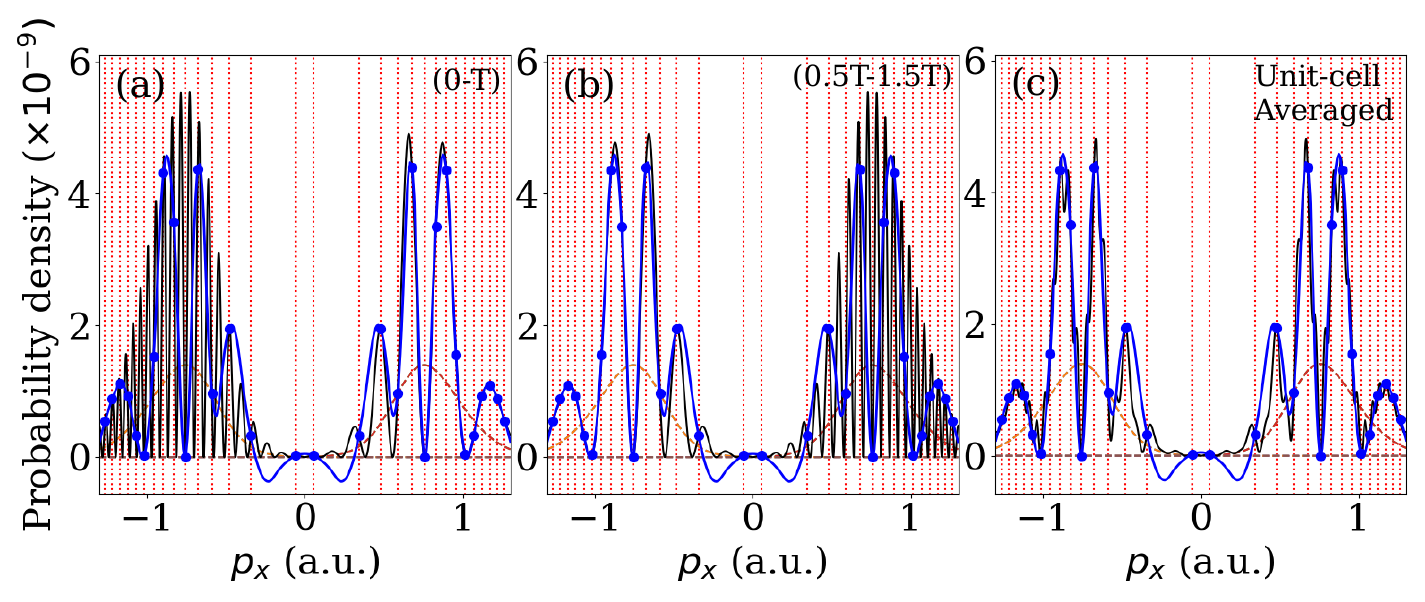}
    \caption{Same as in Fig.~\ref{fig:bi2}  but  for the $\omega$--$3\omega$ bichromatic linearly polarized field with intensity $E_1^2=1.5\times 10^{14}$~W/cm$^2$, fundamental wavelength of 800~nm, and the ratio of the component intensities $\xi=0.1$. The relative phase is $\phi=0^\circ$.}
    \label{fig:bi3phi0}
\end{figure}
Similar analysis can be performed for the $\omega$--$3\omega$ driving field. This field configuration possesses the half-cycle symmetry and there are six saddle-point solutions per optical cycle. The number of the saddle-point solutions (six) is larger than in the case of the monochromatic linearly polarized field. Consequently, we expect it to be more difficult to recover the proper symmetry of the momentum distribution by shifting the time window from which the saddle-point solutions are chosen.  In Fig.~\ref{fig:bi3phi0} we present the probability density calculated as a coherent sum of the partial contributions of the saddle-point solutions from $0\leq\operatorname{Re}t_0\leq  T$ (a) and $T/2\leq\operatorname{Re}t_0\leq 3T/2$ (b), together with the probability density calculated using the unit-cell averaging (black solid lines) for $p_y=0$. These results are obtained without imposing the energy-conservation condition. The red dotted lines represent the values of the momentum allowed by the energy-conservation condition $n\omega=E_\vp+U_p+I_p$ ($n$ is an integer), while the dashed lines correspond to the partial contributions of individual saddle-point solutions. For our $\omega$--$3\omega$ field, there are six saddle-point solutions two of which usually lead to the negligible contributions to the probability density. The significance of the other four solutions depends on the values of the driving-field parameters.  The $\omega$--$3\omega$ driving field possesses the half-cycle symmetry so that the probability density should be invariant with respect to the transformation $p_x\rightarrow -p_x$. The distributions shown in Figs.~\ref{fig:bi3phi0}(a) and (b) do not satisfy the symmetry property required by the field [cf. the black solid lines in the $p_x<0$ half-plane with the corresponding lines in the $p_x>0$ half-plane in Figs.~\ref{fig:bi3phi0}(a) and (b)]. In addition to the absence of the correct symmetry, there are many nonnegligible values of the probability density which correspond to the values of the momentum for which the energy-conservation condition is not satisfied. When the unit-cell averaging is performed, the correct symmetry property of the spectrum is retrieved. In addition, for these values of the laser-field parameters, the nonnegligible values of the probability density which are not supported by the energy-conservation condition are mostly eliminated. The values of the probability density and the corresponding momentum $p_x$ which are supported by the energy-conservation condition are denoted by the blue dots which are connected by the interpolating polynomial. These values are the same regardless of the choice of the time window or whether the unit-cell averaging is performed or not.     

\begin{figure*}[!htbp]
\centerline{\includegraphics[trim={0.cm 0.cm 0.cm 0.cm},clip,width=0.47\textwidth]{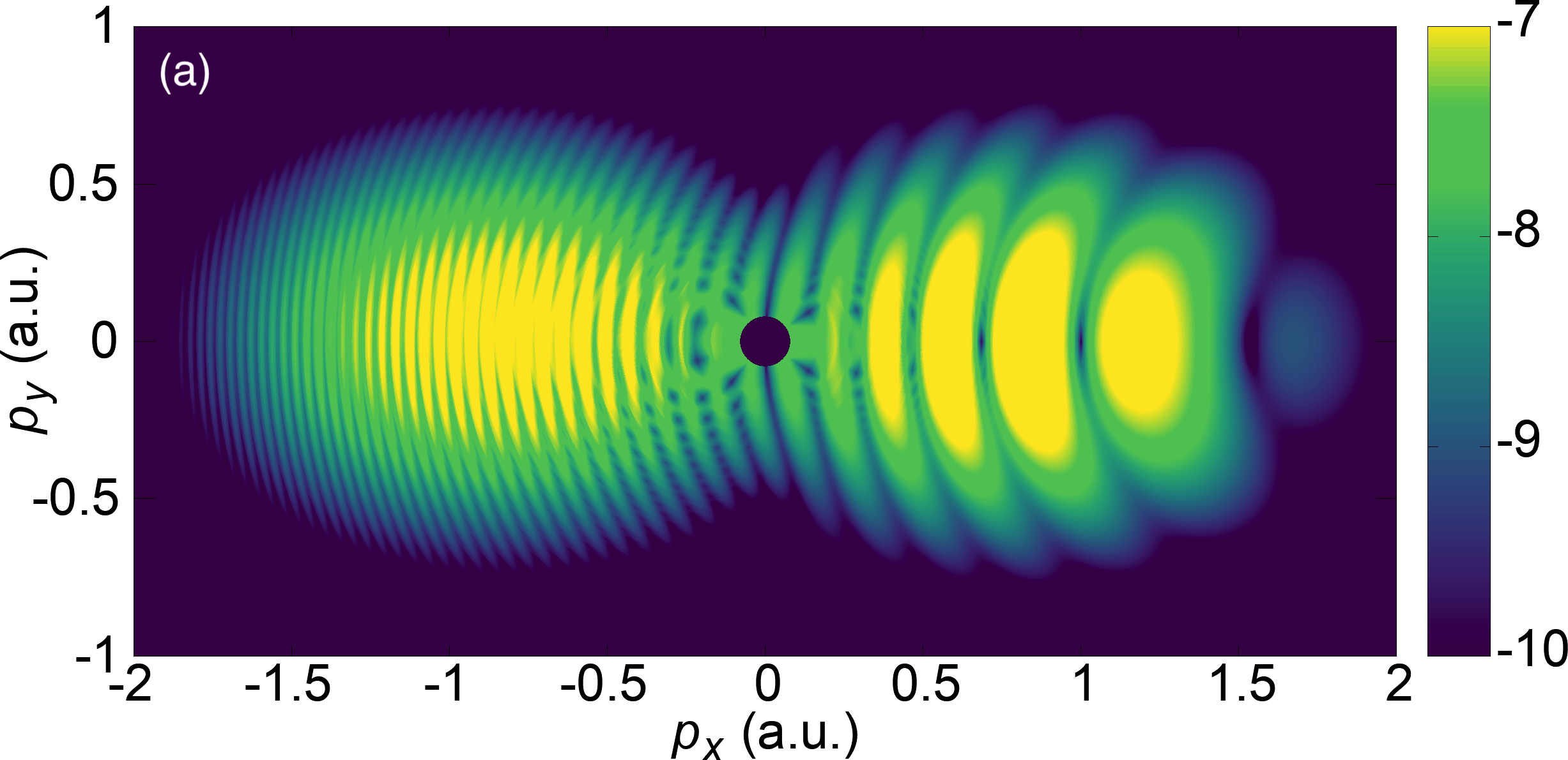}\quad \includegraphics[trim={0.cm 0.cm 0.cm 0.cm},clip,width=0.47\textwidth]{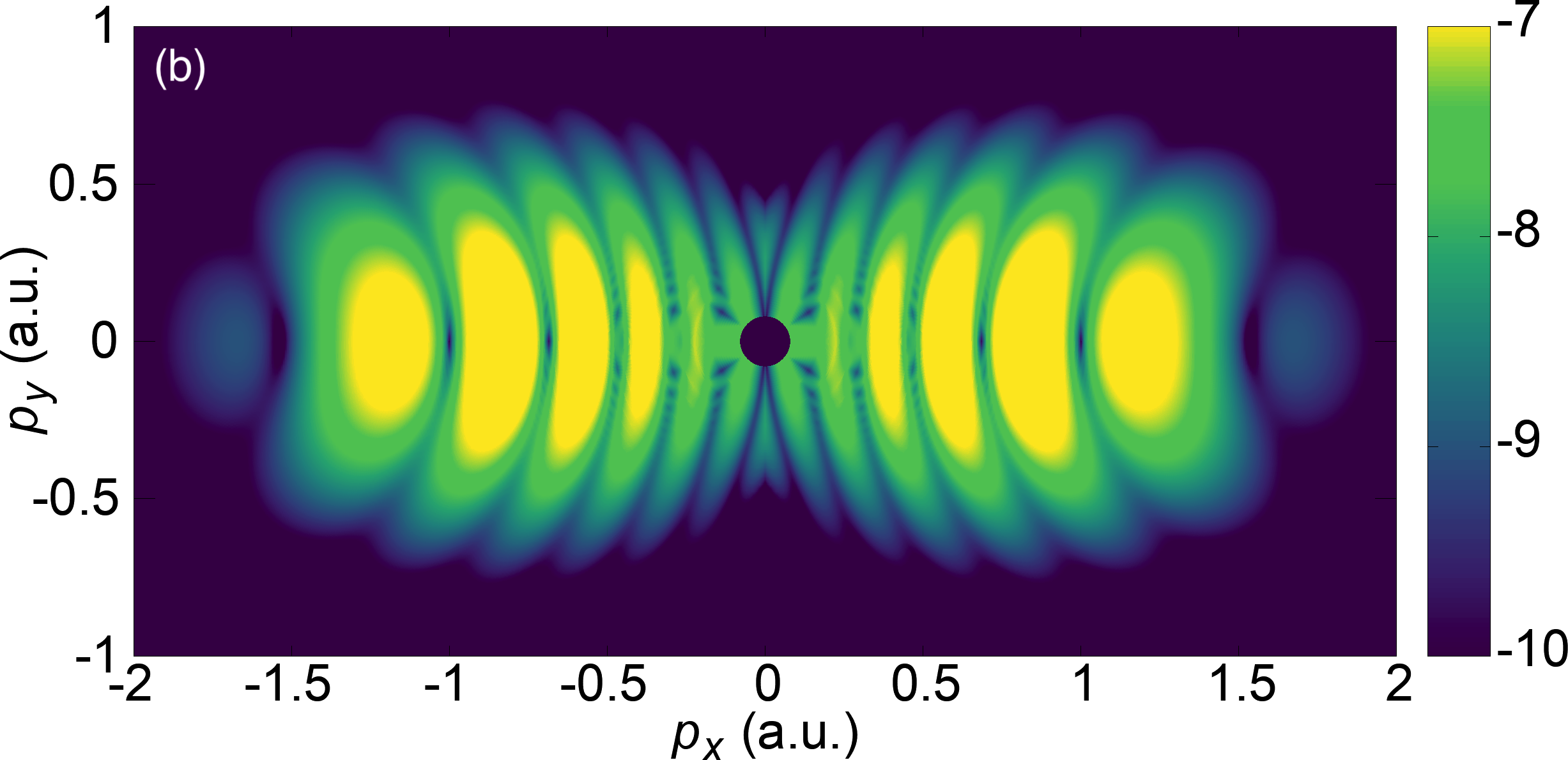}}
\centerline{\includegraphics[trim={0.cm 0.cm 0.cm 0.cm},clip,width=0.47\textwidth]{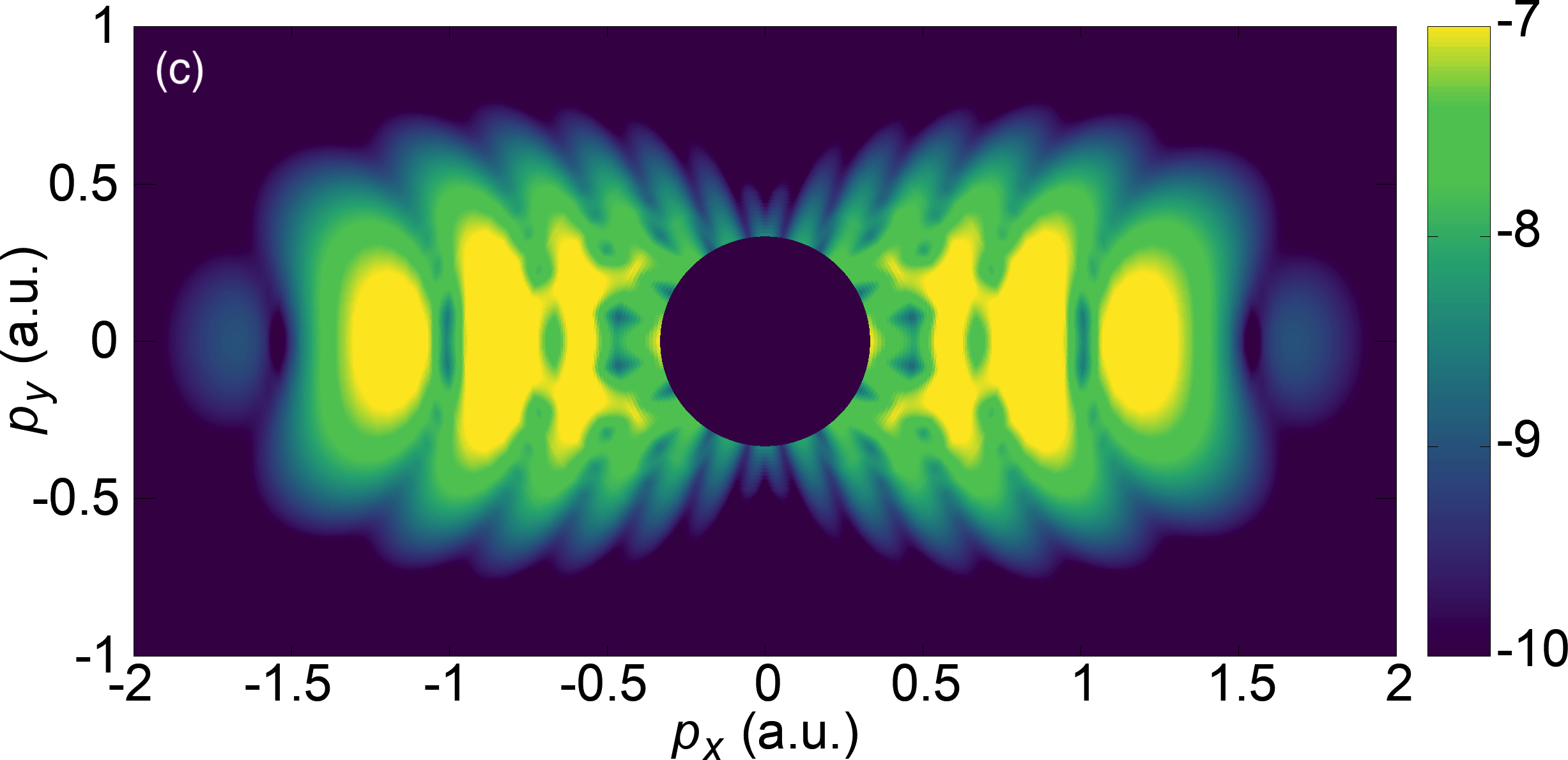}\quad\includegraphics[trim={0.cm 0.cm 0.cm 0.cm},clip,width=0.47\textwidth]{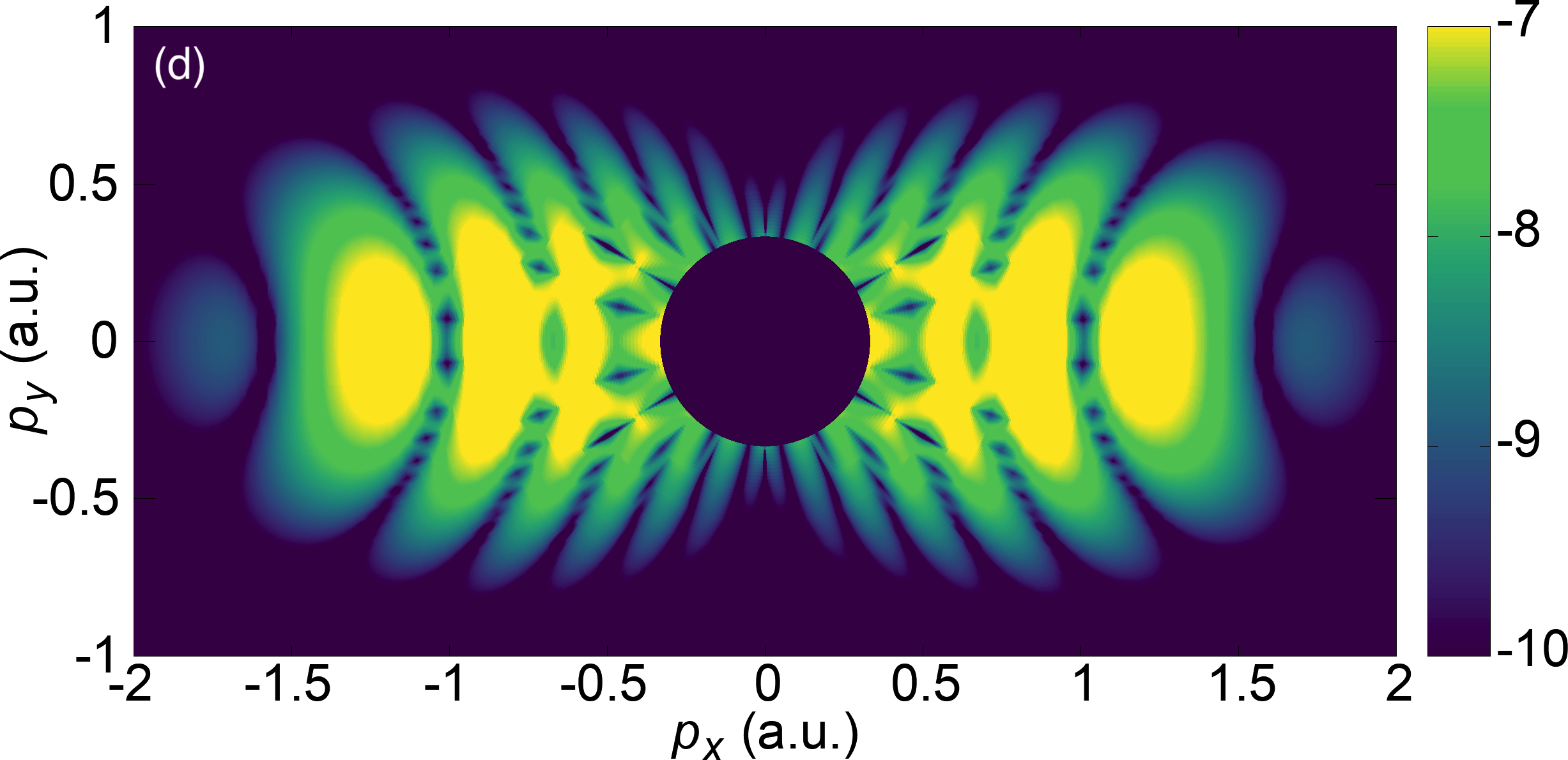}}
\caption{Photoelectron momentum distribution calculated as a coherent sum of the partial contributions of the saddle-point solutions with $0\leq\operatorname{Re}t_0\leq  T$ (a) without imposing the energy-conservation condition, together with the analogous result obtained by imposing the energy-conservation condition (c) and using the numerical integration (d). Photoelectron momentum distribution calculated as a coherent sum of the partial contributions of the saddle-point solutions with $0\leq\operatorname{Re}t_0\leq  T$ for $p_x>0$ and with $T/2\leq\operatorname{Re}t_0\leq 3T/2$ for $p_x<0$ (b). The driving field is the same as in Fig.~\ref{fig:bi3phi0}.} \label{fig:w3wphi0pmd}
\end{figure*}
Let us now investigate the photoelectron momentum distributions. In Fig.~\ref{fig:w3wphi0pmd}(a) we present the photoelectron momentum distribution calculated as a coherent sum of the partial contributions of the saddle-point solutions with $0\leq\operatorname{Re}t_0\leq  T$ and without imposing the energy-conservation condition (in $n\omega=E_\vp+U_p+I_p$, $n$ can have an arbitrary value). The results presented in Fig.~\ref{fig:w3wphi0pmd}(c) are obtained by imposing the energy conservation condition, while Fig.~\ref{fig:w3wphi0pmd}(d) displays the photoelectron momentum distribution calculated via numerical integration. The photoelectron momentum distribution calculated without taking into account the energy-conservation condition is not in agreement with the exact results [cf. the Figs.~\ref{fig:w3wphi0pmd}(a) and (d)]. It neither exhibits the symmetry property imposed by the driving field nor the correct interference pattern. This happens due to the fact that there are many non-vanishing values of the photoelectron yield for which $n\omega=E_\vp+U_p+I_p$ cannot be satisfied with any integer value of $n$. On the other hand, when the energy-conservation condition is imposed, the agreement between the exact results and the results calculated by the saddle-point method is excellent [cf. the Figs.~\ref{fig:w3wphi0pmd}(c) and (d)]. By comparing the momentum distributions displayed in Figs.~\ref{fig:w3wphi0pmd}(a) and (d) we conclude that the disagreement is particularly significant for $p_x<0$. The momentum distribution in this half-plane can artificially be 'corrected' by shifting the real parts of the saddle-point solutions which correspond to $p_x<0$ for a specific value of time. In particular, in Fig.~\ref{fig:w3wphi0pmd}(b) we present the photoelectron momentum distribution calculated as a coherent sum of the partial contributions of the saddle-point solutions with $0\leq\operatorname{Re}t_0\leq T$ for $p_x>0$ and with $T/2\leq\operatorname{Re}t_0\leq 3T/2$ for $p_x<0$. The momentum distribution obtained in this way possesses the correct symmetry property (invariant with respect to the reflection $p_x\rightarrow -p_x$), and the agreement with the exact results is much better. 

\begin{figure}[!htbp]
    \centering
   \includegraphics[trim={0.3cm 1.cm 2.cm 2.cm},clip,width=0.47\textwidth]{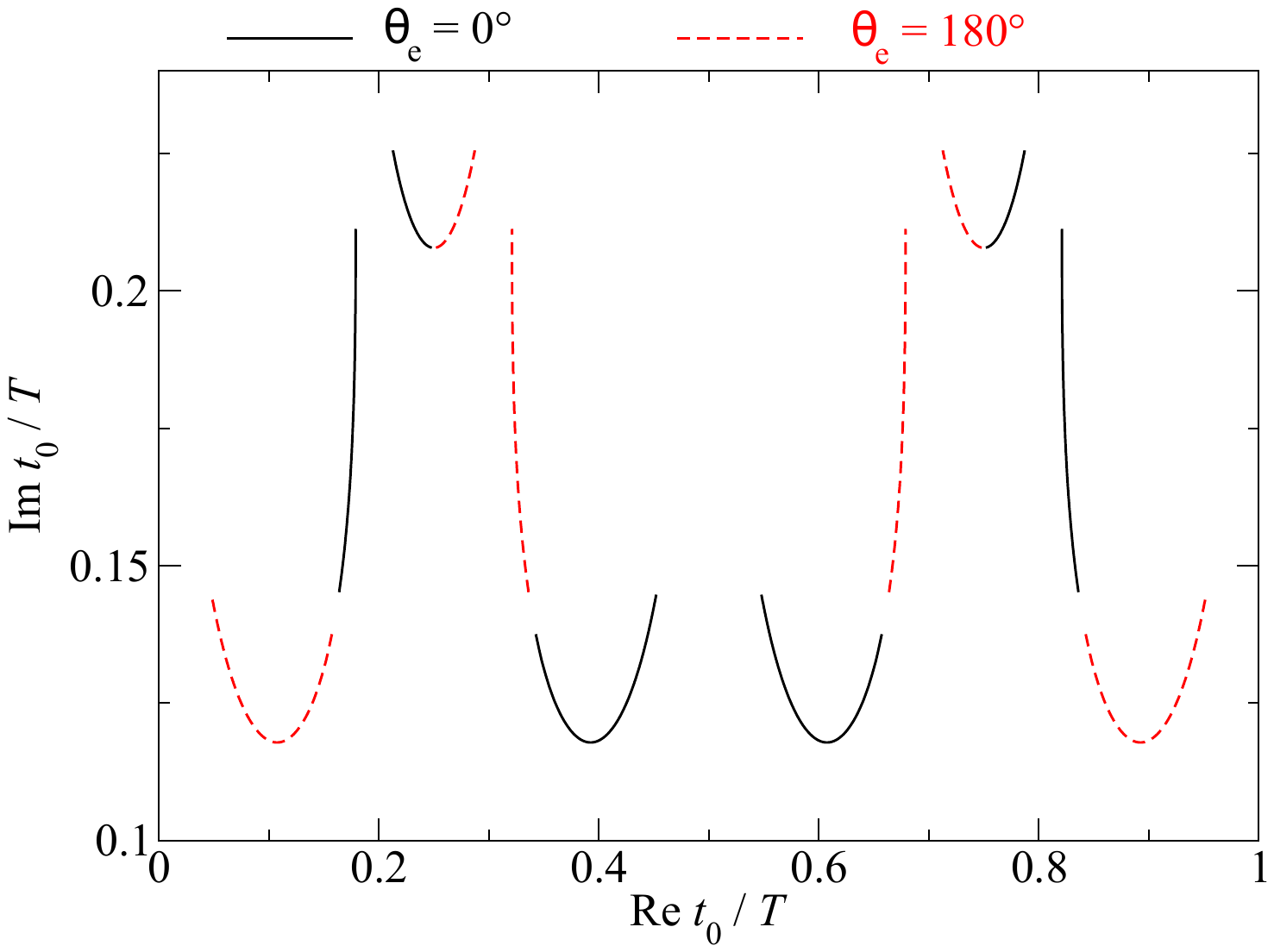}
    \caption{Saddle-point solutions for the ATI induced by the $\omega$--$3\omega$ linearly polarized field with the same parameters as in Fig.~\ref{fig:bi3phi0}. The photoelectrons are emitted in the directions $\theta_e=0^\circ$ (black solid lines) and $\theta_e=180^\circ$ (red dashed lines), and the photoelectron energy changes from $0.01U_p$ to $7U_p$ along the curves.}
    \label{fig:w3wsolphi0}
\end{figure}
In order to explain why this time shift recovers the invariance with respect to $p_x\rightarrow -p_x$ we have to analyze the relative relationship between the saddle-point solutions for $p_x>0$ and $p_x<0$. In Fig.~\ref{fig:w3wsolphi0} we present the saddle-point solutions for the same field as in Fig.~\ref{fig:bi3phi0} and for $\theta_e=0^\circ$ (black solid lines) and $\theta_e=180^\circ$ (red dashed lines). The photoelectron energy changes from $0.01U_p$ to $7U_p$ along the curves. The solutions with the smallest imaginary part lead to the most prominent contributions to the probability density. For both cases, $\theta_e=0^\circ$ and $\theta_e=180^\circ$, there are two saddle-point solutions which have to be taken into consideration.  However, for $\theta_e=0^\circ$ these two solutions are next to each other (see the black solid lines around $\operatorname{Re} t_0=0.4 T$ and $\operatorname{Re} t_0=0.6 T$), while for $\theta_e=180^\circ$ the one dominant solution is around $\operatorname{Re} t_0=0.1 T$, while the other is around $\operatorname{Re} t_0=0.9 T$. Similar conclusions can be derived for other saddle-point solutions even though their contribution to the probability density is small. By shifting the real part of the first three solutions for $\theta_e=180^\circ$ by $T/2$ (i.e., by choosing the time window $T/2\leq\operatorname{Re}t_0\leq 3T/2$), we artificially achieve the same relative relationship between the solutions for $\theta_e=0^\circ$ and $\theta_e=180^\circ$. The reason why we decided to shift the solutions which correspond to the $\theta_e=180^\circ$ and not the solution which corresponds to $\theta_e=0^\circ$ is because, by comparing the results with the numerical integration, we concluded that the agreement is better for $p_x>0$ so that we need to arrange the solutions for $\theta_e=180^\circ$ in the same way as they are arranged for $\theta_e=0^\circ$. 

In conclusion, while it is indeed feasible to recuperate the accurate symmetry property of the momentum distribution and to attain a superior agreement with the exact results by shifting the real part of the selected number of saddle-point solutions, it is imperative to recognise that the most accurate methodology for calculating the momentum distribution entails the incorporation of the energy-conservation condition. In this case, the choice of the time window becomes irrelevant.

\begin{figure}[!htbp]
    \centering
    \includegraphics[width=\linewidth]{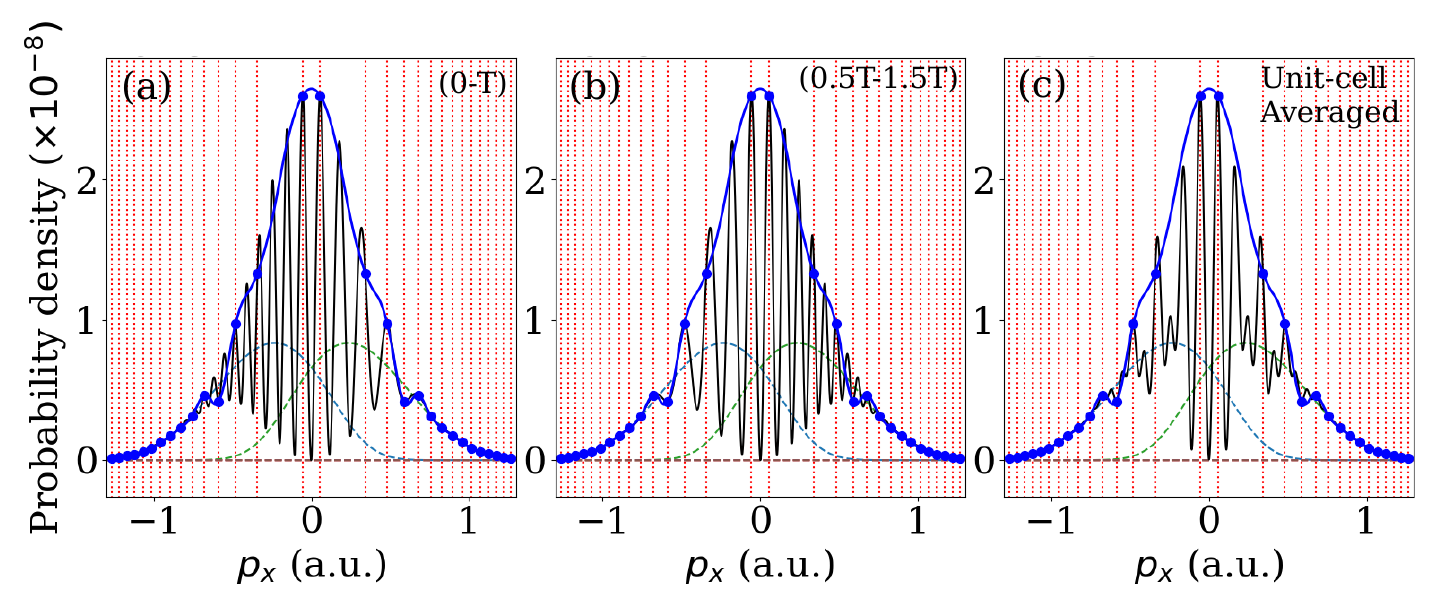}
    \caption{Same as in Fig.~\ref{fig:bi3phi0}  but  for the $\omega$--$3\omega$ bichromatic linearly polarized field with the relative phase $\phi=90^\circ$. The partial  contributions of the saddle-point solutions are now represented by the blue and green dashed lines.}
    \label{fig:bi3phiPi2}
\end{figure}
Finally, in order to investigate how the presented analysis depends on the relative phase between the laser-field components, in Fig.~\ref{fig:bi3phiPi2} we present the results for the probability density analogous to those presented in Fig.~\ref{fig:bi3phi0} but for the $\omega$--$3\omega$ field with the relative phase $\phi=90^\circ$. The conclusions regarding the probability density calculated using the saddle-point solutions from the time window $0\leq\operatorname{Re}t_0\leq  T$ and  $T/2\leq\operatorname{Re}t_0\leq 3T/2$ are the same as for the $\omega$--$3\omega$ field with the relative phase $\phi=0^\circ$. However, in the case of the field with the relative phase $\phi=0^\circ$, the agreement between the results obtained via unit-cell averaging and by imposing the energy-conservation condition was reasonably good [see the black solid line and the blue dots in Fig.~\ref{fig:bi3phi0}(c)]. This is not the case for the $\omega$--$3\omega$ field with the relative phase $\phi=90^\circ$ [see the black solid line and the blue dots in Fig.~\ref{fig:bi3phiPi2}(c)]. In particular, the unit-cell-averaged results still exhibit many nonnegligible values of the probability density which correspond to the values of the momentum for which the energy-conservation condition is not satisfied. This means, that the agreement between the unit-cell-averaged results and the results obtained by imposing the energy-conservation condition $n\omega=E_\vp+U_p+I_p$ with integer $n$ for $\omega$--$3\omega$ field with the relative phase $\phi=0^\circ$ is the exemption rather than the rule.

\section{\label{sec:conc}Conclusions}

The easiest way to calculate the transition amplitude for the above-threshold ionization process is by employing the saddle-point method. In this case, the transition amplitude is in the form of the sum of the partial contributions of different saddle-point solutions and the photoelectron energy is a continuous parameter. The saddle-point solutions are complex, and for the long driving field with a flat envelope, their real parts are within one optical cycle. By adding the laser-field period $T$ to the real part of the saddle-point solution, we practically obtain the same solution in a sense that its partial contribution to the ionization probability is the same as the partial contribution of the original solution. However, the coherent sum of the partial contributions of the saddle-point solutions depends on the time window in which the real parts of the saddle-point solutions are situated.

In the first part of our paper, we have investigated the dependence of the shape and the symmetry properties of the photoelectron spectrum on the choice of this time window. We have found that, in general, the obtained photoelectron spectrum does not satisfy the symmetry properties imposed by the driving field and that it exhibits an oscillatory pattern which is not in agreement with the results obtained by the numerical integration. We have discussed that the correct symmetry property of the spectrum can be recovered by different averaging procedures, the examples being the unit-cell averaging and the procedure which includes averaging of the results from the two time windows separated by a certain value of time. We have also shown that the oscillatory pattern can partially be corrected by shifting the real parts of the saddle-point solutions from one half plane in such a way that the mutual positions of the saddle-point solutions from both half planes are the same. However, all these procedures lead only to an approximate agreement with the numerically integrated results. In order to reproduce the results obtained through numerical integration, it is necessary to impose the energy-conservation condition in the saddle-point method. This entails calculating the ionization probability only for those values of the photoelectron energy for which the integer number of photons is absorbed, rather than calculating it for an arbitrary value of the energy.  

In the second part of our paper, we have conducted a similar analysis for the $\omega$--$2\omega$ and $\omega$--$3\omega$ bichromatic linearly polarized fields with a similar conclusions as for the monochromatic case. For these field configurations, the results obtained as a coherent sum of the partial contributions of different saddle-point solutions are not in agreement with the results obtained by the numerical integration if the energy-conservation condition is not imposed. This remains true regardless of the time window from which the saddle-point solutions are chosen. 
Furthermore, the symmetry of the momentum distribution may be recovered by either modifying the time window for specific solutions or by conducting averaging techniques such as unit-cell averaging. However, the agreement between the results calculated using the saddle-point method and the numerical integration is only observed when the energy-conservation condition is imposed. This holds regardless of the time window in which the saddle-point solutions are positioned and without the necessity for subsequent averaging.   

In practice, the laser field incident on a specific region of space is not ideally monochromatic and will contain a range of frequency components. On top of this, there will exist a non-uniform spatial intensity profile within the laser focus. Both of these effects will lead to a broadening of the ATI peaks \cite{Boning2019} and a failure of the energy conservation condition. However, monochromatic fields are still frequently utilized to a great extent in the theoretical study of strong-field ionization processes. It is questionable whether the broadening of the ATI peaks, which can be observed due to only considering ionization from a restricted subset of the saddle point times, could accurately reflect the precise way in which an experimentally determined distribution depends upon the subtle details of the experimental setup.  

Finally, the present results leave a few open questions. First, do the present findings also hold for theoretical approaches in which the residual binding potential is fully incorporated in the electron's continuum propagation? 
It is difficult to evaluate this in the same way as above since numerical integration approaches do not exist for Coulomb-distorted methods, which are generally reliant upon making semiclassical approximations to approximate a path integral form of the transition amplitude \cite{Lai2015a}. This provides additional challenges in the process of establishing precisely what effect the saddle-point approximation has within such theories, and is the main justification as to why the SFA has been used throughout this article. 
On the one hand, it was shown that the inter-cycle interference condition giving the ATI peaks and the Dirac Delta comb in Eq.~\eqref{eq:deltacomb}  remains the same for Coulomb-distorted approaches \cite{Maxwell2017}, which would be one argument supporting the validity of the condition in this article. Nonetheless, for Coulomb-distorted theories the field-dressed momentum is no longer conserved, during the electron propagation, in contrast to the SFA framework. 
Second, unit cell averaging has been successfully used in order to compare CQSFA calculations with experiments, with excellent agreement even for subtle features such as modulations in spider-like holographic fringes \cite{Werby2021}. Although an additional filtering was used in both experiment and theory, this filtering did not influence these modulations and had the sole aim of removing high-frequency oscillations, such as ATI rings. Would the present condition work better, and if so, what would be the improvements and the reasons behind it? These issues remain to be understood, and may play an important role in the modelling of strong-field ionization and photoelectron holography. 

\begin{acknowledgments}
We gratefully acknowledge support by the Ministry for Science, Higher Education and Youth, Canton Sarajevo, Bosnia and Herzegovina and funding by grants No.EP/T019530/1 and EP/T517793/1, from the UK Engineering and Physical Sciences Research Council (EPSRC). D. Habibovi\' c thanks UCL, where this work was partly carried out, for its kind hospitality. 
\end{acknowledgments}


\begin{thebibliography}{40}
\expandafter\ifx\csname natexlab\endcsname\relax\def\natexlab#1{#1}\fi
\expandafter\ifx\csname bibnamefont\endcsname\relax
  \def\bibnamefont#1{#1}\fi
\expandafter\ifx\csname bibfnamefont\endcsname\relax
  \def\bibfnamefont#1{#1}\fi
\expandafter\ifx\csname citenamefont\endcsname\relax
  \def\citenamefont#1{#1}\fi
\expandafter\ifx\csname url\endcsname\relax
  \def\url#1{\texttt{#1}}\fi
\expandafter\ifx\csname urlprefix\endcsname\relax\def\urlprefix{URL }\fi
\providecommand{\bibinfo}[2]{#2}
\providecommand{\eprint}[2][]{\url{#2}}

\bibitem[{\citenamefont{Agostini et~al.}(1979)\citenamefont{Agostini, Fabre,
  Mainfray, Petite, and Rahman}}]{Agostini1979}
\bibinfo{author}{\bibfnamefont{P.}~\bibnamefont{Agostini}},
  \bibinfo{author}{\bibfnamefont{F.}~\bibnamefont{Fabre}},
  \bibinfo{author}{\bibfnamefont{G.}~\bibnamefont{Mainfray}},
  \bibinfo{author}{\bibfnamefont{G.}~\bibnamefont{Petite}}, \bibnamefont{and}
  \bibinfo{author}{\bibfnamefont{N.~K.} \bibnamefont{Rahman}},
  \bibinfo{journal}{Phys. Rev. Lett.} \textbf{\bibinfo{volume}{42}},
  \bibinfo{pages}{1127} (\bibinfo{year}{1979}).

\bibitem[{\citenamefont{Milo\v{s}evi\'c and Ehlotzky}(2003)}]{Milosevic2003}
\bibinfo{author}{\bibfnamefont{D.~B.} \bibnamefont{Milo\v{s}evi\'c}}
  \bibnamefont{and} \bibinfo{author}{\bibfnamefont{F.}~\bibnamefont{Ehlotzky}},
  \bibinfo{journal}{Adv. At. Mol. Opt. Phy.} \textbf{\bibinfo{volume}{49}},
  \bibinfo{pages}{373} (\bibinfo{year}{2003}).

\bibitem[{\citenamefont{{Becker} et~al.}(2002)\citenamefont{{Becker},
  {Grasbon}, {Kopold}, {Milo{\v s}evi{\'c}}, {Paulus}, and
  {Walther}}}]{Becker2002}
\bibinfo{author}{\bibfnamefont{W.}~\bibnamefont{{Becker}}},
  \bibinfo{author}{\bibfnamefont{F.}~\bibnamefont{{Grasbon}}},
  \bibinfo{author}{\bibfnamefont{R.}~\bibnamefont{{Kopold}}},
  \bibinfo{author}{\bibfnamefont{D.~B.} \bibnamefont{{Milo{\v s}evi{\'c}}}},
  \bibinfo{author}{\bibfnamefont{G.~G.} \bibnamefont{{Paulus}}},
  \bibnamefont{and}
  \bibinfo{author}{\bibfnamefont{H.}~\bibnamefont{{Walther}}},
  \bibinfo{journal}{Adv. At. Mol. Opt. Phy.} \textbf{\bibinfo{volume}{48}},
  \bibinfo{pages}{35} (\bibinfo{year}{2002}).

\bibitem[{\citenamefont{Agostini and DiMauro}(2004)}]{Agostini2004}
\bibinfo{author}{\bibfnamefont{P.}~\bibnamefont{Agostini}} \bibnamefont{and}
  \bibinfo{author}{\bibfnamefont{L.~F.} \bibnamefont{DiMauro}},
  \bibinfo{journal}{Rep. Prog. Phys.} \textbf{\bibinfo{volume}{67}},
  \bibinfo{pages}{813} (\bibinfo{year}{2004}).

\bibitem[{\citenamefont{Scrinzi et~al.}(2005)\citenamefont{Scrinzi, Ivanov,
  Kienberger, and Villeneuve}}]{Scrinzi2006}
\bibinfo{author}{\bibfnamefont{A.}~\bibnamefont{Scrinzi}},
  \bibinfo{author}{\bibfnamefont{M.~Y.} \bibnamefont{Ivanov}},
  \bibinfo{author}{\bibfnamefont{R.}~\bibnamefont{Kienberger}},
  \bibnamefont{and} \bibinfo{author}{\bibfnamefont{D.~M.}
  \bibnamefont{Villeneuve}}, \bibinfo{journal}{J. Phys. B}
  \textbf{\bibinfo{volume}{39}}, \bibinfo{pages}{R1} (\bibinfo{year}{2005}).

\bibitem[{\citenamefont{Lein}(2007)}]{Lein2007}
\bibinfo{author}{\bibfnamefont{M.}~\bibnamefont{Lein}}, \bibinfo{journal}{J.
  Phys. B} \textbf{\bibinfo{volume}{40}}, \bibinfo{pages}{R135}
  (\bibinfo{year}{2007}).

\bibitem[{\citenamefont{Krausz and Ivanov}(2009)}]{Krausz2009}
\bibinfo{author}{\bibfnamefont{F.}~\bibnamefont{Krausz}} \bibnamefont{and}
  \bibinfo{author}{\bibfnamefont{M.}~\bibnamefont{Ivanov}},
  \bibinfo{journal}{Rev. Mod. Phys.} \textbf{\bibinfo{volume}{81}},
  \bibinfo{pages}{163} (\bibinfo{year}{2009}).

\bibitem[{\citenamefont{Popruzhenko}(2014)}]{Popruzhenko2014}
\bibinfo{author}{\bibfnamefont{S.~V.} \bibnamefont{Popruzhenko}},
  \bibinfo{journal}{J. Phys. B} \textbf{\bibinfo{volume}{47}},
  \bibinfo{pages}{204001} (\bibinfo{year}{2014}).

\bibitem[{\citenamefont{Amini et~al.}(2019)\citenamefont{Amini, Biegert,
  Calegari, Chac{\'{o}}n, Ciappina, Dauphin, Efimov, Figueira~de
  Morisson~Faria, Giergiel, Gniewek et~al.}}]{Amini2019}
\bibinfo{author}{\bibfnamefont{K.}~\bibnamefont{Amini}},
  \bibinfo{author}{\bibfnamefont{J.}~\bibnamefont{Biegert}},
  \bibinfo{author}{\bibfnamefont{F.}~\bibnamefont{Calegari}},
  \bibinfo{author}{\bibfnamefont{A.}~\bibnamefont{Chac{\'{o}}n}},
  \bibinfo{author}{\bibfnamefont{M.~F.} \bibnamefont{Ciappina}},
  \bibinfo{author}{\bibfnamefont{A.}~\bibnamefont{Dauphin}},
  \bibinfo{author}{\bibfnamefont{D.~K.} \bibnamefont{Efimov}},
  \bibinfo{author}{\bibfnamefont{C.}~\bibnamefont{Figueira~de Morisson~Faria}},
  \bibinfo{author}{\bibfnamefont{K.}~\bibnamefont{Giergiel}},
  \bibinfo{author}{\bibfnamefont{P.}~\bibnamefont{Gniewek}},
  \bibnamefont{et~al.}, \bibinfo{journal}{Rep. Prog. Phys.}
  \textbf{\bibinfo{volume}{82}}, \bibinfo{pages}{116001}
  (\bibinfo{year}{2019}).

\bibitem[{\citenamefont{de~Morisson~Faria and Maxwell}(2020)}]{Faria2020}
\bibinfo{author}{\bibfnamefont{C.~F.} \bibnamefont{de~Morisson~Faria}}
  \bibnamefont{and} \bibinfo{author}{\bibfnamefont{A.~S.}
  \bibnamefont{Maxwell}}, \bibinfo{journal}{Rep. Prog. Phys.}
  \textbf{\bibinfo{volume}{83}}, \bibinfo{pages}{034401}
  (\bibinfo{year}{2020}).

\bibitem[{\citenamefont{Sali{\`e}res et~al.}(2001)\citenamefont{Sali{\`e}res,
  Carr{\'e}, Le~D{\'e}roff, Grasbon, Paulus, Walther, Kopold, Becker, Milo{\v
  s}evi{\'c}, Sanpera et~al.}}]{Salieres2001}
\bibinfo{author}{\bibfnamefont{P.}~\bibnamefont{Sali{\`e}res}},
  \bibinfo{author}{\bibfnamefont{B.}~\bibnamefont{Carr{\'e}}},
  \bibinfo{author}{\bibfnamefont{L.}~\bibnamefont{Le~D{\'e}roff}},
  \bibinfo{author}{\bibfnamefont{F.}~\bibnamefont{Grasbon}},
  \bibinfo{author}{\bibfnamefont{G.~G.} \bibnamefont{Paulus}},
  \bibinfo{author}{\bibfnamefont{H.}~\bibnamefont{Walther}},
  \bibinfo{author}{\bibfnamefont{R.}~\bibnamefont{Kopold}},
  \bibinfo{author}{\bibfnamefont{W.}~\bibnamefont{Becker}},
  \bibinfo{author}{\bibfnamefont{D.~B.} \bibnamefont{Milo{\v s}evi{\'c}}},
  \bibinfo{author}{\bibfnamefont{A.}~\bibnamefont{Sanpera}},
  \bibnamefont{et~al.}, \bibinfo{journal}{Science}
  \textbf{\bibinfo{volume}{292}}, \bibinfo{pages}{902} (\bibinfo{year}{2001}),
  ISSN \bibinfo{issn}{0036-8075}.

\bibitem[{\citenamefont{Milo{\v s}evi{\'c} et~al.}(2024)\citenamefont{Milo{\v
  s}evi{\'c}, Ja{\v s}arevi{\' c}, Habibovi{\' c}, Hasovi{\' c}, {\v C}erki{\'
  c}, and Becker}}]{Milosevic2024TR}
\bibinfo{author}{\bibfnamefont{D.~B.} \bibnamefont{Milo{\v s}evi{\'c}}},
  \bibinfo{author}{\bibfnamefont{A.~S.} \bibnamefont{Ja{\v s}arevi{\' c}}},
  \bibinfo{author}{\bibfnamefont{D.}~\bibnamefont{Habibovi{\' c}}},
  \bibinfo{author}{\bibfnamefont{E.}~\bibnamefont{Hasovi{\' c}}},
  \bibinfo{author}{\bibfnamefont{A.}~\bibnamefont{{\v C}erki{\' c}}},
  \bibnamefont{and} \bibinfo{author}{\bibfnamefont{W.}~\bibnamefont{Becker}},
  \bibinfo{journal}{J. Phys. A} \textbf{\bibinfo{volume}{XX}},
  \bibinfo{pages}{XX} (\bibinfo{year}{2024}).

\bibitem[{\citenamefont{Lewenstein et~al.}(1994)\citenamefont{Lewenstein,
  Balcou, Ivanov, L'Huillier, and Corkum}}]{Lewenstein1994}
\bibinfo{author}{\bibfnamefont{M.}~\bibnamefont{Lewenstein}},
  \bibinfo{author}{\bibfnamefont{P.}~\bibnamefont{Balcou}},
  \bibinfo{author}{\bibfnamefont{M.~Y.} \bibnamefont{Ivanov}},
  \bibinfo{author}{\bibfnamefont{A.}~\bibnamefont{L'Huillier}},
  \bibnamefont{and} \bibinfo{author}{\bibfnamefont{P.~B.}
  \bibnamefont{Corkum}}, \bibinfo{journal}{Phys. Rev. A}
  \textbf{\bibinfo{volume}{49}}, \bibinfo{pages}{2117} (\bibinfo{year}{1994}).

\bibitem[{\citenamefont{Milo\ifmmode \check{s}\else
  \v{s}\fi{}evi\ifmmode~\acute{c}\else \'{c}\fi{} and
  Becker}(2002)}]{Milosevic2002}
\bibinfo{author}{\bibfnamefont{D.~B.} \bibnamefont{Milo\ifmmode \check{s}\else
  \v{s}\fi{}evi\ifmmode~\acute{c}\else \'{c}\fi{}}} \bibnamefont{and}
  \bibinfo{author}{\bibfnamefont{W.}~\bibnamefont{Becker}},
  \bibinfo{journal}{Phys. Rev. A} \textbf{\bibinfo{volume}{66}},
  \bibinfo{pages}{063417} (\bibinfo{year}{2002}).

\bibitem[{\citenamefont{Od\ifmmode~\check{z}\else \v{z}\fi{}ak and Milo\ifmmode
  \check{s}\else \v{s}\fi{}evi\ifmmode~\acute{c}\else
  \'{c}\fi{}}(2005)}]{Odzak2005}
\bibinfo{author}{\bibfnamefont{S.}~\bibnamefont{Od\ifmmode~\check{z}\else
  \v{z}\fi{}ak}} \bibnamefont{and} \bibinfo{author}{\bibfnamefont{D.~B.}
  \bibnamefont{Milo\ifmmode \check{s}\else \v{s}\fi{}evi\ifmmode~\acute{c}\else
  \'{c}\fi{}}}, \bibinfo{journal}{Phys. Rev. A} \textbf{\bibinfo{volume}{72}},
  \bibinfo{pages}{033407} (\bibinfo{year}{2005}).

\bibitem[{\citenamefont{Zhou et~al.}(2021)\citenamefont{Zhou, Guo, Quan, Wei,
  Zhao, Xu, Xiao, Sun, Wang, Lai et~al.}}]{Zhou2021}
\bibinfo{author}{\bibfnamefont{Y.}~\bibnamefont{Zhou}},
  \bibinfo{author}{\bibfnamefont{L.}~\bibnamefont{Guo}},
  \bibinfo{author}{\bibfnamefont{W.}~\bibnamefont{Quan}},
  \bibinfo{author}{\bibfnamefont{M.}~\bibnamefont{Wei}},
  \bibinfo{author}{\bibfnamefont{M.}~\bibnamefont{Zhao}},
  \bibinfo{author}{\bibfnamefont{S.}~\bibnamefont{Xu}},
  \bibinfo{author}{\bibfnamefont{Z.}~\bibnamefont{Xiao}},
  \bibinfo{author}{\bibfnamefont{R.}~\bibnamefont{Sun}},
  \bibinfo{author}{\bibfnamefont{Y.}~\bibnamefont{Wang}},
  \bibinfo{author}{\bibfnamefont{X.}~\bibnamefont{Lai}}, \bibnamefont{et~al.},
  \textbf{\bibinfo{volume}{54}}, \bibinfo{pages}{144008}
  (\bibinfo{year}{2021}).

\bibitem[{\citenamefont{Guo et~al.}(2022)\citenamefont{Guo, Chen, Hu, Lu, Han,
  Zhang, and Chen}}]{Guo2022}
\bibinfo{author}{\bibfnamefont{L.}~\bibnamefont{Guo}},
  \bibinfo{author}{\bibfnamefont{S.}~\bibnamefont{Chen}},
  \bibinfo{author}{\bibfnamefont{S.}~\bibnamefont{Hu}},
  \bibinfo{author}{\bibfnamefont{R.}~\bibnamefont{Lu}},
  \bibinfo{author}{\bibfnamefont{S.}~\bibnamefont{Han}},
  \bibinfo{author}{\bibfnamefont{J.}~\bibnamefont{Zhang}}, \bibnamefont{and}
  \bibinfo{author}{\bibfnamefont{J.}~\bibnamefont{Chen}},
  \textbf{\bibinfo{volume}{55}}, \bibinfo{pages}{225401}
  (\bibinfo{year}{2022}).

\bibitem[{\citenamefont{Boroumand et~al.}(2022)\citenamefont{Boroumand, Thorpe,
  Parks, and Brabec}}]{Boroumand2022}
\bibinfo{author}{\bibfnamefont{N.}~\bibnamefont{Boroumand}},
  \bibinfo{author}{\bibfnamefont{A.}~\bibnamefont{Thorpe}},
  \bibinfo{author}{\bibfnamefont{A.~M.} \bibnamefont{Parks}}, \bibnamefont{and}
  \bibinfo{author}{\bibfnamefont{T.}~\bibnamefont{Brabec}},
  \textbf{\bibinfo{volume}{55}}, \bibinfo{pages}{213001}
  (\bibinfo{year}{2022}).

\bibitem[{\citenamefont{Carlsen et~al.}(2024)\citenamefont{Carlsen, Hansen,
  Madsen, and Maxwell}}]{Carlsen2024}
\bibinfo{author}{\bibfnamefont{M.~B.} \bibnamefont{Carlsen}},
  \bibinfo{author}{\bibfnamefont{E.}~\bibnamefont{Hansen}},
  \bibinfo{author}{\bibfnamefont{L.~B.} \bibnamefont{Madsen}},
  \bibnamefont{and} \bibinfo{author}{\bibfnamefont{A.~S.}
  \bibnamefont{Maxwell}}, \bibinfo{journal}{New Journal of Physics}
  \textbf{\bibinfo{volume}{26}}, \bibinfo{pages}{023025}
  (\bibinfo{year}{2024}).

\bibitem[{\citenamefont{Habibovi{\' c} and Milo{\v
  s}evi{\'c}}(2020)}]{Habibovic2020}
\bibinfo{author}{\bibfnamefont{D.}~\bibnamefont{Habibovi{\' c}}}
  \bibnamefont{and} \bibinfo{author}{\bibfnamefont{D.~B.} \bibnamefont{Milo{\v
  s}evi{\'c}}}, \bibinfo{journal}{Photonics} \textbf{\bibinfo{volume}{7}}
  (\bibinfo{year}{2020}), ISSN \bibinfo{issn}{2304-6732}.

\bibitem[{\citenamefont{Fang and Liu}(2021)}]{Fang2021}
\bibinfo{author}{\bibfnamefont{Y.}~\bibnamefont{Fang}} \bibnamefont{and}
  \bibinfo{author}{\bibfnamefont{Y.}~\bibnamefont{Liu}},
  \bibinfo{journal}{Phys. Rev. A} \textbf{\bibinfo{volume}{103}},
  \bibinfo{pages}{033116} (\bibinfo{year}{2021}).

\bibitem[{\citenamefont{Milo{\v s}evi{\'c} and Habibovi{\'
  c}}(2023)}]{Milosevic2023}
\bibinfo{author}{\bibfnamefont{D.~B.} \bibnamefont{Milo{\v s}evi{\'c}}}
  \bibnamefont{and} \bibinfo{author}{\bibfnamefont{D.}~\bibnamefont{Habibovi{\'
  c}}}, \bibinfo{journal}{Phys. Chem. Chem. Phys.}
  \textbf{\bibinfo{volume}{25}}, \bibinfo{pages}{28848} (\bibinfo{year}{2023}).

\bibitem[{\citenamefont{Shaaran et~al.}(2010)\citenamefont{Shaaran, Nygren, and
  Figueira~de Morisson~Faria}}]{Shaaran2010}
\bibinfo{author}{\bibfnamefont{T.}~\bibnamefont{Shaaran}},
  \bibinfo{author}{\bibfnamefont{M.~T.} \bibnamefont{Nygren}},
  \bibnamefont{and} \bibinfo{author}{\bibfnamefont{C.}~\bibnamefont{Figueira~de
  Morisson~Faria}}, \bibinfo{journal}{Phys. Rev. A}
  \textbf{\bibinfo{volume}{81}}, \bibinfo{pages}{063413}
  (\bibinfo{year}{2010}).

\bibitem[{\citenamefont{Shaaran et~al.}(2012)\citenamefont{Shaaran, Figueira~de
  Morisson~Faria, and Schomerus}}]{Shaaran2012}
\bibinfo{author}{\bibfnamefont{T.}~\bibnamefont{Shaaran}},
  \bibinfo{author}{\bibfnamefont{C.}~\bibnamefont{Figueira~de Morisson~Faria}},
  \bibnamefont{and}
  \bibinfo{author}{\bibfnamefont{H.}~\bibnamefont{Schomerus}},
  \bibinfo{journal}{Phys. Rev. A} \textbf{\bibinfo{volume}{85}},
  \bibinfo{pages}{023423} (\bibinfo{year}{2012}).

\bibitem[{\citenamefont{Maxwell and Figueira~de
  Morisson~Faria}(2015)}]{Maxwell2015}
\bibinfo{author}{\bibfnamefont{A.~S.} \bibnamefont{Maxwell}} \bibnamefont{and}
  \bibinfo{author}{\bibfnamefont{C.}~\bibnamefont{Figueira~de Morisson~Faria}},
  \bibinfo{journal}{Phys. Rev. A} \textbf{\bibinfo{volume}{92}},
  \bibinfo{pages}{023421} (\bibinfo{year}{2015}).

\bibitem[{\citenamefont{Popruzhenko and Bauer}(2008)}]{Popruzhenko2008}
\bibinfo{author}{\bibfnamefont{S.}~\bibnamefont{Popruzhenko}} \bibnamefont{and}
  \bibinfo{author}{\bibfnamefont{D.}~\bibnamefont{Bauer}}, \bibinfo{journal}{J.
  Mod. Opt.} \textbf{\bibinfo{volume}{55}}, \bibinfo{pages}{2573}
  (\bibinfo{year}{2008}).

\bibitem[{\citenamefont{Smirnova et~al.}(2008)\citenamefont{Smirnova, Spanner,
  and Ivanov}}]{Smirnova2008}
\bibinfo{author}{\bibfnamefont{O.}~\bibnamefont{Smirnova}},
  \bibinfo{author}{\bibfnamefont{M.}~\bibnamefont{Spanner}}, \bibnamefont{and}
  \bibinfo{author}{\bibfnamefont{M.}~\bibnamefont{Ivanov}},
  \bibinfo{journal}{Phys. Rev. A} \textbf{\bibinfo{volume}{77}},
  \bibinfo{pages}{033407} (\bibinfo{year}{2008}).

\bibitem[{\citenamefont{Yan et~al.}(2010)\citenamefont{Yan, Popruzhenko,
  Vrakking, and Bauer}}]{Yan2010}
\bibinfo{author}{\bibfnamefont{T.-M.} \bibnamefont{Yan}},
  \bibinfo{author}{\bibfnamefont{S.~V.} \bibnamefont{Popruzhenko}},
  \bibinfo{author}{\bibfnamefont{M.~J.~J.} \bibnamefont{Vrakking}},
  \bibnamefont{and} \bibinfo{author}{\bibfnamefont{D.}~\bibnamefont{Bauer}},
  \bibinfo{journal}{Phys. Rev. Lett.} \textbf{\bibinfo{volume}{105}},
  \bibinfo{pages}{253002} (\bibinfo{year}{2010}).

\bibitem[{\citenamefont{Yan and Bauer}(2012)}]{Yan2012}
\bibinfo{author}{\bibfnamefont{T.-M.} \bibnamefont{Yan}} \bibnamefont{and}
  \bibinfo{author}{\bibfnamefont{D.}~\bibnamefont{Bauer}},
  \bibinfo{journal}{Phys. Rev. A} \textbf{\bibinfo{volume}{86}},
  \bibinfo{pages}{053403} (\bibinfo{year}{2012}).

\bibitem[{\citenamefont{Lai et~al.}(2015)\citenamefont{Lai, Poli, Schomerus,
  and {Figueira de Morisson Faria}}}]{Lai2015}
\bibinfo{author}{\bibfnamefont{X.~Y.} \bibnamefont{Lai}},
  \bibinfo{author}{\bibfnamefont{C.}~\bibnamefont{Poli}},
  \bibinfo{author}{\bibfnamefont{H.}~\bibnamefont{Schomerus}},
  \bibnamefont{and} \bibinfo{author}{\bibfnamefont{C.}~\bibnamefont{{Figueira
  de Morisson Faria}}}, \bibinfo{journal}{Phys. Rev. A}
  \textbf{\bibinfo{volume}{92}}, \bibinfo{pages}{043407}
  (\bibinfo{year}{2015}), ISSN \bibinfo{issn}{10941622}.

\bibitem[{\citenamefont{Pisanty and Ivanov}(2016)}]{Pisanty2016}
\bibinfo{author}{\bibfnamefont{E.}~\bibnamefont{Pisanty}} \bibnamefont{and}
  \bibinfo{author}{\bibfnamefont{M.}~\bibnamefont{Ivanov}},
  \bibinfo{journal}{Phys. Rev. A} \textbf{\bibinfo{volume}{93}},
  \bibinfo{pages}{043408} (\bibinfo{year}{2016}).

\bibitem[{\citenamefont{Lai et~al.}(2017)\citenamefont{Lai, Yu, Huang, Hua,
  Gong, Quan, Figueira~de Morisson~Faria, and Liu}}]{Lai2017}
\bibinfo{author}{\bibfnamefont{X.~Y.} \bibnamefont{Lai}},
  \bibinfo{author}{\bibfnamefont{S.}~\bibnamefont{Yu}},
  \bibinfo{author}{\bibfnamefont{Y.}~\bibnamefont{Huang}},
  \bibinfo{author}{\bibfnamefont{L.}~\bibnamefont{Hua}},
  \bibinfo{author}{\bibfnamefont{C.}~\bibnamefont{Gong}},
  \bibinfo{author}{\bibfnamefont{W.}~\bibnamefont{Quan}},
  \bibinfo{author}{\bibfnamefont{C.}~\bibnamefont{Figueira~de Morisson~Faria}},
  \bibnamefont{and} \bibinfo{author}{\bibfnamefont{X.}~\bibnamefont{Liu}},
  \bibinfo{journal}{Phys. Rev. A} \textbf{\bibinfo{volume}{96}},
  \bibinfo{pages}{013414} (\bibinfo{year}{2017}), ISSN
  \bibinfo{issn}{24699934}, \eprint{1703.04123}.

\bibitem[{\citenamefont{Madsen}(2022{\natexlab{a}})}]{Madsen2022a}
\bibinfo{author}{\bibfnamefont{L.~B.} \bibnamefont{Madsen}},
  \bibinfo{journal}{Phys. Rev. A} \textbf{\bibinfo{volume}{106}},
  \bibinfo{pages}{043118} (\bibinfo{year}{2022}{\natexlab{a}}).

\bibitem[{\citenamefont{Madsen}(2022{\natexlab{b}})}]{Madsen2022b}
\bibinfo{author}{\bibfnamefont{L.~B.} \bibnamefont{Madsen}},
  \bibinfo{journal}{Phys. Rev. A} \textbf{\bibinfo{volume}{105}},
  \bibinfo{pages}{043107} (\bibinfo{year}{2022}{\natexlab{b}}).

\bibitem[{\citenamefont{Ja{\v s}arevi{\' c} et~al.}(2024)\citenamefont{Ja{\v
  s}arevi{\' c}, Habibovi{\' c}, and Milo{\v s}evi{\'c}}}]{Jasarevic2024}
\bibinfo{author}{\bibfnamefont{A.~S.} \bibnamefont{Ja{\v s}arevi{\' c}}},
  \bibinfo{author}{\bibfnamefont{D.}~\bibnamefont{Habibovi{\' c}}},
  \bibnamefont{and} \bibinfo{author}{\bibfnamefont{D.~B.} \bibnamefont{Milo{\v
  s}evi{\'c}}}, \bibinfo{journal}{Phys. Rev. A} \textbf{\bibinfo{volume}{???}},
  \bibinfo{pages}{???} (\bibinfo{year}{2024}).

\bibitem[{\citenamefont{Kahved{\v z}i{\' c} and Gr\"afe}(2022)}]{Kahvedzic2022}
\bibinfo{author}{\bibfnamefont{R.}~\bibnamefont{Kahved{\v z}i{\' c}}}
  \bibnamefont{and} \bibinfo{author}{\bibfnamefont{S.}~\bibnamefont{Gr\"afe}},
  \bibinfo{journal}{Phys. Rev. A} \textbf{\bibinfo{volume}{106}},
  \bibinfo{pages}{043122} (\bibinfo{year}{2022}).

\bibitem[{\citenamefont{Ja{\v{s}}arevi{\'{c}}
  et~al.}(2020)\citenamefont{Ja{\v{s}}arevi{\'{c}}, Hasovi{\'{c}}, Kopold,
  Becker, and Milo{\v{s}}evi{\'{c}}}}]{Jasarevic2020}
\bibinfo{author}{\bibfnamefont{A.}~\bibnamefont{Ja{\v{s}}arevi{\'{c}}}},
  \bibinfo{author}{\bibfnamefont{E.}~\bibnamefont{Hasovi{\'{c}}}},
  \bibinfo{author}{\bibfnamefont{R.}~\bibnamefont{Kopold}},
  \bibinfo{author}{\bibfnamefont{W.}~\bibnamefont{Becker}}, \bibnamefont{and}
  \bibinfo{author}{\bibfnamefont{D.~B.} \bibnamefont{Milo{\v{s}}evi{\'{c}}}},
  \bibinfo{journal}{J. Phys. A} \textbf{\bibinfo{volume}{53}},
  \bibinfo{pages}{125201} (\bibinfo{year}{2020}).

\bibitem[{\citenamefont{Werby et~al.}(2021)\citenamefont{Werby, Maxwell,
  Forbes, Bucksbaum, and Faria}}]{Werby2021}
\bibinfo{author}{\bibfnamefont{N.}~\bibnamefont{Werby}},
  \bibinfo{author}{\bibfnamefont{A.~S.} \bibnamefont{Maxwell}},
  \bibinfo{author}{\bibfnamefont{R.}~\bibnamefont{Forbes}},
  \bibinfo{author}{\bibfnamefont{P.~H.} \bibnamefont{Bucksbaum}},
  \bibnamefont{and} \bibinfo{author}{\bibfnamefont{C.~F. d.~M.}
  \bibnamefont{Faria}}, \bibinfo{journal}{Phys. Rev. A}
  \textbf{\bibinfo{volume}{104}}, \bibinfo{pages}{013109}
  (\bibinfo{year}{2021}).

\bibitem[{\citenamefont{Rook et~al.}(2024)\citenamefont{Rook,
  Habibovi\ifmmode~\acute{c}\else \'{c}\fi{}, Rodriguez, Milo\ifmmode
  \check{s}\else \v{s}\fi{}evi\ifmmode~\acute{c}\else \'{c}\fi{}, and
  Faria}}]{rook2024}
\bibinfo{author}{\bibfnamefont{T.}~\bibnamefont{Rook}},
  \bibinfo{author}{\bibfnamefont{D.}~\bibnamefont{Habibovi\ifmmode~\acute{c}\else
  \'{c}\fi{}}}, \bibinfo{author}{\bibfnamefont{L.~C.} \bibnamefont{Rodriguez}},
  \bibinfo{author}{\bibfnamefont{D.~B.} \bibnamefont{Milo\ifmmode
  \check{s}\else \v{s}\fi{}evi\ifmmode~\acute{c}\else \'{c}\fi{}}},
  \bibnamefont{and} \bibinfo{author}{\bibfnamefont{C.~F. d.~M.}
  \bibnamefont{Faria}}, \bibinfo{journal}{Phys. Rev. A}
  \textbf{\bibinfo{volume}{109}}, \bibinfo{pages}{033115}
  (\bibinfo{year}{2024}).

\bibitem[{\citenamefont{Habibovi\'{c} et~al.}(2021)\citenamefont{Habibovi\'{c},
  Becker, and Milo\v{s}evi\'{c}}}]{Habibovic2021}
\bibinfo{author}{\bibfnamefont{D.}~\bibnamefont{Habibovi\'{c}}},
  \bibinfo{author}{\bibfnamefont{W.}~\bibnamefont{Becker}}, \bibnamefont{and}
  \bibinfo{author}{\bibfnamefont{D.~B.} \bibnamefont{Milo\v{s}evi\'{c}}},
  \bibinfo{journal}{Symmetry} \textbf{\bibinfo{volume}{14}},
  \bibinfo{pages}{1566} (\bibinfo{year}{2021}).

\end{thebibliography}


\begin{thebibliography}{66}%
\makeatletter
\providecommand \@ifxundefined [1]{%
 \@ifx{#1\undefined}
}%
\providecommand \@ifnum [1]{%
 \ifnum #1\expandafter \@firstoftwo
 \else \expandafter \@secondoftwo
 \fi
}%
\providecommand \@ifx [1]{%
 \ifx #1\expandafter \@firstoftwo
 \else \expandafter \@secondoftwo
 \fi
}%
\providecommand \natexlab [1]{#1}%
\providecommand \enquote  [1]{``#1''}%
\providecommand \bibnamefont  [1]{#1}%
\providecommand \bibfnamefont [1]{#1}%
\providecommand \citenamefont [1]{#1}%
\providecommand \href@noop [0]{\@secondoftwo}%
\providecommand \href [0]{\begingroup \@sanitize@url \@href}%
\providecommand \@href[1]{\@@startlink{#1}\@@href}%
\providecommand \@@href[1]{\endgroup#1\@@endlink}%
\providecommand \@sanitize@url [0]{\catcode `\\12\catcode `\$12\catcode `\&12\catcode `\#12\catcode `\^12\catcode `\_12\catcode `\%12\relax}%
\providecommand \@@startlink[1]{}%
\providecommand \@@endlink[0]{}%
\providecommand \url  [0]{\begingroup\@sanitize@url \@url }%
\providecommand \@url [1]{\endgroup\@href {#1}{\urlprefix }}%
\providecommand \urlprefix  [0]{URL }%
\providecommand \Eprint [0]{\href }%
\providecommand \doibase [0]{https://doi.org/}%
\providecommand \selectlanguage [0]{\@gobble}%
\providecommand \bibinfo  [0]{\@secondoftwo}%
\providecommand \bibfield  [0]{\@secondoftwo}%
\providecommand \translation [1]{[#1]}%
\providecommand \BibitemOpen [0]{}%
\providecommand \bibitemStop [0]{}%
\providecommand \bibitemNoStop [0]{.\EOS\space}%
\providecommand \EOS [0]{\spacefactor3000\relax}%
\providecommand \BibitemShut  [1]{\csname bibitem#1\endcsname}%
\let\auto@bib@innerbib\@empty
\bibitem [{\citenamefont {Agostini}\ \emph {et~al.}(1979)\citenamefont {Agostini}, \citenamefont {Fabre}, \citenamefont {Mainfray}, \citenamefont {Petite},\ and\ \citenamefont {Rahman}}]{Agostini1979}%
  \BibitemOpen
  \bibfield  {author} {\bibinfo {author} {\bibfnamefont {P.}~\bibnamefont {Agostini}}, \bibinfo {author} {\bibfnamefont {F.}~\bibnamefont {Fabre}}, \bibinfo {author} {\bibfnamefont {G.}~\bibnamefont {Mainfray}}, \bibinfo {author} {\bibfnamefont {G.}~\bibnamefont {Petite}},\ and\ \bibinfo {author} {\bibfnamefont {N.~K.}\ \bibnamefont {Rahman}},\ }\bibfield  {title} {\bibinfo {title} {Free-free transitions following six-photon ionization of xenon atoms},\ }\href {https://doi.org/10.1103/PhysRevLett.42.1127} {\bibfield  {journal} {\bibinfo  {journal} {Phys. Rev. Lett.}\ }\textbf {\bibinfo {volume} {42}},\ \bibinfo {pages} {1127} (\bibinfo {year} {1979})}\BibitemShut {NoStop}%
\bibitem [{\citenamefont {Milo\v{s}evi\'c}\ and\ \citenamefont {Ehlotzky}(2003)}]{Milosevic2003}%
  \BibitemOpen
  \bibfield  {author} {\bibinfo {author} {\bibfnamefont {D.~B.}\ \bibnamefont {Milo\v{s}evi\'c}}\ and\ \bibinfo {author} {\bibfnamefont {F.}~\bibnamefont {Ehlotzky}},\ }\bibfield  {title} {\bibinfo {title} {{Scattering and Reaction Processes in Powerful Laser Fields}},\ }\href {https://doi.org/https://doi.org/10.1016/S1049-250X(03)80007-1} {\bibfield  {journal} {\bibinfo  {journal} {Adv. At. Mol. Opt. Phy.}\ }\textbf {\bibinfo {volume} {49}},\ \bibinfo {pages} {373} (\bibinfo {year} {2003})}\BibitemShut {NoStop}%
\bibitem [{\citenamefont {Popruzhenko}(2014)}]{Popruzhenko2014a}%
  \BibitemOpen
  \bibfield  {author} {\bibinfo {author} {\bibfnamefont {S.~V.}\ \bibnamefont {Popruzhenko}},\ }\bibfield  {title} {\bibinfo {title} {Keldysh theory of strong field ionization: history, applications, difficulties and perspectives},\ }\href {http://stacks.iop.org/0953-4075/47/i=20/a=204001} {\bibfield  {journal} {\bibinfo  {journal} {J. Phys. B}\ }\textbf {\bibinfo {volume} {47}},\ \bibinfo {pages} {204001} (\bibinfo {year} {2014})}\BibitemShut {NoStop}%
\bibitem [{\citenamefont {Amini}\ \emph {et~al.}(2019)\citenamefont {Amini}, \citenamefont {Biegert}, \citenamefont {Calegari}, \citenamefont {Chac{\'{o}}n}, \citenamefont {Ciappina}, \citenamefont {Dauphin}, \citenamefont {Efimov}, \citenamefont {de~Morisson~Faria}, \citenamefont {Giergiel}, \citenamefont {Gniewek}, \citenamefont {Landsman}, \citenamefont {Lesiuk}, \citenamefont {Mandrysz}, \citenamefont {Maxwell}, \citenamefont {Moszy{\'{n}}ski}, \citenamefont {Ortmann}, \citenamefont {P{\'{e}}rez-Hern{\'{a}}ndez}, \citenamefont {Pic{\'{o}}n}, \citenamefont {Pisanty}, \citenamefont {Prauzner-Bechcicki}, \citenamefont {Sacha}, \citenamefont {Su{\'{a}}rez}, \citenamefont {Zaïr}, \citenamefont {Zakrzewski},\ and\ \citenamefont {Lewenstein}}]{Amini2019}%
  \BibitemOpen
  \bibfield  {author} {\bibinfo {author} {\bibfnamefont {K.}~\bibnamefont {Amini}}, \bibinfo {author} {\bibfnamefont {J.}~\bibnamefont {Biegert}}, \bibinfo {author} {\bibfnamefont {F.}~\bibnamefont {Calegari}}, \bibinfo {author} {\bibfnamefont {A.}~\bibnamefont {Chac{\'{o}}n}}, \bibinfo {author} {\bibfnamefont {M.~F.}\ \bibnamefont {Ciappina}}, \bibinfo {author} {\bibfnamefont {A.}~\bibnamefont {Dauphin}}, \bibinfo {author} {\bibfnamefont {D.~K.}\ \bibnamefont {Efimov}}, \bibinfo {author} {\bibfnamefont {C.~F.}\ \bibnamefont {de~Morisson~Faria}}, \bibinfo {author} {\bibfnamefont {K.}~\bibnamefont {Giergiel}}, \bibinfo {author} {\bibfnamefont {P.}~\bibnamefont {Gniewek}}, \bibinfo {author} {\bibfnamefont {A.~S.}\ \bibnamefont {Landsman}}, \bibinfo {author} {\bibfnamefont {M.}~\bibnamefont {Lesiuk}}, \bibinfo {author} {\bibfnamefont {M.}~\bibnamefont {Mandrysz}}, \bibinfo {author} {\bibfnamefont {A.~S.}\ \bibnamefont {Maxwell}}, \bibinfo {author} {\bibfnamefont {R.}~\bibnamefont {Moszy{\'{n}}ski}}, \bibinfo
  {author} {\bibfnamefont {L.}~\bibnamefont {Ortmann}}, \bibinfo {author} {\bibfnamefont {J.~A.}\ \bibnamefont {P{\'{e}}rez-Hern{\'{a}}ndez}}, \bibinfo {author} {\bibfnamefont {A.}~\bibnamefont {Pic{\'{o}}n}}, \bibinfo {author} {\bibfnamefont {E.}~\bibnamefont {Pisanty}}, \bibinfo {author} {\bibfnamefont {J.}~\bibnamefont {Prauzner-Bechcicki}}, \bibinfo {author} {\bibfnamefont {K.}~\bibnamefont {Sacha}}, \bibinfo {author} {\bibfnamefont {N.}~\bibnamefont {Su{\'{a}}rez}}, \bibinfo {author} {\bibfnamefont {A.}~\bibnamefont {Zaïr}}, \bibinfo {author} {\bibfnamefont {J.}~\bibnamefont {Zakrzewski}},\ and\ \bibinfo {author} {\bibfnamefont {M.}~\bibnamefont {Lewenstein}},\ }\bibfield  {title} {\bibinfo {title} {Symphony on strong field approximation},\ }\href {https://doi.org/10.1088/1361-6633/ab2bb1} {\bibfield  {journal} {\bibinfo  {journal} {Reports on Progress in Physics}\ }\textbf {\bibinfo {volume} {82}},\ \bibinfo {pages} {116001} (\bibinfo {year} {2019})}\BibitemShut {NoStop}%
\bibitem [{\citenamefont {Yan}\ \emph {et~al.}(2010)\citenamefont {Yan}, \citenamefont {Popruzhenko}, \citenamefont {Vrakking},\ and\ \citenamefont {Bauer}}]{Yan2010}%
  \BibitemOpen
  \bibfield  {author} {\bibinfo {author} {\bibfnamefont {T.-M.}\ \bibnamefont {Yan}}, \bibinfo {author} {\bibfnamefont {S.~V.}\ \bibnamefont {Popruzhenko}}, \bibinfo {author} {\bibfnamefont {M.~J.~J.}\ \bibnamefont {Vrakking}},\ and\ \bibinfo {author} {\bibfnamefont {D.}~\bibnamefont {Bauer}},\ }\bibfield  {title} {\bibinfo {title} {Low-energy structures in strong field ionization revealed by quantum orbits},\ }\href {https://doi.org/10.1103/PhysRevLett.105.253002} {\bibfield  {journal} {\bibinfo  {journal} {Phys. Rev. Lett.}\ }\textbf {\bibinfo {volume} {105}},\ \bibinfo {pages} {253002} (\bibinfo {year} {2010})}\BibitemShut {NoStop}%
\bibitem [{\citenamefont {Yan}\ and\ \citenamefont {Bauer}(2012)}]{Yan2012}%
  \BibitemOpen
  \bibfield  {author} {\bibinfo {author} {\bibfnamefont {T.-M.}\ \bibnamefont {Yan}}\ and\ \bibinfo {author} {\bibfnamefont {D.}~\bibnamefont {Bauer}},\ }\bibfield  {title} {\bibinfo {title} {{Sub-barrier Coulomb effects on the interference pattern in tunneling-ionization photoelectron spectra}},\ }\href {https://doi.org/10.1103/PhysRevA.86.053403} {\bibfield  {journal} {\bibinfo  {journal} {Phys. Rev. A}\ }\textbf {\bibinfo {volume} {86}},\ \bibinfo {pages} {053403} (\bibinfo {year} {2012})}\BibitemShut {NoStop}%
\bibitem [{\citenamefont {Li}\ \emph {et~al.}(2014{\natexlab{a}})\citenamefont {Li}, \citenamefont {Geng}, \citenamefont {Liu}, \citenamefont {Deng}, \citenamefont {Wu}, \citenamefont {Peng}, \citenamefont {Gong},\ and\ \citenamefont {Liu}}]{Li2014}%
  \BibitemOpen
  \bibfield  {author} {\bibinfo {author} {\bibfnamefont {M.}~\bibnamefont {Li}}, \bibinfo {author} {\bibfnamefont {J.-W.}\ \bibnamefont {Geng}}, \bibinfo {author} {\bibfnamefont {H.}~\bibnamefont {Liu}}, \bibinfo {author} {\bibfnamefont {Y.}~\bibnamefont {Deng}}, \bibinfo {author} {\bibfnamefont {C.}~\bibnamefont {Wu}}, \bibinfo {author} {\bibfnamefont {L.-Y.}\ \bibnamefont {Peng}}, \bibinfo {author} {\bibfnamefont {Q.}~\bibnamefont {Gong}},\ and\ \bibinfo {author} {\bibfnamefont {Y.}~\bibnamefont {Liu}},\ }\bibfield  {title} {\bibinfo {title} {Classical-quantum correspondence for above-threshold ionization},\ }\href {https://doi.org/10.1103/PhysRevLett.112.113002} {\bibfield  {journal} {\bibinfo  {journal} {Phys. Rev. Lett.}\ }\textbf {\bibinfo {volume} {112}},\ \bibinfo {pages} {113002} (\bibinfo {year} {2014}{\natexlab{a}})}\BibitemShut {NoStop}%
\bibitem [{\citenamefont {Geng}\ \emph {et~al.}(2014)\citenamefont {Geng}, \citenamefont {Qin}, \citenamefont {Li}, \citenamefont {Xiong}, \citenamefont {Liu}, \citenamefont {Gong},\ and\ \citenamefont {Peng}}]{Geng2014}%
  \BibitemOpen
  \bibfield  {author} {\bibinfo {author} {\bibfnamefont {J.~W.}\ \bibnamefont {Geng}}, \bibinfo {author} {\bibfnamefont {L.}~\bibnamefont {Qin}}, \bibinfo {author} {\bibfnamefont {M.}~\bibnamefont {Li}}, \bibinfo {author} {\bibfnamefont {W.~H.}\ \bibnamefont {Xiong}}, \bibinfo {author} {\bibfnamefont {Y.}~\bibnamefont {Liu}}, \bibinfo {author} {\bibfnamefont {Q.}~\bibnamefont {Gong}},\ and\ \bibinfo {author} {\bibfnamefont {L.~Y.}\ \bibnamefont {Peng}},\ }\bibfield  {title} {\bibinfo {title} {{Nonadiabatic tunneling ionization of atoms in elliptically polarized laser fields}},\ }\href {https://doi.org/10.1088/0953-4075/47/20/204027} {\bibfield  {journal} {\bibinfo  {journal} {J. Phys. B}\ }\textbf {\bibinfo {volume} {47}},\ \bibinfo {pages} {204027} (\bibinfo {year} {2014})}\BibitemShut {NoStop}%
\bibitem [{\citenamefont {Li}\ \emph {et~al.}(2014{\natexlab{b}})\citenamefont {Li}, \citenamefont {Sun}, \citenamefont {Xie}, \citenamefont {Shao}, \citenamefont {Deng}, \citenamefont {Wu}, \citenamefont {Gong},\ and\ \citenamefont {Liu}}]{Li2014c}%
  \BibitemOpen
  \bibfield  {author} {\bibinfo {author} {\bibfnamefont {M.}~\bibnamefont {Li}}, \bibinfo {author} {\bibfnamefont {X.}~\bibnamefont {Sun}}, \bibinfo {author} {\bibfnamefont {X.}~\bibnamefont {Xie}}, \bibinfo {author} {\bibfnamefont {Y.}~\bibnamefont {Shao}}, \bibinfo {author} {\bibfnamefont {Y.}~\bibnamefont {Deng}}, \bibinfo {author} {\bibfnamefont {C.}~\bibnamefont {Wu}}, \bibinfo {author} {\bibfnamefont {Q.}~\bibnamefont {Gong}},\ and\ \bibinfo {author} {\bibfnamefont {Y.}~\bibnamefont {Liu}},\ }\bibfield  {title} {\bibinfo {title} {{Revealing backward rescattering photoelectron interference of molecules in strong infrared laser fields}},\ }\href {https://doi.org/10.1038/srep08519} {\bibfield  {journal} {\bibinfo  {journal} {Sci. Rep.}\ }\textbf {\bibinfo {volume} {5}},\ \bibinfo {pages} {8519} (\bibinfo {year} {2014}{\natexlab{b}})}\BibitemShut {NoStop}%
\bibitem [{\citenamefont {Richter}\ \emph {et~al.}(2015)\citenamefont {Richter}, \citenamefont {Kunitski}, \citenamefont {Sch\"offler}, \citenamefont {Jahnke}, \citenamefont {Schmidt}, \citenamefont {Li}, \citenamefont {Liu},\ and\ \citenamefont {D\"orner}}]{richter2015streaking}%
  \BibitemOpen
  \bibfield  {author} {\bibinfo {author} {\bibfnamefont {M.}~\bibnamefont {Richter}}, \bibinfo {author} {\bibfnamefont {M.}~\bibnamefont {Kunitski}}, \bibinfo {author} {\bibfnamefont {M.}~\bibnamefont {Sch\"offler}}, \bibinfo {author} {\bibfnamefont {T.}~\bibnamefont {Jahnke}}, \bibinfo {author} {\bibfnamefont {L.~P.~H.}\ \bibnamefont {Schmidt}}, \bibinfo {author} {\bibfnamefont {M.}~\bibnamefont {Li}}, \bibinfo {author} {\bibfnamefont {Y.}~\bibnamefont {Liu}},\ and\ \bibinfo {author} {\bibfnamefont {R.}~\bibnamefont {D\"orner}},\ }\bibfield  {title} {\bibinfo {title} {Streaking temporal double-slit interference by an orthogonal two-color laser field},\ }\href {https://doi.org/10.1103/PhysRevLett.114.143001} {\bibfield  {journal} {\bibinfo  {journal} {Phys. Rev. Lett.}\ }\textbf {\bibinfo {volume} {114}},\ \bibinfo {pages} {143001} (\bibinfo {year} {2015})}\BibitemShut {NoStop}%
\bibitem [{\citenamefont {Xie}\ \emph {et~al.}(2016)\citenamefont {Xie}, \citenamefont {Li}, \citenamefont {Li}, \citenamefont {Zhou},\ and\ \citenamefont {Lu}}]{Xie2016}%
  \BibitemOpen
  \bibfield  {author} {\bibinfo {author} {\bibfnamefont {H.}~\bibnamefont {Xie}}, \bibinfo {author} {\bibfnamefont {M.}~\bibnamefont {Li}}, \bibinfo {author} {\bibfnamefont {Y.}~\bibnamefont {Li}}, \bibinfo {author} {\bibfnamefont {Y.}~\bibnamefont {Zhou}},\ and\ \bibinfo {author} {\bibfnamefont {P.}~\bibnamefont {Lu}},\ }\bibfield  {title} {\bibinfo {title} {Intra-half-cycle interference of low-energy photoelectron in strong midinfrared laser fields},\ }\href {https://doi.org/10.1364/OE.24.027726} {\bibfield  {journal} {\bibinfo  {journal} {Opt. Express}\ }\textbf {\bibinfo {volume} {24}},\ \bibinfo {pages} {27726} (\bibinfo {year} {2016})}\BibitemShut {NoStop}%
\bibitem [{\citenamefont {Li}\ \emph {et~al.}(2016)\citenamefont {Li}, \citenamefont {Geng}, \citenamefont {Han}, \citenamefont {Liu}, \citenamefont {Peng}, \citenamefont {Gong},\ and\ \citenamefont {Liu}}]{Li2016PRA}%
  \BibitemOpen
  \bibfield  {author} {\bibinfo {author} {\bibfnamefont {M.}~\bibnamefont {Li}}, \bibinfo {author} {\bibfnamefont {J.~W.}\ \bibnamefont {Geng}}, \bibinfo {author} {\bibfnamefont {M.}~\bibnamefont {Han}}, \bibinfo {author} {\bibfnamefont {M.~M.}\ \bibnamefont {Liu}}, \bibinfo {author} {\bibfnamefont {L.~Y.}\ \bibnamefont {Peng}}, \bibinfo {author} {\bibfnamefont {Q.}~\bibnamefont {Gong}},\ and\ \bibinfo {author} {\bibfnamefont {Y.}~\bibnamefont {Liu}},\ }\bibfield  {title} {\bibinfo {title} {{Subcycle nonadiabatic strong-field tunneling ionization}},\ }\href {https://doi.org/10.1103/PhysRevA.93.013402} {\bibfield  {journal} {\bibinfo  {journal} {Phys. Rev. A}\ }\textbf {\bibinfo {volume} {93}},\ \bibinfo {pages} {013402} (\bibinfo {year} {2016})}\BibitemShut {NoStop}%
\bibitem [{\citenamefont {Liu}\ \emph {et~al.}(2016)\citenamefont {Liu}, \citenamefont {Li}, \citenamefont {Wu}, \citenamefont {Gong}, \citenamefont {Staudte},\ and\ \citenamefont {Liu}}]{Liu2016}%
  \BibitemOpen
  \bibfield  {author} {\bibinfo {author} {\bibfnamefont {M.-M.}\ \bibnamefont {Liu}}, \bibinfo {author} {\bibfnamefont {M.}~\bibnamefont {Li}}, \bibinfo {author} {\bibfnamefont {C.}~\bibnamefont {Wu}}, \bibinfo {author} {\bibfnamefont {Q.}~\bibnamefont {Gong}}, \bibinfo {author} {\bibfnamefont {A.}~\bibnamefont {Staudte}},\ and\ \bibinfo {author} {\bibfnamefont {Y.}~\bibnamefont {Liu}},\ }\bibfield  {title} {\bibinfo {title} {Phase structure of strong-field tunneling wave packets from molecules},\ }\href {https://doi.org/10.1103/PhysRevLett.116.163004} {\bibfield  {journal} {\bibinfo  {journal} {Phys. Rev. Lett.}\ }\textbf {\bibinfo {volume} {116}},\ \bibinfo {pages} {163004} (\bibinfo {year} {2016})}\BibitemShut {NoStop}%
\bibitem [{\citenamefont {Shvetsov-Shilovski}\ \emph {et~al.}(2016)\citenamefont {Shvetsov-Shilovski}, \citenamefont {Lein}, \citenamefont {Madsen}, \citenamefont {R\"as\"anen}, \citenamefont {Lemell}, \citenamefont {Burgd\"orfer}, \citenamefont {Arb\'o},\ and\ \citenamefont {T\H{o}k\'esi}}]{Lein2016}%
  \BibitemOpen
  \bibfield  {author} {\bibinfo {author} {\bibfnamefont {N.~I.}\ \bibnamefont {Shvetsov-Shilovski}}, \bibinfo {author} {\bibfnamefont {M.}~\bibnamefont {Lein}}, \bibinfo {author} {\bibfnamefont {L.~B.}\ \bibnamefont {Madsen}}, \bibinfo {author} {\bibfnamefont {E.}~\bibnamefont {R\"as\"anen}}, \bibinfo {author} {\bibfnamefont {C.}~\bibnamefont {Lemell}}, \bibinfo {author} {\bibfnamefont {J.}~\bibnamefont {Burgd\"orfer}}, \bibinfo {author} {\bibfnamefont {D.~G.}\ \bibnamefont {Arb\'o}},\ and\ \bibinfo {author} {\bibfnamefont {K.}~\bibnamefont {T\H{o}k\'esi}},\ }\bibfield  {title} {\bibinfo {title} {Semiclassical two-step model for strong-field ionization},\ }\href {https://doi.org/10.1103/PhysRevA.94.013415} {\bibfield  {journal} {\bibinfo  {journal} {Phys. Rev. A}\ }\textbf {\bibinfo {volume} {94}},\ \bibinfo {pages} {013415} (\bibinfo {year} {2016})}\BibitemShut {NoStop}%
\bibitem [{\citenamefont {Shvetsov-Shilovski}(2021)}]{shvetsov2021semiclassical}%
  \BibitemOpen
  \bibfield  {author} {\bibinfo {author} {\bibfnamefont {N.}~\bibnamefont {Shvetsov-Shilovski}},\ }\bibfield  {title} {\bibinfo {title} {Semiclassical two-step model for ionization by a strong laser pulse: further developments and applications},\ }\href {https://doi.org/10.1140/epjd/s10053-021-00134-3} {\bibfield  {journal} {\bibinfo  {journal} {Eur. Phys. J. D}\ }\textbf {\bibinfo {volume} {75}},\ \bibinfo {pages} {1} (\bibinfo {year} {2021})}\BibitemShut {NoStop}%
\bibitem [{\citenamefont {Shvetsov-Shilovski}\ and\ \citenamefont {Lein}(2018)}]{Shilovski2018}%
  \BibitemOpen
  \bibfield  {author} {\bibinfo {author} {\bibfnamefont {N.~I.}\ \bibnamefont {Shvetsov-Shilovski}}\ and\ \bibinfo {author} {\bibfnamefont {M.}~\bibnamefont {Lein}},\ }\bibfield  {title} {\bibinfo {title} {Effects of the coulomb potential in interference patterns of strong-field holography with photoelectrons},\ }\href {https://doi.org/10.1103/PhysRevA.97.013411} {\bibfield  {journal} {\bibinfo  {journal} {Phys. Rev. A}\ }\textbf {\bibinfo {volume} {97}},\ \bibinfo {pages} {013411} (\bibinfo {year} {2018})}\BibitemShut {NoStop}%
\bibitem [{\citenamefont {Lai}\ \emph {et~al.}(2017)\citenamefont {Lai}, \citenamefont {Yu}, \citenamefont {Huang}, \citenamefont {Hua}, \citenamefont {Gong}, \citenamefont {Quan}, \citenamefont {Faria},\ and\ \citenamefont {Liu}}]{Lai2017}%
  \BibitemOpen
  \bibfield  {author} {\bibinfo {author} {\bibfnamefont {X.~Y.}\ \bibnamefont {Lai}}, \bibinfo {author} {\bibfnamefont {S.}~\bibnamefont {Yu}}, \bibinfo {author} {\bibfnamefont {Y.}~\bibnamefont {Huang}}, \bibinfo {author} {\bibfnamefont {L.}~\bibnamefont {Hua}}, \bibinfo {author} {\bibfnamefont {C.}~\bibnamefont {Gong}}, \bibinfo {author} {\bibfnamefont {W.}~\bibnamefont {Quan}}, \bibinfo {author} {\bibfnamefont {C.~F. D.~M.}\ \bibnamefont {Faria}},\ and\ \bibinfo {author} {\bibfnamefont {X.}~\bibnamefont {Liu}},\ }\bibfield  {title} {\bibinfo {title} {{Near-threshold photoelectron holography beyond the strong-field approximation}},\ }\href {https://doi.org/10.1103/PhysRevA.96.013414} {\bibfield  {journal} {\bibinfo  {journal} {Phys. Rev. A}\ }\textbf {\bibinfo {volume} {96}},\ \bibinfo {pages} {013414} (\bibinfo {year} {2017})},\ \Eprint {https://arxiv.org/abs/1703.04123} {arXiv:1703.04123} \BibitemShut {NoStop}%
\bibitem [{\citenamefont {Maxwell}\ \emph {et~al.}(2017)\citenamefont {Maxwell}, \citenamefont {Al-Jawahiry}, \citenamefont {Das},\ and\ \citenamefont {Faria}}]{Maxwell2017}%
  \BibitemOpen
  \bibfield  {author} {\bibinfo {author} {\bibfnamefont {A.~S.}\ \bibnamefont {Maxwell}}, \bibinfo {author} {\bibfnamefont {A.}~\bibnamefont {Al-Jawahiry}}, \bibinfo {author} {\bibfnamefont {T.}~\bibnamefont {Das}},\ and\ \bibinfo {author} {\bibfnamefont {C.~F. d.~M.}\ \bibnamefont {Faria}},\ }\bibfield  {title} {\bibinfo {title} {{Coulomb-corrected quantum interference in above-threshold ionization: Working towards multi-trajectory electron holography}},\ }\href {https://doi.org/10.1103/PhysRevA.96.023420} {\bibfield  {journal} {\bibinfo  {journal} {Phys. Rev. A}\ }\textbf {\bibinfo {volume} {96}},\ \bibinfo {pages} {023420} (\bibinfo {year} {2017})},\ \Eprint {https://arxiv.org/abs/1705.01518} {arXiv:1705.01518} \BibitemShut {NoStop}%
\bibitem [{\citenamefont {Maxwell}\ \emph {et~al.}(2018{\natexlab{a}})\citenamefont {Maxwell}, \citenamefont {Al-Jawahiry}, \citenamefont {Lai},\ and\ \citenamefont {{Figueira de Morisson Faria}}}]{Maxwell2017a}%
  \BibitemOpen
  \bibfield  {author} {\bibinfo {author} {\bibfnamefont {A.~S.}\ \bibnamefont {Maxwell}}, \bibinfo {author} {\bibfnamefont {A.}~\bibnamefont {Al-Jawahiry}}, \bibinfo {author} {\bibfnamefont {X.~Y.}\ \bibnamefont {Lai}},\ and\ \bibinfo {author} {\bibfnamefont {C.}~\bibnamefont {{Figueira de Morisson Faria}}},\ }\bibfield  {title} {\bibinfo {title} {{Analytic quantum-interference conditions in Coulomb corrected photoelectron holography}},\ }\href {https://doi.org/10.1088/1361-6455/aa9e81} {\bibfield  {journal} {\bibinfo  {journal} {J. Phys. B}\ }\textbf {\bibinfo {volume} {51}},\ \bibinfo {pages} {044004} (\bibinfo {year} {2018}{\natexlab{a}})}\BibitemShut {NoStop}%
\bibitem [{\citenamefont {Maxwell}\ and\ \citenamefont {{Figueira de Morisson Faria}}(2018)}]{Maxwell2018}%
  \BibitemOpen
  \bibfield  {author} {\bibinfo {author} {\bibfnamefont {A.~S.}\ \bibnamefont {Maxwell}}\ and\ \bibinfo {author} {\bibfnamefont {C.}~\bibnamefont {{Figueira de Morisson Faria}}},\ }\bibfield  {title} {\bibinfo {title} {{Coulomb-free and Coulomb-distorted recolliding quantum orbits in photoelectron holography}},\ }\href {http://arxiv.org/abs/1802.00789} {\bibfield  {journal} {\bibinfo  {journal} {J. Phys. B}\ }\textbf {\bibinfo {volume} {51}},\ \bibinfo {pages} {124001} (\bibinfo {year} {2018})},\ \Eprint {https://arxiv.org/abs/1802.00789} {arXiv:1802.00789} \BibitemShut {NoStop}%
\bibitem [{\citenamefont {Maxwell}\ \emph {et~al.}(2018{\natexlab{b}})\citenamefont {Maxwell}, \citenamefont {Popruzhenko},\ and\ \citenamefont {Faria}}]{Maxwell2018b}%
  \BibitemOpen
  \bibfield  {author} {\bibinfo {author} {\bibfnamefont {A.~S.}\ \bibnamefont {Maxwell}}, \bibinfo {author} {\bibfnamefont {S.~V.}\ \bibnamefont {Popruzhenko}},\ and\ \bibinfo {author} {\bibfnamefont {C.~F. d.~M.}\ \bibnamefont {Faria}},\ }\bibfield  {title} {\bibinfo {title} {Treating branch cuts in quantum trajectory models for photoelectron holography},\ }\href {https://doi.org/10.1103/PhysRevA.98.063423} {\bibfield  {journal} {\bibinfo  {journal} {Phys. Rev. A}\ }\textbf {\bibinfo {volume} {98}},\ \bibinfo {pages} {063423} (\bibinfo {year} {2018}{\natexlab{b}})}\BibitemShut {NoStop}%
\bibitem [{\citenamefont {Rook}\ and\ \citenamefont {de~Morisson~Faria}(2022)}]{Rook2022}%
  \BibitemOpen
  \bibfield  {author} {\bibinfo {author} {\bibfnamefont {T.}~\bibnamefont {Rook}}\ and\ \bibinfo {author} {\bibfnamefont {C.~F.}\ \bibnamefont {de~Morisson~Faria}},\ }\bibfield  {title} {\bibinfo {title} {Exploring symmetries in photoelectron holography with two-color linearly polarized fields},\ }\href {https://doi.org/10.1088/1361-6455/ac7bbf} {\bibfield  {journal} {\bibinfo  {journal} {J. Phys. B}\ }\textbf {\bibinfo {volume} {55}},\ \bibinfo {pages} {165601} (\bibinfo {year} {2022})}\BibitemShut {NoStop}%
\bibitem [{\citenamefont {Cruz~Rodriguez}\ \emph {et~al.}(2023)\citenamefont {Cruz~Rodriguez}, \citenamefont {Rook}, \citenamefont {Augstein}, \citenamefont {Maxwell},\ and\ \citenamefont {Figueira~de Morisson~Faria}}]{rodriguez2023}%
  \BibitemOpen
  \bibfield  {author} {\bibinfo {author} {\bibfnamefont {L.}~\bibnamefont {Cruz~Rodriguez}}, \bibinfo {author} {\bibfnamefont {T.}~\bibnamefont {Rook}}, \bibinfo {author} {\bibfnamefont {B.~B.}\ \bibnamefont {Augstein}}, \bibinfo {author} {\bibfnamefont {A.~S.}\ \bibnamefont {Maxwell}},\ and\ \bibinfo {author} {\bibfnamefont {C.}~\bibnamefont {Figueira~de Morisson~Faria}},\ }\bibfield  {title} {\bibinfo {title} {Forward and hybrid path-integral methods in photoelectron holography: Sub-barrier corrections, initial sampling, and momentum mapping},\ }\href {https://doi.org/10.1103/PhysRevA.108.033114} {\bibfield  {journal} {\bibinfo  {journal} {Phys. Rev. A}\ }\textbf {\bibinfo {volume} {108}},\ \bibinfo {pages} {033114} (\bibinfo {year} {2023})}\BibitemShut {NoStop}%
\bibitem [{\citenamefont {Rook}\ \emph {et~al.}(2024)\citenamefont {Rook}, \citenamefont {Habibovi\ifmmode~\acute{c}\else \'{c}\fi{}}, \citenamefont {Rodriguez}, \citenamefont {Milo\ifmmode \check{s}\else \v{s}\fi{}evi\ifmmode~\acute{c}\else \'{c}\fi{}},\ and\ \citenamefont {Faria}}]{Rook2024}%
  \BibitemOpen
  \bibfield  {author} {\bibinfo {author} {\bibfnamefont {T.}~\bibnamefont {Rook}}, \bibinfo {author} {\bibfnamefont {D.}~\bibnamefont {Habibovi\ifmmode~\acute{c}\else \'{c}\fi{}}}, \bibinfo {author} {\bibfnamefont {L.~C.}\ \bibnamefont {Rodriguez}}, \bibinfo {author} {\bibfnamefont {D.~B.}\ \bibnamefont {Milo\ifmmode \check{s}\else \v{s}\fi{}evi\ifmmode~\acute{c}\else \'{c}\fi{}}},\ and\ \bibinfo {author} {\bibfnamefont {C.~F. d.~M.}\ \bibnamefont {Faria}},\ }\bibfield  {title} {\bibinfo {title} {Impact of the continuum coulomb interaction in quantum-orbit-based treatments of high-order above-threshold ionization},\ }\href {https://doi.org/10.1103/PhysRevA.109.033115} {\bibfield  {journal} {\bibinfo  {journal} {Phys. Rev. A}\ }\textbf {\bibinfo {volume} {109}},\ \bibinfo {pages} {033115} (\bibinfo {year} {2024})}\BibitemShut {NoStop}%
\bibitem [{\citenamefont {{Becker}}\ \emph {et~al.}(2002)\citenamefont {{Becker}}, \citenamefont {{Grasbon}}, \citenamefont {{Kopold}}, \citenamefont {{Milo{\v s}evi{\'c}}}, \citenamefont {{Paulus}},\ and\ \citenamefont {{Walther}}}]{Becker2002}%
  \BibitemOpen
  \bibfield  {author} {\bibinfo {author} {\bibfnamefont {W.}~\bibnamefont {{Becker}}}, \bibinfo {author} {\bibfnamefont {F.}~\bibnamefont {{Grasbon}}}, \bibinfo {author} {\bibfnamefont {R.}~\bibnamefont {{Kopold}}}, \bibinfo {author} {\bibfnamefont {D.~B.}\ \bibnamefont {{Milo{\v s}evi{\'c}}}}, \bibinfo {author} {\bibfnamefont {G.~G.}\ \bibnamefont {{Paulus}}},\ and\ \bibinfo {author} {\bibfnamefont {H.}~\bibnamefont {{Walther}}},\ }\bibfield  {title} {\bibinfo {title} {{Above-threshold ionization: From classical features to quantum effects}},\ }\href {https://doi.org/10.1016/S1049-250X(02)80006-4} {\bibfield  {journal} {\bibinfo  {journal} {Adv. At. Mol. Opt. Phy.}\ }\textbf {\bibinfo {volume} {48}},\ \bibinfo {pages} {35} (\bibinfo {year} {2002})}\BibitemShut {NoStop}%
\bibitem [{\citenamefont {Agostini}\ and\ \citenamefont {DiMauro}(2004)}]{Agostini2004}%
  \BibitemOpen
  \bibfield  {author} {\bibinfo {author} {\bibfnamefont {P.}~\bibnamefont {Agostini}}\ and\ \bibinfo {author} {\bibfnamefont {L.~F.}\ \bibnamefont {DiMauro}},\ }\bibfield  {title} {\bibinfo {title} {The physics of attosecond light pulses},\ }\href {https://doi.org/10.1088/0034-4885/67/6/R01} {\bibfield  {journal} {\bibinfo  {journal} {Rep. Prog. Phys.}\ }\textbf {\bibinfo {volume} {67}},\ \bibinfo {pages} {813} (\bibinfo {year} {2004})}\BibitemShut {NoStop}%
\bibitem [{\citenamefont {Scrinzi}\ \emph {et~al.}(2005)\citenamefont {Scrinzi}, \citenamefont {Ivanov}, \citenamefont {Kienberger},\ and\ \citenamefont {Villeneuve}}]{Scrinzi2006}%
  \BibitemOpen
  \bibfield  {author} {\bibinfo {author} {\bibfnamefont {A.}~\bibnamefont {Scrinzi}}, \bibinfo {author} {\bibfnamefont {M.~Y.}\ \bibnamefont {Ivanov}}, \bibinfo {author} {\bibfnamefont {R.}~\bibnamefont {Kienberger}},\ and\ \bibinfo {author} {\bibfnamefont {D.~M.}\ \bibnamefont {Villeneuve}},\ }\bibfield  {title} {\bibinfo {title} {Attosecond physics},\ }\href {https://doi.org/10.1088/0953-4075/39/1/R01} {\bibfield  {journal} {\bibinfo  {journal} {J. Phys. B}\ }\textbf {\bibinfo {volume} {39}},\ \bibinfo {pages} {R1} (\bibinfo {year} {2005})}\BibitemShut {NoStop}%
\bibitem [{\citenamefont {Lein}(2007)}]{Lein2007}%
  \BibitemOpen
  \bibfield  {author} {\bibinfo {author} {\bibfnamefont {M.}~\bibnamefont {Lein}},\ }\bibfield  {title} {\bibinfo {title} {Molecular imaging using recolliding electrons},\ }\href {http://stacks.iop.org/0953-4075/40/i=16/a=R01} {\bibfield  {journal} {\bibinfo  {journal} {J. Phys. B}\ }\textbf {\bibinfo {volume} {40}},\ \bibinfo {pages} {R135} (\bibinfo {year} {2007})}\BibitemShut {NoStop}%
\bibitem [{\citenamefont {Krausz}\ and\ \citenamefont {Ivanov}(2009)}]{Krausz2009}%
  \BibitemOpen
  \bibfield  {author} {\bibinfo {author} {\bibfnamefont {F.}~\bibnamefont {Krausz}}\ and\ \bibinfo {author} {\bibfnamefont {M.}~\bibnamefont {Ivanov}},\ }\bibfield  {title} {\bibinfo {title} {Attosecond physics},\ }\href {https://doi.org/10.1103/RevModPhys.81.163} {\bibfield  {journal} {\bibinfo  {journal} {Rev. Mod. Phys.}\ }\textbf {\bibinfo {volume} {81}},\ \bibinfo {pages} {163} (\bibinfo {year} {2009})}\BibitemShut {NoStop}%
\bibitem [{\citenamefont {Faria}\ and\ \citenamefont {Maxwell}(2020)}]{Faria2020}%
  \BibitemOpen
  \bibfield  {author} {\bibinfo {author} {\bibfnamefont {C.~F. d.~M.}\ \bibnamefont {Faria}}\ and\ \bibinfo {author} {\bibfnamefont {A.~S.}\ \bibnamefont {Maxwell}},\ }\bibfield  {title} {\bibinfo {title} {It is all about phases: ultrafast holographic photoelectron imaging},\ }\href {https://doi.org/10.1088/1361-6633/ab5c91} {\bibfield  {journal} {\bibinfo  {journal} {Rep. Prog. Phys.}\ }\textbf {\bibinfo {volume} {83}},\ \bibinfo {pages} {034401} (\bibinfo {year} {2020})}\BibitemShut {NoStop}%
\bibitem [{\citenamefont {Sali{\`e}res}\ \emph {et~al.}(2001)\citenamefont {Sali{\`e}res}, \citenamefont {Carr{\'e}}, \citenamefont {Le~D{\'e}roff}, \citenamefont {Grasbon}, \citenamefont {Paulus}, \citenamefont {Walther}, \citenamefont {Kopold}, \citenamefont {Becker}, \citenamefont {Milo{\v s}evi{\'c}}, \citenamefont {Sanpera},\ and\ \citenamefont {Lewenstein}}]{Salieres2001}%
  \BibitemOpen
  \bibfield  {author} {\bibinfo {author} {\bibfnamefont {P.}~\bibnamefont {Sali{\`e}res}}, \bibinfo {author} {\bibfnamefont {B.}~\bibnamefont {Carr{\'e}}}, \bibinfo {author} {\bibfnamefont {L.}~\bibnamefont {Le~D{\'e}roff}}, \bibinfo {author} {\bibfnamefont {F.}~\bibnamefont {Grasbon}}, \bibinfo {author} {\bibfnamefont {G.~G.}\ \bibnamefont {Paulus}}, \bibinfo {author} {\bibfnamefont {H.}~\bibnamefont {Walther}}, \bibinfo {author} {\bibfnamefont {R.}~\bibnamefont {Kopold}}, \bibinfo {author} {\bibfnamefont {W.}~\bibnamefont {Becker}}, \bibinfo {author} {\bibfnamefont {D.~B.}\ \bibnamefont {Milo{\v s}evi{\'c}}}, \bibinfo {author} {\bibfnamefont {A.}~\bibnamefont {Sanpera}},\ and\ \bibinfo {author} {\bibfnamefont {M.}~\bibnamefont {Lewenstein}},\ }\bibfield  {title} {\bibinfo {title} {Feynman{\textquoteright}s path-integral approach for intense-laser-atom interactions},\ }\href {https://doi.org/10.1126/science.108836} {\bibfield  {journal} {\bibinfo  {journal} {Science}\ }\textbf {\bibinfo {volume} {292}},\
  \bibinfo {pages} {902} (\bibinfo {year} {2001})}\BibitemShut {NoStop}%
\bibitem [{\citenamefont {Milo{\v s}evi{\'c}}\ \emph {et~al.}(2024)\citenamefont {Milo{\v s}evi{\'c}}, \citenamefont {Ja{\v s}arevi{\' c}}, \citenamefont {Habibovi{\' c}}, \citenamefont {Hasovi{\' c}}, \citenamefont {{\v C}erki{\' c}},\ and\ \citenamefont {Becker}}]{Milosevic2024TR}%
  \BibitemOpen
  \bibfield  {author} {\bibinfo {author} {\bibfnamefont {D.~B.}\ \bibnamefont {Milo{\v s}evi{\'c}}}, \bibinfo {author} {\bibfnamefont {A.~S.}\ \bibnamefont {Ja{\v s}arevi{\' c}}}, \bibinfo {author} {\bibfnamefont {D.}~\bibnamefont {Habibovi{\' c}}}, \bibinfo {author} {\bibfnamefont {E.}~\bibnamefont {Hasovi{\' c}}}, \bibinfo {author} {\bibfnamefont {A.}~\bibnamefont {{\v C}erki{\' c}}},\ and\ \bibinfo {author} {\bibfnamefont {W.}~\bibnamefont {Becker}},\ }\bibfield  {title} {\bibinfo {title} {Asymptotic methods applied to integrals occurring in strong-laser-field processes},\ }\href {https://doi.org/10.1088/1751-8121/ad7212} {\bibfield  {journal} {\bibinfo  {journal} {J. Phys. A}\ }\textbf {\bibinfo {volume} {57}},\ \bibinfo {pages} {393001} (\bibinfo {year} {2024})}\BibitemShut {NoStop}%
\bibitem [{\citenamefont {Lewenstein}\ \emph {et~al.}(1994)\citenamefont {Lewenstein}, \citenamefont {Balcou}, \citenamefont {Ivanov}, \citenamefont {L'Huillier},\ and\ \citenamefont {Corkum}}]{Lewenstein1994}%
  \BibitemOpen
  \bibfield  {author} {\bibinfo {author} {\bibfnamefont {M.}~\bibnamefont {Lewenstein}}, \bibinfo {author} {\bibfnamefont {P.}~\bibnamefont {Balcou}}, \bibinfo {author} {\bibfnamefont {M.~Y.}\ \bibnamefont {Ivanov}}, \bibinfo {author} {\bibfnamefont {A.}~\bibnamefont {L'Huillier}},\ and\ \bibinfo {author} {\bibfnamefont {P.~B.}\ \bibnamefont {Corkum}},\ }\bibfield  {title} {\bibinfo {title} {Theory of high-harmonic generation by low-frequency laser fields},\ }\href {https://doi.org/10.1103/PhysRevA.49.2117} {\bibfield  {journal} {\bibinfo  {journal} {Phys. Rev. A}\ }\textbf {\bibinfo {volume} {49}},\ \bibinfo {pages} {2117} (\bibinfo {year} {1994})}\BibitemShut {NoStop}%
\bibitem [{\citenamefont {Milo\ifmmode \check{s}\else \v{s}\fi{}evi\ifmmode~\acute{c}\else \'{c}\fi{}}\ and\ \citenamefont {Becker}(2002)}]{Milosevic2002}%
  \BibitemOpen
  \bibfield  {author} {\bibinfo {author} {\bibfnamefont {D.~B.}\ \bibnamefont {Milo\ifmmode \check{s}\else \v{s}\fi{}evi\ifmmode~\acute{c}\else \'{c}\fi{}}}\ and\ \bibinfo {author} {\bibfnamefont {W.}~\bibnamefont {Becker}},\ }\bibfield  {title} {\bibinfo {title} {Role of long quantum orbits in high-order harmonic generation},\ }\href {https://doi.org/10.1103/PhysRevA.66.063417} {\bibfield  {journal} {\bibinfo  {journal} {Phys. Rev. A}\ }\textbf {\bibinfo {volume} {66}},\ \bibinfo {pages} {063417} (\bibinfo {year} {2002})}\BibitemShut {NoStop}%
\bibitem [{\citenamefont {Od\ifmmode~\check{z}\else \v{z}\fi{}ak}\ and\ \citenamefont {Milo\ifmmode \check{s}\else \v{s}\fi{}evi\ifmmode~\acute{c}\else \'{c}\fi{}}(2005)}]{Odzak2005}%
  \BibitemOpen
  \bibfield  {author} {\bibinfo {author} {\bibfnamefont {S.}~\bibnamefont {Od\ifmmode~\check{z}\else \v{z}\fi{}ak}}\ and\ \bibinfo {author} {\bibfnamefont {D.~B.}\ \bibnamefont {Milo\ifmmode \check{s}\else \v{s}\fi{}evi\ifmmode~\acute{c}\else \'{c}\fi{}}},\ }\bibfield  {title} {\bibinfo {title} {High-order harmonic generation in the presence of a static electric field},\ }\href {https://doi.org/10.1103/PhysRevA.72.033407} {\bibfield  {journal} {\bibinfo  {journal} {Phys. Rev. A}\ }\textbf {\bibinfo {volume} {72}},\ \bibinfo {pages} {033407} (\bibinfo {year} {2005})}\BibitemShut {NoStop}%
\bibitem [{\citenamefont {Zhou}\ \emph {et~al.}(2021)\citenamefont {Zhou}, \citenamefont {Guo}, \citenamefont {Quan}, \citenamefont {Wei}, \citenamefont {Zhao}, \citenamefont {Xu}, \citenamefont {Xiao}, \citenamefont {Sun}, \citenamefont {Wang}, \citenamefont {Lai}, \citenamefont {Chen},\ and\ \citenamefont {Liu}}]{Zhou2021}%
  \BibitemOpen
  \bibfield  {author} {\bibinfo {author} {\bibfnamefont {Y.}~\bibnamefont {Zhou}}, \bibinfo {author} {\bibfnamefont {L.}~\bibnamefont {Guo}}, \bibinfo {author} {\bibfnamefont {W.}~\bibnamefont {Quan}}, \bibinfo {author} {\bibfnamefont {M.}~\bibnamefont {Wei}}, \bibinfo {author} {\bibfnamefont {M.}~\bibnamefont {Zhao}}, \bibinfo {author} {\bibfnamefont {S.}~\bibnamefont {Xu}}, \bibinfo {author} {\bibfnamefont {Z.}~\bibnamefont {Xiao}}, \bibinfo {author} {\bibfnamefont {R.}~\bibnamefont {Sun}}, \bibinfo {author} {\bibfnamefont {Y.}~\bibnamefont {Wang}}, \bibinfo {author} {\bibfnamefont {X.}~\bibnamefont {Lai}}, \bibinfo {author} {\bibfnamefont {J.}~\bibnamefont {Chen}},\ and\ \bibinfo {author} {\bibfnamefont {X.}~\bibnamefont {Liu}},\ }\bibfield  {title} {\bibinfo {title} {Carrier-envelope phase dependence of high-order above-threshold ionization by few-cycle laser pulses},\ }\href {https://doi.org/10.1088/1361-6455/ac19f7} {\bibfield  {journal} {\bibinfo  {journal} {J. Phys. B}\ }\textbf {\bibinfo {volume}
  {54}},\ \bibinfo {pages} {144008} (\bibinfo {year} {2021})}\BibitemShut {NoStop}%
\bibitem [{\citenamefont {Guo}\ \emph {et~al.}(2022)\citenamefont {Guo}, \citenamefont {Chen}, \citenamefont {Hu}, \citenamefont {Lu}, \citenamefont {Han}, \citenamefont {Zhang},\ and\ \citenamefont {Chen}}]{Guo2022}%
  \BibitemOpen
  \bibfield  {author} {\bibinfo {author} {\bibfnamefont {L.}~\bibnamefont {Guo}}, \bibinfo {author} {\bibfnamefont {S.}~\bibnamefont {Chen}}, \bibinfo {author} {\bibfnamefont {S.}~\bibnamefont {Hu}}, \bibinfo {author} {\bibfnamefont {R.}~\bibnamefont {Lu}}, \bibinfo {author} {\bibfnamefont {S.}~\bibnamefont {Han}}, \bibinfo {author} {\bibfnamefont {J.}~\bibnamefont {Zhang}},\ and\ \bibinfo {author} {\bibfnamefont {J.}~\bibnamefont {Chen}},\ }\bibfield  {title} {\bibinfo {title} {Investigation of the middle-energy region of the above-threshold ionization spectrum in a few-cycle laser field},\ }\href {https://doi.org/10.1088/1361-6455/ac92f0} {\bibfield  {journal} {\bibinfo  {journal} {J. Phys. B}\ }\textbf {\bibinfo {volume} {55}},\ \bibinfo {pages} {225401} (\bibinfo {year} {2022})}\BibitemShut {NoStop}%
\bibitem [{\citenamefont {Boroumand}\ \emph {et~al.}(2022)\citenamefont {Boroumand}, \citenamefont {Thorpe}, \citenamefont {Parks},\ and\ \citenamefont {Brabec}}]{Boroumand2022}%
  \BibitemOpen
  \bibfield  {author} {\bibinfo {author} {\bibfnamefont {N.}~\bibnamefont {Boroumand}}, \bibinfo {author} {\bibfnamefont {A.}~\bibnamefont {Thorpe}}, \bibinfo {author} {\bibfnamefont {A.~M.}\ \bibnamefont {Parks}},\ and\ \bibinfo {author} {\bibfnamefont {T.}~\bibnamefont {Brabec}},\ }\bibfield  {title} {\bibinfo {title} {Keldysh ionization theory of atoms: mathematical details},\ }\href {https://doi.org/10.1088/1361-6455/ac9205} {\bibfield  {journal} {\bibinfo  {journal} {J. Phys. B}\ }\textbf {\bibinfo {volume} {55}},\ \bibinfo {pages} {213001} (\bibinfo {year} {2022})}\BibitemShut {NoStop}%
\bibitem [{\citenamefont {Habibovi{\' c}}\ and\ \citenamefont {Milo{\v s}evi{\'c}}(2020)}]{Habibovic2020}%
  \BibitemOpen
  \bibfield  {author} {\bibinfo {author} {\bibfnamefont {D.}~\bibnamefont {Habibovi{\' c}}}\ and\ \bibinfo {author} {\bibfnamefont {D.~B.}\ \bibnamefont {Milo{\v s}evi{\'c}}},\ }\bibfield  {title} {\bibinfo {title} {Ellipticity of high-order harmonics generated by aligned homonuclear diatomic molecules exposed to an orthogonal two-color laser field},\ }\href {https://doi.org/10.3390/photonics7040110} {\bibfield  {journal} {\bibinfo  {journal} {Photonics}\ }\textbf {\bibinfo {volume} {7}},\ \bibinfo {pages} {110} (\bibinfo {year} {2020})}\BibitemShut {NoStop}%
\bibitem [{\citenamefont {Fang}\ and\ \citenamefont {Liu}(2021)}]{Fang2021}%
  \BibitemOpen
  \bibfield  {author} {\bibinfo {author} {\bibfnamefont {Y.}~\bibnamefont {Fang}}\ and\ \bibinfo {author} {\bibfnamefont {Y.}~\bibnamefont {Liu}},\ }\bibfield  {title} {\bibinfo {title} {Optimal control over high-order-harmonic ellipticity in two-color cross-linearly-polarized laser fields},\ }\href {https://doi.org/10.1103/PhysRevA.103.033116} {\bibfield  {journal} {\bibinfo  {journal} {Phys. Rev. A}\ }\textbf {\bibinfo {volume} {103}},\ \bibinfo {pages} {033116} (\bibinfo {year} {2021})}\BibitemShut {NoStop}%
\bibitem [{\citenamefont {Milo{\v s}evi{\'c}}\ and\ \citenamefont {Habibovi{\' c}}(2023)}]{Milosevic2023}%
  \BibitemOpen
  \bibfield  {author} {\bibinfo {author} {\bibfnamefont {D.~B.}\ \bibnamefont {Milo{\v s}evi{\'c}}}\ and\ \bibinfo {author} {\bibfnamefont {D.}~\bibnamefont {Habibovi{\' c}}},\ }\bibfield  {title} {\bibinfo {title} {High-order harmonic generation by aligned homonuclear diatomic cations},\ }\href {https://doi.org/10.1039/D3CP02447D} {\bibfield  {journal} {\bibinfo  {journal} {Phys. Chem. Chem. Phys.}\ }\textbf {\bibinfo {volume} {25}},\ \bibinfo {pages} {28848} (\bibinfo {year} {2023})}\BibitemShut {NoStop}%
\bibitem [{\citenamefont {Figueira~de Morisson~Faria}\ \emph {et~al.}(2004)\citenamefont {Figueira~de Morisson~Faria}, \citenamefont {Schomerus}, \citenamefont {Liu},\ and\ \citenamefont {Becker}}]{Faria2004}%
  \BibitemOpen
  \bibfield  {author} {\bibinfo {author} {\bibfnamefont {C.}~\bibnamefont {Figueira~de Morisson~Faria}}, \bibinfo {author} {\bibfnamefont {H.}~\bibnamefont {Schomerus}}, \bibinfo {author} {\bibfnamefont {X.}~\bibnamefont {Liu}},\ and\ \bibinfo {author} {\bibfnamefont {W.}~\bibnamefont {Becker}},\ }\bibfield  {title} {\bibinfo {title} {Electron-electron dynamics in laser-induced nonsequential double ionization},\ }\href {https://doi.org/10.1103/PhysRevA.69.043405} {\bibfield  {journal} {\bibinfo  {journal} {Phys. Rev. A}\ }\textbf {\bibinfo {volume} {69}},\ \bibinfo {pages} {043405} (\bibinfo {year} {2004})}\BibitemShut {NoStop}%
\bibitem [{\citenamefont {Faria}\ and\ \citenamefont {Lewenstein}(2005)}]{Faria2005}%
  \BibitemOpen
  \bibfield  {author} {\bibinfo {author} {\bibfnamefont {C.~F. D.~M.}\ \bibnamefont {Faria}}\ and\ \bibinfo {author} {\bibfnamefont {M.}~\bibnamefont {Lewenstein}},\ }\bibfield  {title} {\bibinfo {title} {Bound-state corrections in laser-induced nonsequential double ionization},\ }\href {https://doi.org/10.1088/0953-4075/38/17/014} {\bibfield  {journal} {\bibinfo  {journal} {J. Phys. B}\ }\textbf {\bibinfo {volume} {38}},\ \bibinfo {pages} {3251} (\bibinfo {year} {2005})}\BibitemShut {NoStop}%
\bibitem [{\citenamefont {Shaaran}\ \emph {et~al.}(2010)\citenamefont {Shaaran}, \citenamefont {Nygren},\ and\ \citenamefont {Figueira~de Morisson~Faria}}]{Shaaran2010}%
  \BibitemOpen
  \bibfield  {author} {\bibinfo {author} {\bibfnamefont {T.}~\bibnamefont {Shaaran}}, \bibinfo {author} {\bibfnamefont {M.~T.}\ \bibnamefont {Nygren}},\ and\ \bibinfo {author} {\bibfnamefont {C.}~\bibnamefont {Figueira~de Morisson~Faria}},\ }\bibfield  {title} {\bibinfo {title} {Laser-induced nonsequential double ionization at and above the recollision-excitation-tunneling threshold},\ }\href {https://doi.org/10.1103/PhysRevA.81.063413} {\bibfield  {journal} {\bibinfo  {journal} {Phys. Rev. A}\ }\textbf {\bibinfo {volume} {81}},\ \bibinfo {pages} {063413} (\bibinfo {year} {2010})}\BibitemShut {NoStop}%
\bibitem [{\citenamefont {Shaaran}\ \emph {et~al.}(2012)\citenamefont {Shaaran}, \citenamefont {Figueira~de Morisson~Faria},\ and\ \citenamefont {Schomerus}}]{Shaaran2012}%
  \BibitemOpen
  \bibfield  {author} {\bibinfo {author} {\bibfnamefont {T.}~\bibnamefont {Shaaran}}, \bibinfo {author} {\bibfnamefont {C.}~\bibnamefont {Figueira~de Morisson~Faria}},\ and\ \bibinfo {author} {\bibfnamefont {H.}~\bibnamefont {Schomerus}},\ }\bibfield  {title} {\bibinfo {title} {Causality and quantum interference in time-delayed laser-induced nonsequential double ionization},\ }\href {https://doi.org/10.1103/PhysRevA.85.023423} {\bibfield  {journal} {\bibinfo  {journal} {Phys. Rev. A}\ }\textbf {\bibinfo {volume} {85}},\ \bibinfo {pages} {023423} (\bibinfo {year} {2012})}\BibitemShut {NoStop}%
\bibitem [{\citenamefont {Maxwell}\ and\ \citenamefont {Faria}(2015)}]{Maxwell2015}%
  \BibitemOpen
  \bibfield  {author} {\bibinfo {author} {\bibfnamefont {A.~S.}\ \bibnamefont {Maxwell}}\ and\ \bibinfo {author} {\bibfnamefont {C.~F. d.~M.}\ \bibnamefont {Faria}},\ }\bibfield  {title} {\bibinfo {title} {Quantum interference in time-delayed nonsequential double ionization},\ }\href {https://doi.org/10.1103/PhysRevA.92.023421} {\bibfield  {journal} {\bibinfo  {journal} {Phys. Rev. A}\ }\textbf {\bibinfo {volume} {92}},\ \bibinfo {pages} {023421} (\bibinfo {year} {2015})}\BibitemShut {NoStop}%
\bibitem [{\citenamefont {Hashim}\ \emph {et~al.}(2024)\citenamefont {Hashim}, \citenamefont {Tenney},\ and\ \citenamefont {Figueira~de Morisson~Faria}}]{Hashim2024}%
  \BibitemOpen
  \bibfield  {author} {\bibinfo {author} {\bibfnamefont {S.}~\bibnamefont {Hashim}}, \bibinfo {author} {\bibfnamefont {R.}~\bibnamefont {Tenney}},\ and\ \bibinfo {author} {\bibfnamefont {C.}~\bibnamefont {Figueira~de Morisson~Faria}},\ }\bibfield  {title} {\bibinfo {title} {Detangling the quantum tapestry of intrachannel interference in below-threshold nonsequential double ionization with few-cycle laser pulses},\ }\href {https://doi.org/10.1103/PhysRevA.109.063110} {\bibfield  {journal} {\bibinfo  {journal} {Phys. Rev. A}\ }\textbf {\bibinfo {volume} {109}},\ \bibinfo {pages} {063110} (\bibinfo {year} {2024})}\BibitemShut {NoStop}%
\bibitem [{\citenamefont {Madsen}(2022{\natexlab{a}})}]{Madsen2022a}%
  \BibitemOpen
  \bibfield  {author} {\bibinfo {author} {\bibfnamefont {L.~B.}\ \bibnamefont {Madsen}},\ }\bibfield  {title} {\bibinfo {title} {Disappearance and reappearance of above-threshold-ionization peaks},\ }\href {https://doi.org/10.1103/PhysRevA.106.043118} {\bibfield  {journal} {\bibinfo  {journal} {Phys. Rev. A}\ }\textbf {\bibinfo {volume} {106}},\ \bibinfo {pages} {043118} (\bibinfo {year} {2022}{\natexlab{a}})}\BibitemShut {NoStop}%
\bibitem [{\citenamefont {Madsen}(2022{\natexlab{b}})}]{Madsen2022b}%
  \BibitemOpen
  \bibfield  {author} {\bibinfo {author} {\bibfnamefont {L.~B.}\ \bibnamefont {Madsen}},\ }\bibfield  {title} {\bibinfo {title} {Nondipole effects in tunneling ionization by intense laser pulses},\ }\href {https://doi.org/10.1103/PhysRevA.105.043107} {\bibfield  {journal} {\bibinfo  {journal} {Phys. Rev. A}\ }\textbf {\bibinfo {volume} {105}},\ \bibinfo {pages} {043107} (\bibinfo {year} {2022}{\natexlab{b}})}\BibitemShut {NoStop}%
\bibitem [{\citenamefont {Kahved{\v z}i{\' c}}\ and\ \citenamefont {Gr\"afe}(2022)}]{Kahvedzic2022}%
  \BibitemOpen
  \bibfield  {author} {\bibinfo {author} {\bibfnamefont {R.}~\bibnamefont {Kahved{\v z}i{\' c}}}\ and\ \bibinfo {author} {\bibfnamefont {S.}~\bibnamefont {Gr\"afe}},\ }\bibfield  {title} {\bibinfo {title} {Shift of the photoelectron momentum against the radiation pressure force in linearly polarized intense midinfrared laser fields},\ }\href {https://doi.org/10.1103/PhysRevA.106.043122} {\bibfield  {journal} {\bibinfo  {journal} {Phys. Rev. A}\ }\textbf {\bibinfo {volume} {106}},\ \bibinfo {pages} {043122} (\bibinfo {year} {2022})}\BibitemShut {NoStop}%
\bibitem [{\citenamefont {Ja\ifmmode \check{s}\else \v{s}\fi{}arevi\ifmmode~\acute{c}\else \'{c}\fi{}}\ \emph {et~al.}(2024)\citenamefont {Ja\ifmmode \check{s}\else \v{s}\fi{}arevi\ifmmode~\acute{c}\else \'{c}\fi{}}, \citenamefont {Habibovi\ifmmode~\acute{c}\else \'{c}\fi{}},\ and\ \citenamefont {Milo\ifmmode \check{s}\else \v{s}\fi{}evi\ifmmode~\acute{c}\else \'{c}\fi{}}}]{Jasarevic2024}%
  \BibitemOpen
  \bibfield  {author} {\bibinfo {author} {\bibfnamefont {A.~S.}\ \bibnamefont {Ja\ifmmode \check{s}\else \v{s}\fi{}arevi\ifmmode~\acute{c}\else \'{c}\fi{}}}, \bibinfo {author} {\bibfnamefont {D.}~\bibnamefont {Habibovi\ifmmode~\acute{c}\else \'{c}\fi{}}},\ and\ \bibinfo {author} {\bibfnamefont {D.~B.}\ \bibnamefont {Milo\ifmmode \check{s}\else \v{s}\fi{}evi\ifmmode~\acute{c}\else \'{c}\fi{}}},\ }\bibfield  {title} {\bibinfo {title} {Quantum orbits in atomic ionization beyond the dipole approximation},\ }\href {https://doi.org/10.1103/PhysRevA.110.023111} {\bibfield  {journal} {\bibinfo  {journal} {Phys. Rev. A}\ }\textbf {\bibinfo {volume} {110}},\ \bibinfo {pages} {023111} (\bibinfo {year} {2024})}\BibitemShut {NoStop}%
\bibitem [{\citenamefont {Popruzhenko}\ and\ \citenamefont {Bauer}(2008)}]{Popruzhenko2008}%
  \BibitemOpen
  \bibfield  {author} {\bibinfo {author} {\bibfnamefont {S.}~\bibnamefont {Popruzhenko}}\ and\ \bibinfo {author} {\bibfnamefont {D.}~\bibnamefont {Bauer}},\ }\bibfield  {title} {\bibinfo {title} {Strong field approximation for systems with {C}oulomb interaction},\ }\href {https://doi.org/10.1080/09500340802161881} {\bibfield  {journal} {\bibinfo  {journal} {J. Mod. Opt.}\ }\textbf {\bibinfo {volume} {55}},\ \bibinfo {pages} {2573} (\bibinfo {year} {2008})}\BibitemShut {NoStop}%
\bibitem [{\citenamefont {Torlina}\ and\ \citenamefont {Smirnova}(2012)}]{Torlina2012}%
  \BibitemOpen
  \bibfield  {author} {\bibinfo {author} {\bibfnamefont {L.}~\bibnamefont {Torlina}}\ and\ \bibinfo {author} {\bibfnamefont {O.}~\bibnamefont {Smirnova}},\ }\bibfield  {title} {\bibinfo {title} {Time-dependent analytical {$R$}-matrix approach for strong-field dynamics. {I}. one-electron systems},\ }\href {https://doi.org/10.1103/PhysRevA.86.043408} {\bibfield  {journal} {\bibinfo  {journal} {Phys. Rev. A}\ }\textbf {\bibinfo {volume} {86}},\ \bibinfo {pages} {043408} (\bibinfo {year} {2012})}\BibitemShut {NoStop}%
\bibitem [{\citenamefont {Kaushal}\ and\ \citenamefont {Smirnova}(2013)}]{Kaushal2013}%
  \BibitemOpen
  \bibfield  {author} {\bibinfo {author} {\bibfnamefont {J.}~\bibnamefont {Kaushal}}\ and\ \bibinfo {author} {\bibfnamefont {O.}~\bibnamefont {Smirnova}},\ }\bibfield  {title} {\bibinfo {title} {{Nonadiabatic Coulomb effects in strong-field ionization in circularly polarized laser fields}},\ }\href {https://doi.org/10.1103/PhysRevA.88.013421} {\bibfield  {journal} {\bibinfo  {journal} {Phys. Rev. A}\ }\textbf {\bibinfo {volume} {88}},\ \bibinfo {pages} {013421} (\bibinfo {year} {2013})},\ \Eprint {https://arxiv.org/abs/1302.2609} {arXiv:1302.2609} \BibitemShut {NoStop}%
\bibitem [{\citenamefont {Brennecke}\ \emph {et~al.}(2020)\citenamefont {Brennecke}, \citenamefont {Eicke},\ and\ \citenamefont {Lein}}]{brennecke2020gouy}%
  \BibitemOpen
  \bibfield  {author} {\bibinfo {author} {\bibfnamefont {S.}~\bibnamefont {Brennecke}}, \bibinfo {author} {\bibfnamefont {N.}~\bibnamefont {Eicke}},\ and\ \bibinfo {author} {\bibfnamefont {M.}~\bibnamefont {Lein}},\ }\bibfield  {title} {\bibinfo {title} {Gouy's phase anomaly in electron waves produced by strong-field ionization},\ }\href {https://doi.org/10.1103/PhysRevLett.124.153202} {\bibfield  {journal} {\bibinfo  {journal} {Phys. Rev. Lett.}\ }\textbf {\bibinfo {volume} {124}},\ \bibinfo {pages} {153202} (\bibinfo {year} {2020})}\BibitemShut {NoStop}%
\bibitem [{\citenamefont {Lai}\ \emph {et~al.}(2015{\natexlab{a}})\citenamefont {Lai}, \citenamefont {Poli}, \citenamefont {Schomerus},\ and\ \citenamefont {{Figueira de Morisson Faria}}}]{Lai2015}%
  \BibitemOpen
  \bibfield  {author} {\bibinfo {author} {\bibfnamefont {X.~Y.}\ \bibnamefont {Lai}}, \bibinfo {author} {\bibfnamefont {C.}~\bibnamefont {Poli}}, \bibinfo {author} {\bibfnamefont {H.}~\bibnamefont {Schomerus}},\ and\ \bibinfo {author} {\bibfnamefont {C.}~\bibnamefont {{Figueira de Morisson Faria}}},\ }\bibfield  {title} {\bibinfo {title} {{Influence of the Coulomb potential on above-threshold ionization: A quantum-orbit analysis beyond the strong-field approximation}},\ }\href {https://doi.org/10.1103/PhysRevA.92.043407} {\bibfield  {journal} {\bibinfo  {journal} {Phys. Rev. A}\ }\textbf {\bibinfo {volume} {92}},\ \bibinfo {pages} {043407} (\bibinfo {year} {2015}{\natexlab{a}})}\BibitemShut {NoStop}%
\bibitem [{\citenamefont {Carlsen}\ \emph {et~al.}(2024)\citenamefont {Carlsen}, \citenamefont {Hansen}, \citenamefont {Madsen},\ and\ \citenamefont {Maxwell}}]{Carlsen2024}%
  \BibitemOpen
  \bibfield  {author} {\bibinfo {author} {\bibfnamefont {M.~B.}\ \bibnamefont {Carlsen}}, \bibinfo {author} {\bibfnamefont {E.}~\bibnamefont {Hansen}}, \bibinfo {author} {\bibfnamefont {L.~B.}\ \bibnamefont {Madsen}},\ and\ \bibinfo {author} {\bibfnamefont {A.~S.}\ \bibnamefont {Maxwell}},\ }\bibfield  {title} {\bibinfo {title} {Advanced momentum sampling and maslov phases for a precise semiclassical model of strong-field ionization},\ }\href {https://doi.org/10.1088/1367-2630/ad2410} {\bibfield  {journal} {\bibinfo  {journal} {New J. Phys.}\ }\textbf {\bibinfo {volume} {26}},\ \bibinfo {pages} {023025} (\bibinfo {year} {2024})}\BibitemShut {NoStop}%
\bibitem [{\citenamefont {Maxwell}\ \emph {et~al.}(2020)\citenamefont {Maxwell}, \citenamefont {Faria}, \citenamefont {Lai}, \citenamefont {Sun},\ and\ \citenamefont {Liu}}]{Maxwell2019}%
  \BibitemOpen
  \bibfield  {author} {\bibinfo {author} {\bibfnamefont {A.~S.}\ \bibnamefont {Maxwell}}, \bibinfo {author} {\bibfnamefont {C.~F. d.~M.}\ \bibnamefont {Faria}}, \bibinfo {author} {\bibfnamefont {X.}~\bibnamefont {Lai}}, \bibinfo {author} {\bibfnamefont {R.}~\bibnamefont {Sun}},\ and\ \bibinfo {author} {\bibfnamefont {X.}~\bibnamefont {Liu}},\ }\bibfield  {title} {\bibinfo {title} {Spiral-like holographic structures: Unwinding interference carpets of coulomb-distorted orbits in strong-field ionization},\ }\href {https://doi.org/10.1103/PhysRevA.102.033111} {\bibfield  {journal} {\bibinfo  {journal} {Phys. Rev. A}\ }\textbf {\bibinfo {volume} {102}},\ \bibinfo {pages} {033111} (\bibinfo {year} {2020})}\BibitemShut {NoStop}%
\bibitem [{\citenamefont {Kang}\ \emph {et~al.}(2020)\citenamefont {Kang}, \citenamefont {Maxwell}, \citenamefont {Trabert}, \citenamefont {Lai}, \citenamefont {Eckart}, \citenamefont {Kunitski}, \citenamefont {Sch{\"o}ffler}, \citenamefont {Jahnke}, \citenamefont {Bian}, \citenamefont {D{\"o}rner},\ and\ \citenamefont {Faria}}]{Kang2020}%
  \BibitemOpen
  \bibfield  {author} {\bibinfo {author} {\bibfnamefont {H.}~\bibnamefont {Kang}}, \bibinfo {author} {\bibfnamefont {A.~S.}\ \bibnamefont {Maxwell}}, \bibinfo {author} {\bibfnamefont {D.}~\bibnamefont {Trabert}}, \bibinfo {author} {\bibfnamefont {X.}~\bibnamefont {Lai}}, \bibinfo {author} {\bibfnamefont {S.}~\bibnamefont {Eckart}}, \bibinfo {author} {\bibfnamefont {M.}~\bibnamefont {Kunitski}}, \bibinfo {author} {\bibfnamefont {M.}~\bibnamefont {Sch{\"o}ffler}}, \bibinfo {author} {\bibfnamefont {T.}~\bibnamefont {Jahnke}}, \bibinfo {author} {\bibfnamefont {X.}~\bibnamefont {Bian}}, \bibinfo {author} {\bibfnamefont {R.}~\bibnamefont {D{\"o}rner}},\ and\ \bibinfo {author} {\bibfnamefont {C.~F. D.~M.}\ \bibnamefont {Faria}},\ }\bibfield  {title} {\bibinfo {title} {Holographic detection of parity in atomic and molecular orbitals},\ }\href {https://doi.org/10.1103/PhysRevA.102.013109} {\bibfield  {journal} {\bibinfo  {journal} {Phys. Rev. A}\ }\textbf {\bibinfo {volume} {102}},\ \bibinfo {pages} {013109} (\bibinfo
  {year} {2020})}\BibitemShut {NoStop}%
\bibitem [{\citenamefont {Werby}\ \emph {et~al.}(2021)\citenamefont {Werby}, \citenamefont {Maxwell}, \citenamefont {Forbes}, \citenamefont {Bucksbaum},\ and\ \citenamefont {Faria}}]{Werby2021}%
  \BibitemOpen
  \bibfield  {author} {\bibinfo {author} {\bibfnamefont {N.}~\bibnamefont {Werby}}, \bibinfo {author} {\bibfnamefont {A.~S.}\ \bibnamefont {Maxwell}}, \bibinfo {author} {\bibfnamefont {R.}~\bibnamefont {Forbes}}, \bibinfo {author} {\bibfnamefont {P.~H.}\ \bibnamefont {Bucksbaum}},\ and\ \bibinfo {author} {\bibfnamefont {C.~F. d.~M.}\ \bibnamefont {Faria}},\ }\bibfield  {title} {\bibinfo {title} {Dissecting subcycle interference in photoelectron holography},\ }\href {https://doi.org/10.1103/PhysRevA.104.013109} {\bibfield  {journal} {\bibinfo  {journal} {Phys. Rev. A}\ }\textbf {\bibinfo {volume} {104}},\ \bibinfo {pages} {013109} (\bibinfo {year} {2021})}\BibitemShut {NoStop}%
\bibitem [{\citenamefont {Werby}\ \emph {et~al.}(2022)\citenamefont {Werby}, \citenamefont {Maxwell}, \citenamefont {Forbes}, \citenamefont {Faria},\ and\ \citenamefont {Bucksbaum}}]{Werby2022}%
  \BibitemOpen
  \bibfield  {author} {\bibinfo {author} {\bibfnamefont {N.}~\bibnamefont {Werby}}, \bibinfo {author} {\bibfnamefont {A.~S.}\ \bibnamefont {Maxwell}}, \bibinfo {author} {\bibfnamefont {R.}~\bibnamefont {Forbes}}, \bibinfo {author} {\bibfnamefont {C.~F. d.~M.}\ \bibnamefont {Faria}},\ and\ \bibinfo {author} {\bibfnamefont {P.~H.}\ \bibnamefont {Bucksbaum}},\ }\bibfield  {title} {\bibinfo {title} {Probing two-path electron quantum interference in strong-field ionization with time-correlation filtering},\ }\href {https://doi.org/10.1103/PhysRevA.106.033118} {\bibfield  {journal} {\bibinfo  {journal} {Phys. Rev. A}\ }\textbf {\bibinfo {volume} {106}},\ \bibinfo {pages} {033118} (\bibinfo {year} {2022})}\BibitemShut {NoStop}%
\bibitem [{\citenamefont {Arbo}\ \emph {et~al.}(2012)\citenamefont {Arbo}, \citenamefont {Ishikawa}, \citenamefont {Persson},\ and\ \citenamefont {Burgdorfer}}]{Arbo2012}%
  \BibitemOpen
  \bibfield  {author} {\bibinfo {author} {\bibfnamefont {D.~G.}\ \bibnamefont {Arbo}}, \bibinfo {author} {\bibfnamefont {K.~L.}\ \bibnamefont {Ishikawa}}, \bibinfo {author} {\bibfnamefont {E.}~\bibnamefont {Persson}},\ and\ \bibinfo {author} {\bibfnamefont {J.}~\bibnamefont {Burgdorfer}},\ }\bibfield  {title} {\bibinfo {title} {{Doubly differential diffraction at a time grating in above-threshold ionization: Intracycle and intercycle interferences}},\ }\href {https://doi.org/10.1016/j.nimb.2011.10.030} {\bibfield  {journal} {\bibinfo  {journal} {Nucl. Instrum. Methods B}\ }\textbf {\bibinfo {volume} {279}},\ \bibinfo {pages} {24} (\bibinfo {year} {2012})},\ \Eprint {https://arxiv.org/abs/0912.5470} {arXiv:0912.5470} \BibitemShut {NoStop}%
\bibitem [{\citenamefont {Ja{\v{s}}arevi{\'{c}}}\ \emph {et~al.}(2020)\citenamefont {Ja{\v{s}}arevi{\'{c}}}, \citenamefont {Hasovi{\'{c}}}, \citenamefont {Kopold}, \citenamefont {Becker},\ and\ \citenamefont {Milo{\v{s}}evi{\'{c}}}}]{Jasarevic2020}%
  \BibitemOpen
  \bibfield  {author} {\bibinfo {author} {\bibfnamefont {A.}~\bibnamefont {Ja{\v{s}}arevi{\'{c}}}}, \bibinfo {author} {\bibfnamefont {E.}~\bibnamefont {Hasovi{\'{c}}}}, \bibinfo {author} {\bibfnamefont {R.}~\bibnamefont {Kopold}}, \bibinfo {author} {\bibfnamefont {W.}~\bibnamefont {Becker}},\ and\ \bibinfo {author} {\bibfnamefont {D.~B.}\ \bibnamefont {Milo{\v{s}}evi{\'{c}}}},\ }\bibfield  {title} {\bibinfo {title} {Application of the saddle-point method to strong-laser-field ionization},\ }\href {https://doi.org/10.1088/1751-8121/ab749b} {\bibfield  {journal} {\bibinfo  {journal} {J. Phys. A}\ }\textbf {\bibinfo {volume} {53}},\ \bibinfo {pages} {125201} (\bibinfo {year} {2020})}\BibitemShut {NoStop}%
\bibitem [{\citenamefont {Habibovi\'{c}}\ \emph {et~al.}(2021)\citenamefont {Habibovi\'{c}}, \citenamefont {Becker},\ and\ \citenamefont {Milo\v{s}evi\'{c}}}]{Habibovic2021}%
  \BibitemOpen
  \bibfield  {author} {\bibinfo {author} {\bibfnamefont {D.}~\bibnamefont {Habibovi\'{c}}}, \bibinfo {author} {\bibfnamefont {W.}~\bibnamefont {Becker}},\ and\ \bibinfo {author} {\bibfnamefont {D.~B.}\ \bibnamefont {Milo\v{s}evi\'{c}}},\ }\bibfield  {title} {\bibinfo {title} {Symmetries and selection rules of the spectra of photoelectrons and high-order harmonics generated by field-driven atoms and molecules},\ }\href {https://doi.org/10.3390/sym13091566} {\bibfield  {journal} {\bibinfo  {journal} {Symmetry}\ }\textbf {\bibinfo {volume} {13}},\ \bibinfo {pages} {1566} (\bibinfo {year} {2021})}\BibitemShut {NoStop}%
\bibitem [{\citenamefont {B\"oning}\ \emph {et~al.}(2019)\citenamefont {B\"oning}, \citenamefont {Paufler},\ and\ \citenamefont {Fritzsche}}]{Boning2019}%
  \BibitemOpen
  \bibfield  {author} {\bibinfo {author} {\bibfnamefont {B.}~\bibnamefont {B\"oning}}, \bibinfo {author} {\bibfnamefont {W.}~\bibnamefont {Paufler}},\ and\ \bibinfo {author} {\bibfnamefont {S.}~\bibnamefont {Fritzsche}},\ }\bibfield  {title} {\bibinfo {title} {Nondipole strong-field approximation for spatially structured laser fields},\ }\href {https://doi.org/10.1103/PhysRevA.99.053404} {\bibfield  {journal} {\bibinfo  {journal} {Phys. Rev. A}\ }\textbf {\bibinfo {volume} {99}},\ \bibinfo {pages} {053404} (\bibinfo {year} {2019})}\BibitemShut {NoStop}%
\bibitem [{\citenamefont {Lai}\ \emph {et~al.}(2015{\natexlab{b}})\citenamefont {Lai}, \citenamefont {Poli}, \citenamefont {Schomerus},\ and\ \citenamefont {{Figueira De Morisson Faria}}}]{Lai2015a}%
  \BibitemOpen
  \bibfield  {author} {\bibinfo {author} {\bibfnamefont {X.~Y.}\ \bibnamefont {Lai}}, \bibinfo {author} {\bibfnamefont {C.}~\bibnamefont {Poli}}, \bibinfo {author} {\bibfnamefont {H.}~\bibnamefont {Schomerus}},\ and\ \bibinfo {author} {\bibfnamefont {C.}~\bibnamefont {{Figueira De Morisson Faria}}},\ }\bibfield  {title} {\bibinfo {title} {{Influence of the Coulomb potential on above-threshold ionization: A quantum-orbit analysis beyond the strong-field approximation}},\ }\href {https://doi.org/10.1103/PhysRevA.92.043407} {\bibfield  {journal} {\bibinfo  {journal} {Phys. Rev. A}\ }\textbf {\bibinfo {volume} {92}},\ \bibinfo {pages} {1} (\bibinfo {year} {2015}{\natexlab{b}})}\BibitemShut {NoStop}%
\end{thebibliography}%
\end{document}